\documentclass{SciPost}

\hypersetup{
    colorlinks,
    linkcolor={red!50!black},
    citecolor={blue!50!black},
    urlcolor={blue!80!black}
}

\usepackage[bitstream-charter]{mathdesign}
\usepackage{float}
\usepackage{subcaption}

\DeclareSymbolFont{usualmathcal}{OMS}{cmsy}{m}{n}
\DeclareSymbolFontAlphabet{\mathcal}{usualmathcal}

\begin{document}

\begin{center}{\Large \textbf{\color{scipostdeepblue}{
%%%%%%%%%% TODO: Write your article's title here
Local equations for the generalized Lotka-Volterra model on sparse asymmetric graphs\\
%%%%%%%%%% END TODO: TITLE
}}}\end{center}

\begin{center}\textbf{
%%%%%%%%%% TODO: AUTHORS
% Write the author list here. 
% Use (full) first name (+ middle name initials) + surname format.
% Separate subsequent authors by a comma, omit comma and use "and" for the last author.
% Mark the corresponding author(s) with a superscript symbol in this order
% \star, \dagger, \ddagger, \circ, \S, \P, \parallel, ...
D. Machado\textsuperscript{1, 2, 3, $\star$},
P. Valigi\textsuperscript{1, 4},
T. Tonolo\textsuperscript{5, 6} and
M. C. Angelini\textsuperscript{1, 7, $\dagger$}
%%%%%%%%%% END TODO: AUTHORS
}\end{center}

\begin{center}
%%%%%%%%%% TODO: AFFILIATIONS
% Write all affiliations here.
% Format: institute, city, country
{\bf 1} Dipartimento di Fisica, Sapienza Università di Roma, P.le Aldo Moro 5, 00185 Rome, Italy
\\
{\bf 2} Group of Complex Systems and Statistical Physics and  Department of Theoretical Physics. Physics Faculty, University of Havana. CP10400, La Habana, Cuba
\\
{\bf 3} CNR - Nanotec, unit\`a di Roma, P.le Aldo Moro 5, 00185 Rome, Italy
\\
{\bf 4} Department of Mathematical Sciences, University of Bath, BA2 7AY Bath, United Kingdom
\\
{\bf 5} Gran Sasso Science Institute, Viale F. Crispi 7, 67100 L’Aquila, Italy
\\
{\bf 6} INFN-Laboratori Nazionali del Gran Sasso, Via G. Acitelli 22, 67100 Assergi (AQ), Italy
\\
{\bf 7} Istituto Nazionale di Fisica Nucleare, Sezione di Roma I, P.le A. Moro 5, 00185 Rome, Italy
%%%%%%%%%% END TODO: AFFILIATIONS
%%%%%%%%%% TODO: EMAIL
% Provide email address of corresponding author(s)
\\[\baselineskip]
$\star$ \href{mailto:david.machado@uniroma1.it}{\small david.machado@uniroma1.it}\,,\quad
$\dagger$ \href{mailto:mariachiara.angelini@uniroma1.it}{\small mariachiara.angelini@uniroma1.it}
%%%%%%%%%% END TODO: EMAIL
\end{center}

\section*{\color{scipostdeepblue}{Abstract}}
\textbf{\boldmath{%
%%%%%%%%%% TODO: ABSTRACT
% Write your abstract here.
Real ecosystems are characterized by sparse and asymmetric interactions, posing a major challenge to theoretical analysis. We introduce a method to study the generalized Lotka-Volterra model with stochastic dynamics on sparse graphs. By deriving local Fokker-Planck equations and employing a mean-field closure, we can efficiently compute stationary states for both symmetric and asymmetric interactions. We validate our approach by comparing the results with the direct integration of the dynamical equations and by reproducing known results, and we map the phase diagram for sparse asymmetric networks. Our framework provides a versatile tool for exploring stability in realistic ecological communities and can be generalized to applications in different contexts, such as economics and evolutionary game theory.
%%%%%%%%%% END TODO: ABSTRACT
}}

\vspace{\baselineskip}

\vspace{10pt}
\noindent\rule{\textwidth}{1pt}
\tableofcontents
\noindent\rule{\textwidth}{1pt}
\vspace{10pt}
%%%%%%%%%% END TODO: TOC

%%%%%%%%% TODO: CONTENTS 
% Write your article contents here, starting from first \section.
% An example structure is given below.

\makeatletter
\renewcommand{\doi}[1]{\href{https://doi.org/#1}{doi: #1}}
\makeatother

\newcommand{\iu}{\mkern1mu\mathrm{i}\mkern1mu}
\newcommand{\ec}{\mkern1mu\mathrm{e}\mkern1mu}
\newcommand{\signop}{{\rm sign}}
\newcommand{\pv}[1]{{\color[RGB]{119,7,55}\textbf{[PV: #1]}}}
\newcommand{\david}[1]{{\color{red}{\textbf{[D: #1]}}}}
\newcommand{\mc}[1]{{\color{magenta}{\textbf{[MC: #1]}}}}
\newcommand{\tom}[1]{{\color[RGB]{212,175,55}\textbf{[TT: #1]}}}

\section{Introduction}
The stability of complex ecosystems and the rules governing species coexistence present a central puzzle in theoretical ecology. The generalized Lotka-Volterra (gLV) model has been a cornerstone of this inquiry, also related to models used in evolutionary game theory and in economic theory \cite{moran2019may, galla2013complex, garnier2021new}. Recently, the random, symmetric, and dense version of the gLV model has been analyzed through the lens of equilibrium statistical mechanics. This approach has yielded profound insights, revealing how the phases of single equilibrium, unbounded growth, and multiple equilibria are dictated by a handful of macroscopic parameters: the mean $\mu$ and variance $\sigma$ of inter-species interactions, and the intensity of demographic noise $T$. A key finding is that densely connected networks with random symmetric interactions can exhibit a phase of multiple equilibria at high interaction heterogeneity \cite{Diederich_PRA_1989, altieri2021properties}.

However, real ecological networks are generally neither dense nor symmetric. In real ecosystems, the interactions between two species are almost always asymmetric, the existence of predator-prey couples of species being just an example. For this reason, some other works have tried to go beyond the assumption of symmetry \cite{bunin2017ecological, galla2018dynamically, ros2023generalized, ros2023quenched, poley2023generalized, baron2023breakdown, hatton2024diversity, barbier2018generic, giral2024interplay, pearce2020stabilization}. Moreover, in real ecosystems, a species typically interacts only with a few others \cite{dunne2002food, busiello2017explorability}. In the last years, Random Matrix Theory (RMT) results~\cite{mambuca2022dynamical, valigi2024localsignstability} have shown that the spectra of sparse random graphs exhibit qualitative differences from the dense case, suggesting that this may have implications for the stability of ecological models defined on sparse networks. Accordingly, it would be interesting to study the properties of the gLV model on graphs with finite connectivity. Unfortunately, in this case, the methods used for fully-connected systems  \cite{Diederich_PRA_1989, altieri2021properties, ros2023generalized} cannot be applied, in particular because no central-limit-type arguments hold.

In a very recent paper, the equilibrium properties of the symmetric gLV model on a sparse graph were analyzed \cite{Tonolo2026sparse} using the so-called Belief-Propagation (BP) cavity method. This can, however, be used only when an equilibrium measure exists, preventing its implementation in the case of asymmetric interactions. In general, the description of out-of-equilibrium systems has been even more elusive, and the available techniques deal only with specific limits. When species interact through a fully-connected network~\cite{Opper_PRL_1992, Roy_2019}, we can use Dynamical Mean-Field Theory (DMFT) to describe the temporal evolution of dynamic observables, which has also been extended to non-Gaussian disorder in the interactions \cite{AzaeleMaritan_PRL_2024}. Other generalizations include the limit of very small connectivity \cite{AzaeleMaritan_PRL_2024}, the limit of a large connectivity that grows sublinearly with the number of species in the graph \cite{AguirreLopez_2024}, and the case of unidirectional interactions \cite{metz2025dynamical}. However, until now, there has been no way to systematically analyze gLV models with sparse asymmetric interactions.

In this work, we bridge this gap by analyzing both symmetric and asymmetric, quenched-disordered sparse interactions. We go beyond equilibrium statistical mechanics, introducing a method for the evaluation of the stationary probability distribution for the stochastic differential equations (SDE) that describe the evolution of the species abundance. We start from the usual formulation of the stochastic dynamics in terms of an SDE and derive the equivalent Fokker-Planck equation for the associated time-dependent probability densities. However, solving the full system of partial differential equations defined on the whole graph in a high-dimensional space is a cumbersome task. To overcome this difficulty, we derive local closures that allow us to obtain tractable relations for the stationary distributions. 

The main idea of a local closure is to propose an \emph{ansatz} for the probability densities, in general involving some suitable factorizations that exploit the properties of the interaction graph. It has been used successfully in several contexts, such as the study of epidemics spreading on networks \cite{WangIBMF2003, cator2012second, pastor2015epidemic, CME-SIS}, algorithmic dynamics in hard combinatorial optimization problems \cite{CME-PRL, CDA_FMS_arxiv}, spin-glass dynamics in random graphs \cite{CDA-Pspin}, or the dynamics of the voter model \cite{PairApproxVoter_PRE_2022}. As far as we know, this work constitutes the first application of local closures to Fokker-Planck equations in sparse graphs. The approximate descriptions derived here are what we call local Fokker-Planck equations. 

We will validate our method by applying it to different situations, comparing its prediction with the results obtained from the direct numerical integration of the SDE and recovering known results from previous literature. 
We also show how, starting from the general equations in the asymmetric case, one can recover the BP equations when only symmetric couplings are considered.

The rest of the manuscript is organized as follows. In Section \ref{sec:GLV} we introduce the generalized Lotka-Volterra equations and the underlying networks that we will analyze. In Section \ref{sec:FokkerPlanck}, we present the Fokker-Planck equations that describe the evolution of the probability distributions of the species abundances in time. These are complicated global equations, which in general are not solvable. For this reason, in Section \ref{susec:IBMF_eqs} we introduce a local closure, corresponding to a mean-field approximation for the dynamics, called Individual Based Mean Field (IBMF). Our procedure leads to the main \textit{local} solvable Fokker-Planck equations that we discuss in this article. In Section \ref{subsec:PBMF_and_BP} we go beyond IBMF, introducing a more refined closed local Fokker-Planck equation that we call Pair Based Mean Field (PBMF), showing that the BP equations introduced in Ref.~\cite{Tonolo2026sparse} correspond to the stationary solution of PBMF in the symmetric case.

We validate our methods in Section \ref{sec:IBMFnum}, where we give details on the numerical implementation of IBMF both at null and finite temperatures. We compare the stationary abundances obtained with IBMF with those obtained from simulations in the presence of thermal noise for a single random graph. The results are a useful example to emphasize the strengths of IBMF, and to also point out its limitations. Then, we proceed to apply IBMF to three different scenarios. In Subsection \ref{subsec:IBMF_asym}, we study undirected graphs with asymmetric interactions at null temperature, obtaining the corresponding phase diagram in the plane $(\mu,\sigma)$ for the first time, as far as we know. We thus generalize the results for the fully-connected asymmetric case in Ref.~\cite{bunin2017ecological} and for the sparse case with symmetric interactions in Ref.~\cite{Tonolo2026sparse}. In Subsection \ref{subsec:IBMF_directed}, we apply the IBMF closure to directed graphs with null variance in the couplings and null temperature, confirming and extending the results of Ref.~\cite{StavPLOS2022}, which were obtained there with a completely different method. In Subsection \ref{subsec:IBMFsym} we move back to undirected graphs, but this time with symmetric couplings at finite $T$. This setting helps us study the performance of IBMF in the presence of thermal noise in a systematic way. We compare the results with Ref.~\cite{Tonolo2026sparse}, where the BP method is used to exactly solve the model. We identify the limitations of IBMF, which stops converging to a single equilibrium as soon as the exact species abundances found by BP start developing a non-Gaussian distribution tilted towards extinctions. Finally, in Section \ref{sec:discussion}, we draw our conclusions.

\section{The model} \label{sec:GLV}

Let us introduce the generalized Lotka-Volterra (gLV) equations that we will study in the rest of this article. They describe the dynamics of an ecosystem with $N$ interacting species. To each of them, we associate a positive real variable $n_i$, interpreted as the abundance of the $i$-th species, with $i=1, \ldots, N$. In general, a single species will not interact with all the others, but instead with a subset of the species known as the neighborhood of $i$. The interactions occur in a graph $G(V, E)$, where $V$ is the set of vertices, each representing a species, and $E$ is the set of edges. 

To keep the definitions as general as needed, for now the reader should think of $G$ as a directed graph. If the presence of species $i$ influences the growth of species $j$, we add the directed edge $i \to j$. It is possible to have $i \to j$ in the graph without having the edge in the opposite direction ($j \to i$). We define the in-neighborhood $\partial i^{-}$ of $i$ as the set of in-neighbors $j$ such that the edge $j \to i$ exists in the graph. For simplicity, graphs $G$ with self-loops will not be considered here

The gLV equation for the abundance of the $i$-th species can be written as:

\begin{equation}
\frac{dn_i}{dt} = \frac{r_i}{K_i} \, n_i(K_i - n_i - \sum_{j \in \partial i^{-}} \alpha_{ij} n_j) + \xi_i(t) + \lambda \: ,
 \label{eq:LVmodel}
\end{equation}
where $n_i \geq 0$ is the abundance of the $i$-th species, and the real parameters $r_i$ and $K_i$ are known as the intrinsic growth rate and carrying capacity, respectively. To simplify the setting, we will take $r_i=K_i=1$ in what follows, but the reader will find no difficulties in generalizing our results to consider other values of these constants. 

The term $\xi_i(t)$ in Eq.~\eqref{eq:LVmodel} is a noise term, which has average $\langle \xi_i(t)\rangle = 0$ and second moments $\langle \xi_i(t_1) \, \xi_j(t_2) \rangle = 2 \, T \, n_i \, \delta_{i,j} \, \delta(t_1-t_2)$, where $T$ is known as temperature of the noise. This thermal noise is referred to as \emph{demographic}~\cite{loreau2013biodiversity, biroli2018marginally, altieri2021properties, larroya2023demographic, lorenzana2024interactions} and accounts for death and birth processes.
The parameter $\lambda$, known as \emph{immigration rate}, acts as a small source term that allows extinct species to come back should conditions become favorable to them~\cite{Roy_2019, arnoulx2024many}. Its effect will be clarified later.

The couplings $\alpha_{ij}$ are real numbers that set the type and strength of the interactions. The value of $\alpha_{ij}$ encodes the way that species $j$ affects the evolution of species $i$, and therefore corresponds to the edge $j \to i$ on the graph. In the case where for all $j \in \partial i^{-}$ we also have the edge $i \to j$, the graph is known as undirected \footnote{Sometimes the term undirected is used for graphs with a symmetric interaction matrix. Here, we prefer to use the term for graphs with a symmetric adjacency matrix, whatever the interaction matrix on top of it will be.}. Having $\alpha_{ij}$ and $\alpha_{ji}$ simultaneously positive means that the two species $i$ and $j$ have a competitive interaction, where the presence of individuals of species $j$ is prejudicial for the individuals of species $i$, and vice versa. When they are both negative, we have a mutualistic interaction, and the species are beneficial to each other. On the other hand, when the interaction is positive for one species and negative for the other, we have a predator-prey or antagonistic interaction. Finally, in the presence of directed interactions, we can also have commensalism and amensalism, whereby one species benefits or is harmed by the interaction, while the other is unaffected.

Given that the edge $j \to i$ already exists in the graph, we can add noise to the interactions by drawing $\alpha_{ij}$ at random from some probability distribution. Following many other works \cite{Roy_2019, AltieriSciPost2022, Tonolo2026sparse}, we choose the Gaussian distribution $ \alpha_{ij} \sim  \mathcal{N}(\mu, \sigma)$, with mean $\mu$ and variance $\sigma^{2}$. The reader should note that, as a particular case, we can set the interaction strengths $\alpha_{ij}$ to be homogeneous by choosing $\sigma=0$. In that case, we get $\alpha_{ij}=\mu$ for all the edges $j \to i$.

Our methodology, derived below in Section \ref{sec:FokkerPlanck}, applies to graphs with directed and/or undirected interactions. We demonstrate this by including results for three different scenarios in Section \ref{sec:IBMFnum}. In Subsection \ref{subsec:IBMF_directed} we study a case where, with high probability, the edges $j \to i$ and $i \to j$ are not simultaneously present. We follow the same model used in Ref.~\cite{StavPRE2024} to study the gLV dynamics with asymmetric interactions. To construct the network, for each species $i$ we select the incoming edges by going over all possible $j \neq i$, and adding the edge $j \neq i$ with probability $c / N$, where $c>0$. The in-neighbors of $i$ are chosen independently of the in-neighbors of $j$, and in the limit when the number of species is large, it is highly improbable that we find $i \to j$ and $j \to i$ simultaneously in the graph. The result is a graph where the degree follows a Poisson distribution with mean $c$, and where most interactions are directed. Finally, for each edge $j \to i$ we draw $\alpha_{ij}$ from the Gaussian distribution $\mathcal{N}(\mu, \sigma)$. In Subsection \ref{subsec:IBMF_directed} below, we include results for different values of $\mu$ but only two values of $\sigma$ ($\sigma=0$ and $\sigma=0.15$).

In Fig.~\ref{fig:individual_abundances}, and in Subsections \ref{subsec:IBMF_asym} and \ref{subsec:IBMFsym}, the species interact over an undirected random regular graph, whose edges are randomly selected such that each species (vertex) has the same number of neighbors, denoted by $c$ and called connectivity. After the graph is built, we need to choose the interaction strengths $\alpha_{ij}$ and $\alpha_{ji}$ for each edge. In Subsection \ref{subsec:IBMF_asym}, we draw $\alpha_{ij}$ independently of $\alpha_{ji}$ using the Gaussian distribution $\mathcal{N}(\mu, \sigma)$, for different values of $\mu$ and $\sigma$. Notice that this creates asymmetric interactions where, in general, we have $\alpha_{ij}\neq\alpha_{ji}$ whenever $\sigma \neq 0$. In Subsection \ref{subsec:IBMFsym} we study the case $\sigma=0$, where we always get $\alpha_{ij}=\alpha_{ji}=\mu$ and the interactions are symmetric.

\section{Local Fokker-Planck equations} \label{sec:FokkerPlanck}

Given a graph of interactions, Eq.~\eqref{eq:LVmodel} gives the temporal evolution of the abundances in a stochastic process with thermal noise $\xi(t)$. Sampling different realizations of the initial conditions and of $\xi(t)$, one gets the probability distribution $P_t(\vec{n})$ of the vector $\vec{n}=(n_1, \ldots, n_N)$ at time $t$. This quantity obeys a Fokker-Planck equation that can be derived from Ito's rule \cite{AltieriSciPost2022}:

\begin{eqnarray}
 & & \frac{\partial P_t(\vec{n})}{\partial t} =  T \sum_{i=1}^{N} \frac{\partial^{2}}{\partial n_i^{2}} \big\{ n_i P_t(\vec{n}) \big\}    -\sum_{i=1}^{N}  \frac{\partial }{\partial n_i} \Big\{ \big[n_i(1 - n_i - \sum_{j \in \partial i^{-}} \alpha_{ij} n_j) + \lambda \big] P_t(\vec{n})  \Big\} , \label{eq:FokkerPlanck}
\end{eqnarray}

The first and second terms on the right-hand side of Eq.~\eqref{eq:FokkerPlanck} are the usual diffusion and drift terms of the Fokker-Planck equation, respectively. They encode the evolution of a species subject to Eq.~\eqref{eq:LVmodel}. The deterministic growth ratio $n_i(1-n_i-\sum_{j \in \partial i^{-}} \alpha_{ij} \,n_j)+\lambda$ experienced by species $i$ goes into the drift term. The thermal noise with temperature $T$ gives birth to the diffusion term. 

Contrary to the usual case where the variables are defined in the interval $(-\infty, +\infty)$, each abundance $n_i$ is defined in the interval $[0, +\infty)$. As a natural consequence of this fact, the procedure to obtain the Fokker-Planck equation from Ito's rule gives some extra surface terms that are not considered in Eq. \eqref{eq:FokkerPlanck} (see Appendix \ref{app:derivation_gen_FP} for more details). We thus impose a reflecting boundary condition \cite{Gardiner2009stochastic}, the proper boundary conditions at $n_i=0$, guaranteeing that the current of probability density through the border is always zero. More specifically, by enforcing that $(T-\lambda) \lim_{n_i \to 0} P(\vec{n})=0$ for all species $i$, we neglect all the surface terms, and Eq. \eqref{eq:FokkerPlanck} follows.

In any case, solving Eq.~\eqref{eq:FokkerPlanck} is a cumbersome task mainly because $P_t(\vec{n})$ is a highly dimensional object. The abundances are defined on the space $(0, +\infty)^{N}$ and the time can be in general defined in the space $(-\infty, \infty)$. Even when we consider a single species ($N=1$), finding $P_t(n)$ at any time $t$ is not simple. However, we can obtain its stationary solution (see Appendix \ref{app:FokkerPlanck_single}), which will be useful for us later. It reads:

\begin{equation}
P_{\infty}(n) = \frac{1}{Z} \, n^{\beta \lambda - 1} \, \exp \Big\{ -\frac{\beta}{2} \,(n - 1)^{2} \Big\},
 \label{eq:sol_single_var}
\end{equation}
where $Z$ is a normalization constant and $\beta \equiv 1/T$.

Eq.~\eqref{eq:sol_single_var} clarifies the role of the parameter $\lambda$ in the model. The integral $\int_{0}^{\infty} dn\,  P_{\infty}(n)$ is finite if and only if $\lambda > 0$. Otherwise, the divergence at $n=0$ dominates the integral, which would be divergent. In other words, the existence of $\lambda>0$ allows the density $P_{\infty}(n)$ to be normalizable. On the other hand, when $\lambda=0$ and $T>0$ the species are doomed to go extinct for large times.
 
\subsection{Individual Based Mean Field} \label{susec:IBMF_eqs}
To find solvable equations, we need to simplify Eq.~\eqref{eq:FokkerPlanck}. In this subsection, we obtain the first local Fokker-Planck equation for the gLV model. Let us marginalize Eq.~\eqref{eq:FokkerPlanck} over all the abundances except $n_i$ to obtain the differential equations for the local probabilities, which are defined as $P_t(n_i) = \int_0^{\infty} [\,\prod_{k\neq i} dn_k] P_t(\vec{n})$:

\begin{eqnarray}
 & & \frac{\partial P_t(n_i)}{\partial t} =  T \frac{\partial^{2}}{\partial n_i^{2}} \big\{ n_i P_t(n_i) \big\} \!    -\frac{\partial }{\partial n_i} \Big\{ \big[ n_i \big(1 - n_i - \!\! \sum_{j \in \partial i^{-}} \! \alpha_{ij} \, m_{j \to i}(n_i, t) \big) + \lambda \big] P_t(n_i) \Big\} ,  \label{eq:der_Pi_final}
\end{eqnarray}
where $m_{j \to i}(n_i)$ is the conditional average 

\begin{equation}
 m_{j \to i}(n_i, t) \equiv \int_{0}^{\infty} dn_j \,  n_j P_t(n_j \mid n_i). \label{eq:def_mji}
\end{equation}

As in the Fokker-Planck equation for the whole system, the local version in Eq.~\eqref{eq:der_Pi_final} has two different contributions. The first line in the equation shows the diffusion term. On the other hand, after averaging over the rest of species, the single species $i$ senses an effective drift $n_i \big(1 - n_i - \!\! \sum_{j \in \partial i^{-}} \! \alpha_{ij} \, m_{j \to i}(n_i, t) \big) + \lambda$, where $n_j$ is substituted by its conditional average $m_{j \to i}(n_i, t)$. To get Eq.~\eqref{eq:der_Pi_final}, we again need the reflecting boundary conditions of the full Fokker-Planck equation \eqref{eq:FokkerPlanck}, which translate to enforcing that $(T-\lambda) \lim_{n_i \to 0} P(n_i)=0$ for all $i$. For more details on the derivation, the reader is referred to Section 1 of the Supplemental Materials (SM).  

We have not introduced any approximation so far. To solve Eq.~\eqref{eq:der_Pi_final}, one would also need to obtain all the functions $m_{j \to i}(n_i)$, but from its definition (Eq.~\eqref{eq:def_mji}) it is evident that this is equivalent to getting the solution for the pair probabilities $P_t(n_i, n_j)$. Indeed, to compute the conditional probability density $P_t(n_j \mid n_i)$, we need the pair $P_t(n_i, n_j)$ and the single-site $P_t(n_i)$ probabilities. As we will show in the next section, the local Fokker-Planck equation for $P_t(n_i, n_j)$ depends, in turn, on probabilities $P_t(n_i, n_j, n_k)$ defined over three species. After iterating this process, we get a hierarchy of equations that never closes until we recover the full Eq.~\eqref{eq:FokkerPlanck}. Therefore, solving Eq.~\eqref{eq:der_Pi_final} has the same level of difficulty as solving Eq.~\eqref{eq:FokkerPlanck}.

To overcome this problem, we need to introduce an approximation that allows us to get a \emph{closed} system of differential equations for the $P_t(n_i)$, \textit{i.e}, one that can be solved without going up in the hierarchy. The first step that one could take in that direction is to assume $m_{j \to i}(n_i, t)$ is independent of $n_i$ and write $m_{j\to i}(n_i, t) \approx m_{j}(t)$, where

\begin{equation}
 m_{j}(t) \equiv \int_{0}^{\infty} dn_j \,  n_j P_t(n_j). \label{eq:def_mj}
\end{equation}

Doing this is equivalent to assuming that the pair probabilities are all factorized such that $P_t(n_i, n_j) \approx P_t(n_i)P_t(n_j)$. Therefore, we are trivializing the correlations in the system. Nevertheless, this approximation allows us to close the system of differential equations since all the information that we need is in the single-species distributions $P_t(n_i)$. We get:

\begin{eqnarray}
 & & \frac{\partial P_t(n_i)}{\partial t}  = T \frac{\partial^{2}\big\{ n_i P_t(n_i) \big\}}{\partial n_i^{2}}  -\frac{\partial }{\partial n_i} \Big\{ \big[ n_i \big(1 - n_i - \!\! \sum_{j \in \partial i^{-}} \! \alpha_{ij} \, m_j(t) \big) + \lambda \big] P_t(n_i) \Big\}    .    \label{eq:IBMF_diff_eq}
\end{eqnarray}

These local Fokker-Planck equations form a dynamic closure that can in principle be solved, and that we will call Individual Based Mean Field (IBMF) in what follows. This name has been used before in the literature, particularly in the study of epidemic spreading throughout a network \cite{WangIBMF2003, pastor2015epidemic}, to identify an approximation that factorizes the pair probabilities distribution as explained above. Finding the stationary solution of Eq.~\eqref{eq:IBMF_diff_eq} has the same level of difficulty as for an isolated variable (see Eq.~\eqref{eq:sol_single_var} and Appendix \ref{app:FokkerPlanck_single}). The result is:

\begin{equation}
P_{\infty}(n_i) = \frac{1}{Z_i} \, n_i^{\beta \lambda - 1} \, \exp \Big\{ -\frac{\beta}{2} \,(n_i - M_i)^{2} \Big\},
 \label{eq:sol_IBMF}
\end{equation}
where 

\begin{eqnarray}
    Z_i \!\!\!\! &=& \!\!\!\! \int_{0}^{\infty} dn_i \, n_i^{\beta \lambda-1} \, \exp \Big\{ -\frac{\beta}{2} \,(n_i - M_i)^{2} \Big\} \label{eq:Z_IBMF} \\
    M_i \!\!\!\! &=& \!\!\!\! 1 - \sum_{j \in \partial i^{-}} \alpha_{ij}\, m_{j}(\infty) \:\:. \label{eq:M_IBMF}
\end{eqnarray}

This is a mean-field solution to the problem derived for sparse graphs with any type of interactions. In fact, a similar probability density has been recently introduced in Ref.~\cite{Pasqualini_2025} for the case of fully-connected models, a scenario where mean-field assumptions like these are more commonly used. The presence of other species modifies the center $M_i$ of the Gaussian in Eq.~\eqref{eq:sol_IBMF} to make $n_i$ align with the average effect of its neighbors. To evaluate the stationary solution of IBMF, we need to design an algorithm capable of computing the averages $m_j(\infty)$, which we will denote by $m_j$ for simplicity. Exploiting Eq.~\eqref{eq:sol_IBMF}, we can write:

\begin{equation}
m_i =  \frac{1}{Z_i} \int_{0}^{\infty} dn_i \, n_i^{\beta \lambda} \, \exp \Big\{ -\frac{\beta}{2} \,(n_i - M_i)^{2} \Big\} .
 \label{eq:m_IBMF}
\end{equation}

As said above, $Z_i$ and $M_i$ are functions of the averages $m_j$, with $j \in \partial i^{-}$. After making an initial guess for the average abundances $m_i$, with $i =1, \ldots, N$, we can use Eq.~\eqref{eq:m_IBMF} to update their values. Then, we iterate until all $m_i$ converge to the IBMF's prediction for the stationary average abundances. In practice, we employ numerical tricks, such as adding damping and using sequential updates, to aid this iterative process in reaching convergence (see Appendices \ref{app:parallel_up} and \ref{app:damping}).

By inspecting the stationary distribution in Eq. \eqref{eq:sol_IBMF}, we see that $P_{\infty}(0)=0$ for $\lambda > T$, which is consistent with the reflecting boundary condition imposed in obtaining the Fokker-Planck equations. For $0<\lambda<T$, on the contrary, we have that $P_{\infty}(n_i)$ diverges when $n_i$ approaches zero. Interestingly, the stationary distribution in Eq. \eqref{eq:sol_IBMF} is still normalizable for any $\lambda > 0$, independently of the temperature. Therefore, we find no obstacle in assuming Eq. \eqref{eq:sol_IBMF} to be a valid stationary solution of IBMF for any $\lambda > 0$. Our numerical implementation of the iterative process described above indeed converges both when $\lambda > T$ and when $0 < \lambda < T$.

In the process of obtaining the stationary solution of IBMF, we did not assume any particular structure of the graph, nor any specific type of interaction. It is in principle applicable to any directed or undirected graph, with symmetric or asymmetric interactions. However, one should expect better results when correlations are weak enough for the factorization $P_{\infty}(n_i, n_j) \approx P_{\infty}(n_i) \, P_{\infty}(n_j)$ to approximately hold. %The latter is intuitively more likely to happen in the case of asymmetric interactions.

\subsection{Continuous Belief Propagation} \label{subsec:PBMF_and_BP}

To go beyond IBMF, we need to include non-trivial correlations between pairs of interacting species. Therefore, it is reasonable to go up one level in the hierarchy and write the local Fokker-Planck equations for the pair probabilities $P_t(n_i, n_j) = \int_0^{\infty} [\,\prod_{k\neq i, j} dn_k] P_t(\vec{n})$. After marginalizing Eq.~\eqref{eq:FokkerPlanck} over the abundances of all the species (the details can be found in Section 1 of the SM), except for $i$ and $j$, we obtain a local equation that depends on the conditional averages:

\begin{equation}
    m_{k \to i,j}(n_i, n_j, t) \equiv \int_{0}^{\infty} dn_k \, n_k \, P_t(n_k \mid n_i, n_j) . \label{eq:cond_av_3}
\end{equation}

Solving for $m_{k \to i,j}(n_i, n_j, t)$ implies knowing the three-species probabilities $P_t(n_i, n_j, n_k)$, but we need to close the hierarchy at some point. However, if one focuses on random sparse graphs like random regular or Erdős-Rényi, the tree-like structure of the interactions makes it very unlikely that species $k$ interacts with $i$ and $j$ simultaneously, provided that $i$ and $j$ interact. If $k \in \partial i^{-}$, then with high probability $k \not\in \partial j^{-}$ and the only short path connecting $j$ with $k$ in the graph necessarily passes through $i$. We can approximately write the three-species probabilities as: $P_t(n_k, n_i, n_j) \approx P_t(n_k \mid n_i) \, P_t(n_i) \, P_t(n_j \mid n_i)$. Noting that $P_t(n_k \mid n_i, n_j)=P_t(n_k, n_i, n_j) / P_t(n_i, n_j)$ and putting this expression back into Eq. \eqref{eq:cond_av_3}, we get:

\begin{equation}
    m_{k \to i,j}(n_i, n_j, t) \approx \int_{0}^{\infty} dn_k \, n_k \, \frac{P_t(n_k \mid n_i) \, P_t(n_i) \, P_t(n_j \mid n_i)} {P_t(n_j, n_i)}  = \int_{0}^{\infty} dn_k \, n_k \, P_t(n_k \mid n_i) \label{eq:cond_av_approx}
\end{equation}

Recalling the definition of the conditional average $m_{k \to i}(n_i)$ in Eq. \eqref{eq:def_mji}, we can see this is equivalent to $m_{k \to i,j}(n_i, n_j, t) \approx m_{k \to i}(n_i, t)$. In other words, we assume that $m_{k \to i,j}(n_i, n_j, t)$ strongly depends on $n_i$ and only weakly depends on $n_j$. The result (see Section 1 of the SM) is another closed local Fokker-Planck equation that we call Pair Based Mean Field (PBMF), also using a name that is popular in the context of epidemics spreading on networks when the dynamics is described using pair probabilities~\cite{cator2012second, pastor2015epidemic}. 

Although the PBMF is simpler than the full Eq.~\eqref{eq:FokkerPlanck}, it is still difficult to obtain a general solution, even when we focus only on the stationary point. Only in the case of symmetric interactions in undirected graphs ($\alpha_{ij}=\alpha_{ji}$), as we show in Subsection 2.1 of the SM, the proper solution to the local Fokker-Planck equation for pair probabilities in sparse random graphs is Belief Propagation (BP). This technique was already introduced in Ref.~\cite{Tonolo2026sparse} for the gLV model. However, while in that case the abundances of the species are considered as discrete variables with states $n_i=1,2,\ldots$, here we use the continuous version of the model. We solve this issue by proposing a continuous implementation of BP equations for our model, which have the same structure as the ones in Ref.~\cite{Tonolo2026sparse}, but include the proper adjustments to consider continuous $n_i \in [0, +\infty)$. These are:

\begin{eqnarray}
    \eta_{i \to j}(n_i)=\frac{n_i^{\beta \lambda -1}}{z_{i \to j}} \,  \exp \Big\{-\frac{\beta}{2}(n_i^{2} - 2 n_i) \Big\} \prod_{k \in \partial i^{-} \setminus j } \int_{0}^{\infty} dn_k \, \eta_{k \to i}(n_k) \, e^{- \beta \alpha_{ik} n_i \, n_k} \:\:\:, \label{eq:BP_up}
\end{eqnarray}
where $\eta_{i \to j}(n_i)$ is the cavity marginal, or message, that represents the marginal probability density of species $i$ in a modified graph where the edge connecting $i$ and $j$ is removed. The constant $z_{i \to j}$ is a normalization factor. 

Here, we can identify that the local field $h_i$, whose exponential $e^{-\beta \, h_i(n_i)}$ usually appears in front of BP equations, is simply $h_i=n_i^{2}/2-n_i+(T-\lambda) \, \ln(n_i)$. This expression properly considers the immigration rate and the continuous nature of $n_i$. From it, we can obtain the stationary single-site and the pair probabilities as follows:

\begin{eqnarray}
    P_{BP}(n_i) \!\!\!\! &=& \!\!\!\! \frac{n_i^{\beta \lambda -1}}{Z_{i}} \,  \exp \Big\{-\frac{\beta}{2}(n_i^{2} - 2 n_i) \Big\}  \prod_{k \in \partial i^{-} } \int_{0}^{\infty} dn_k \, \eta_{k \to i}(n_k) \, e^{- \beta \alpha_{ik} n_i \, n_k} \label{eq:PiBP} \\
    P_{BP}(n_i, n_j) \!\!\!\! &=& \!\!\!\! \frac{1}{Z_{ij}} \, \eta_{i \to j}(n_i) \, e^{-\beta \alpha_{ij} n_i \, n_j} \, \eta_{j \to i}(n_j) \: , \:\:\:\:\:\: \label{eq:PijBP}
\end{eqnarray}
where $Z_i$ and $Z_{ij}$ are normalization factors and the messages $\eta_{i \to j}(n_i)$ are the fixed point solution of Eq. \eqref{eq:BP_up}.

The reader could wonder what the relation is between BP and the local Fokker-Planck equations that we have been presenting here. Remarkably, it is possible to prove that the expression in Eq.~\eqref{eq:PijBP} for the pair probabilities, together with BP equations (Eq.~\eqref{eq:BP_up}), is a stationary solution of PBMF when the interactions are symmetric. The details of the proof are given in Subsection 2.1 of the SM. To obtain numerical results from BP, we use an iterative algorithm analogous to the one we introduced above for IBMF. Making an initial guess for the messages $\eta_{i \to j}(n_i)$, we can use Eq.~\eqref{eq:BP_up} to update their values at each $n_i$. This procedure is iterated until all $\eta_{i \to j}(n_i)$ converge, and the final messages are used to compute the true marginals $P_{BP}(n_i)$ and $P_{BP}(n_i, n_j)$. For the interested reader, we include details about our specific implementation of BP in Section 4 of the SM.

\section{Numerical results} \label{sec:IBMFnum}

As said above, to use IBMF to obtain the actual values of the averages $m_{i}$, we need to numerically compute integrals of the form:

\begin{equation}
I_{k}(\beta, \lambda, M) = \int_{0}^{\infty} \, dn \, n^{\beta \lambda - 1 + k} \, \exp \Big\{ -\frac{\beta}{2} \, (n - M)^{2} \Big\}  \: ,
 \label{eq:integral_k}
\end{equation}
with the parameter $k$ taking the value $k=0$ in Eq.~\eqref{eq:Z_IBMF}, and $k=1$ in Eq.~\eqref{eq:m_IBMF}.

Luckily, the integral in Eq.~\eqref{eq:integral_k} can be expressed in terms of known special functions, called parabolic cylinder functions (see 9.241 in Ref.~\cite{Gradshteyn7ed}). This is very convenient because we can write the parabolic cylinder functions in terms of the more practical Kummer's confluent hypergeometric function, which can be found already tabulated in different programming languages. The interested reader can find the details in Section 3 of the SM. The code is available at Ref.~\cite{GenLotkaVolterra_SparseGraphs}.

When the temperature is zero or close to zero, the equations can be simplified even further. The exponential in Eq.~\eqref{eq:integral_k} concentrates around its maximum, and provided that $\lambda$ is small, IBMF equations reduce to:

\begin{equation}
 n_i = \max \Big\{0, \, 1 - \sum_{j \in \partial i^{-}} \alpha_{ij} \, n_j \Big\}.\label{eq:IBMFT0}
\end{equation}

At $T=0$, we simply need to iterate Eq.~\eqref{eq:IBMFT0} until convergence. It is important to note that this is not the same as running the zero-temperature simulations of the dynamics, which in turn implies integrating the system of differential equations:
\begin{equation}
\frac{dn_i}{dt} = n_i(1 - n_i - \sum_{j \in \partial i^{-}} \alpha_{ij} n_j) + \lambda \:\:.
 \label{eq:LVmodel_T0}
\end{equation}
Such numerical integration of the gLV dynamical equations at $T=0$ can be performed using the Cash-Karp adaptive Runge-Kutta method~\cite{press1988numerical}. The code is available at Ref.~\cite{GenLotkaVolterra_SparseGraphs}.

For small $\lambda$, the fixed points of Eq.~\eqref{eq:IBMFT0} coincide with the stationary solutions of the exact dynamics in Eq.~\eqref{eq:LVmodel_T0}. Therefore, whenever IBMF converges, the fixed point represents an actual stationary configuration of the dynamics. On the other hand, the non-convergence of IBMF is not guaranteed to be reflected in the behavior of the simulated dynamics. One could think of them as two different algorithms trying to find the same fixed points. If one of the algorithms succeeds, the resulting abundances constitute a fixed point also for the other algorithm. If one of them does not succeed, the other still could. 

However, we show in Subsections \ref{subsec:IBMF_asym} and \ref{subsec:IBMF_directed} (see below) that IBMF can nevertheless be used to predict the relevant phase transitions observed in the simulations at $T=0$. In Appendix \ref{app:parallel_up} and for a random regular graph with homogeneous interactions, we show that the exact result for the transition from single-to-multiple equilibria \cite{StavPLOS2022} arises naturally from IBMF. Furthermore, in Section 2.2 of the SM, we recover the stationary solution of Dynamical Mean Field Theory \cite{Opper_PRL_1992, Roy_2019} in the limit of large connectivity at zero temperature. 

\begin{figure}[t]
\centering
\includegraphics[width=0.6\textwidth]{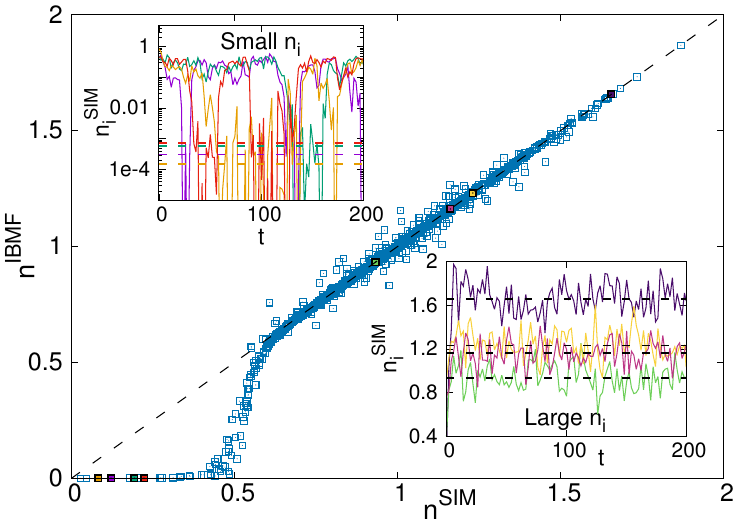}

\caption{Comparing individual abundances from IBMF and simulations in a random regular graph at finite temperature $T=0.015$. The connectivity is $c=3$, and the immigration rate is $\lambda=10^{-6}$. Each $\alpha_{ij}$ is independently drawn from the Gaussian $\mathcal{N}(0, \sigma)$, with $\sigma=0.15$ (the interactions are asymmetric). Each point in the main graphic has coordinates $(n_i^{\text{SIM}}, n_i^{\text{IBMF}})$, where $n_i^{\text{SIM}}$ is the average stationary abundance of species $i$ obtained from 100 simulations of the dynamics, and $n_i^{\text{IBMF}}$ is the prediction of IBMF for the same species. The black dashed line is just the linear function $f(x)=x$. The system has $N=1024$ species, thus there are $1024$ points in the main graphic. The inserted graphic in the top-left corner shows the temporal evolution of four species with small stationary abundances. The corresponding points are marked with the same colors in the main graphic. The horizontal lines are the predictions of IBMF for the same species. The graphic is in semi-log scale. Analogously, the inserted graphic in the bottom-right is done with four species whose abundances are not small. Colored lines show the results of simulations, and the horizontal black lines show the predictions made with IBMF.}
\label{fig:individual_abundances}
\end{figure}

In the presence of thermal noise ($T>0$), we still have a fast implementation of IBMF. It is important to note that, for finite temperatures, IBMF is always a factorized ansatz for the probability density of the abundances. With its stationary solution, we can predict the final average abundance $n_i^{\text{IBMF}}$ for each of the species in a given graph. To illustrate how this works, Fig.~\ref{fig:individual_abundances} compares each $n_i^{\text{IBMF}}$ with the average stationary abundances $n_i^{\text{SIM}}$ obtained from simulations for a specific realisation of a random regular graph in the presence of thermal noise. The numerical integration of the SDE defining the gLV model in Eq.~\eqref{eq:LVmodel} are performed by means of the Milstein method~\cite{milstein1975integration, kloeden_platen1992numerical}. As with the other algorithms, the code is provided in Ref.~\cite{GenLotkaVolterra_SparseGraphs}. The interactions are asymmetric, \textit{i.e.}, we choose $\alpha_{ij}$ independently of $\alpha_{ji}$, each from a Gaussian distribution $\mathcal{N}(0, \sigma)$. As far as we know, this is the first time a theoretical prediction of this kind has been made for sparse graphs.

The main graphic of Fig.~\ref{fig:individual_abundances} shows that IBMF accurately predicts the abundances of the species that are dominant in the ecosystem. For $n_i > 0.6$, the points $(n_i^{\text{SIM}}, n_i^{\text{IBMF}})$ lie around the line $f(x)=x$ of perfect agreement. Most species ($\sim 91\%$) are in this group. In the bottom-right corner of Fig.~\ref{fig:individual_abundances}, the inserted graphic shows the temporal evolution of four of those species observed in a single simulation. The corresponding points in the main graphic are marked using the same colors. The stationary abundances of those species, which were selected at random, oscillate around the corresponding predictions from IBMF (presented in black dashed lines). We observe almost perfect agreement between them.

On the other hand, IBMF consistently underestimates the stationary abundances obtained from simulations for species that are closer to extinction, with $n_i < 0.6$. Very few ($\sim 9 \%$) species are in this group. The results become clearer after analyzing the inserted graphic in the top-left of Fig.~\ref{fig:individual_abundances}. There, we show the temporal evolution of the abundances of four species that are very close to extinction according to IBMF, but whose average stationary abundance from simulations is not as small. The corresponding points in the main graphic are marked using the same colors. 

The inserted graphic shows that these species in the lower bottom corner of Fig.~\ref{fig:individual_abundances} continuously switch between two time-persistent states. After spending some time oscillating around a value of the abundance $n_i^{\text{high}}$ that is not small (from the figure we see that $n_i^{\text{high}}>0.1$), the species suddenly drop down and oscillate for a while around a very small abundance $n_i^{\text{low}}\sim0.001$. This small $n_i^{\text{low}}$ corresponds well to the predictions of IBMF, marked with horizontal dashed lines in the graphic. The real stationary abundance measured in simulations by averaging $n_i$ for long times, however, is somewhere in between $n_i^{\text{low}}$ and $n_i^{\text{high}}$. Instead of mimicking this intermediate value without a clear physical meaning, IBMF gives only the smallest of the two true values $n_i^{\text{low}}$ and $n_i^{\text{high}}$. Although it is only partially right, it definitely allows identifying the species that are going to exhibit this type of dynamics. A similar behavior has been recently found in fully-connected systems with asymmetric interactions and without thermal noise \cite{ThibaultSciPost2025}. In that case, when the variability $\sigma$ in the interaction strength is large enough, all the species switch between two time-persistent states, only one of which is close to extinction. When $\sigma$ is small, all the species reach a fixed point abundance. Remarkably, in our case we have only a few switching species, while the rest reach a fixed point. This species heterogeneity is probably related to the sparsity of interactions and is not present in the fully-connected case for any value of the variability $\sigma$.

Both with IBMF and with simulations, we verified that the corresponding stationary abundances were independent of the initial conditions (and of the realization of the noise in simulations). Moreover, we verified that the fraction of species for which IBMF and simulations give different results remains unchanged when the number of species $N$ is increased (see Appendix \ref{app:FSundirected}). Remarkably, computing the stationary abundances with IBMF is two orders of magnitude faster than running the simulations. By averaging over $100$ different initial conditions, we get the average wall-clock times of $22.0 \pm 0.3 \, \text{ms}$ for IBMF, and of $1390 \pm 60 \, \text{ms}$ for simulations (ms stands for milliseconds).

The results in Fig.~\ref{fig:individual_abundances} clarify the meaning of IBMF and its predictions for a single graph, while also raising new questions on the link between the structure of the interaction graph and the observed non-trivial dynamics. When the interactions are sparse, large fluctuations in individual interaction strengths or the presence of short cycles in the graph can lead some species to display qualitatively distinct dynamics. In other words, adding sparsity opens the possibility for species differentiation. This phenomenon can also depend on the model's parameters. Getting a clear picture will require further work, especially because of the difficulties involved in numerically analyzing the results from simulations in the presence of thermal noise.

In this introductory work, we provide a general and clearer picture of how IBMF works in more controlled scenarios. First, Subsections \ref{subsec:IBMF_asym} and \ref{subsec:IBMF_directed} compare IBMF with simulations at zero temperature, where the results from the latter are easier to interpret. In Subsection \ref{subsec:IBMF_asym}, we study the phase diagram of the model in graphs with asymmetric interactions in undirected random regular graphs. In Subsection \ref{subsec:IBMF_directed}, we revisit a model discussed in Ref.~\cite{StavPRE2024} to predict the probability of observing persistent fluctuations in the dynamics for any given system size. Subsection \ref{subsec:IBMFsym} is devoted, instead, to a case where we include thermal noise. Although IBMF with $T>0$ can be applied to symmetric or asymmetric interactions, we chose to study the model with symmetric and homogeneous interactions. The reason is that, in this case, we can compare the output of IBMF with the results of BP, thereby avoiding the numerical complications associated with studying the phase transitions of the simulated dynamics in the presence of thermal noise. The latter is left for future work.

\subsection{Undirected graphs with asymmetric interactions} \label{subsec:IBMF_asym}

In this Subsection, we apply IBMF to undirected graphs with Gaussian noise in the interactions at zero temperature. We take a random regular graph with a given connectivity $c$, and draw every $\alpha_{ij}$ from a Gaussian with mean $\mu$ and standard deviation $\sigma$ ($\alpha_{ij} \sim \mathcal{N}(\mu, \sigma)$). This means that the coupling in the opposite direction, $\alpha_{ji}$, is independently drawn from the same distribution. Thus, the interactions are generally asymmetric. The larger the standard deviation $\sigma$, the bigger the average difference between $\alpha_{ij}$ and $\alpha_{ji}$. 

\begin{figure}[t]
\centering
\includegraphics[width=0.6\textwidth]{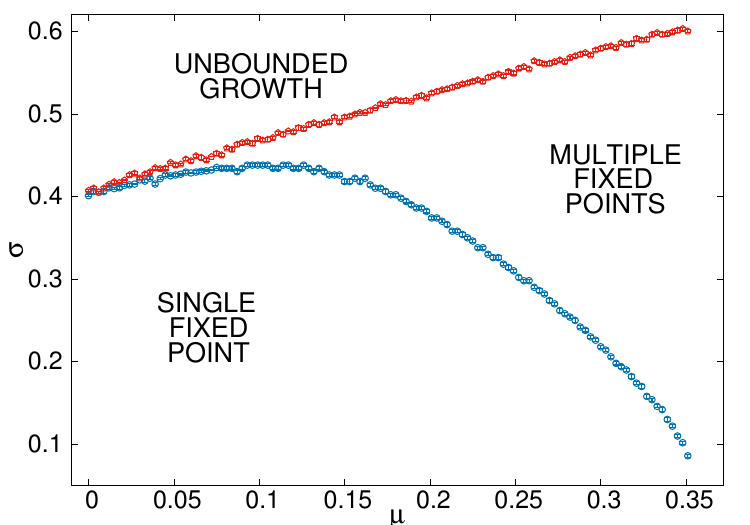}
\caption{Transitions obtained simulating the gLV model for $T=0$, asymmetric interactions ($\alpha_{ij}$ is chosen independently of $\alpha_{ji}$), and $\lambda=10^{-6}$. For several pairs $(\mu, \sigma)$, we run the dynamics for $10000$ different random regular graphs with connectivity $c=3$ and size $N=1024$. The interaction strengths are drawn from the Gaussian distribution: $\alpha_{ij}\sim\mathcal{N}(\mu, \sigma)$. By repeating the simulation $10$ times with different initial conditions for each graph, we identify one of three possible outcomes: i) all realizations converge to the same fixed point, ii) all the realizations converge but the fixed points are different, or iii) the abundances in at least one of the simulations grow and diverge for long times. The blue points mark, for each $\mu$, the maximum value of $\sigma$ at which more than $50\%$ of the samples are of type i). The red points mark, for each $\mu$, the minimum value of $\sigma$ at which more than $50\%$ of the samples are of type iii).}
\label{fig:Langevin_eps_0}
\end{figure}

Fig.~\ref{fig:Langevin_eps_0} shows the phase diagram obtained by simulating the gLV dynamics at $T=0$ and $\mu > 0$, with $\lambda=10^{-6}$ (see Eq. \eqref{eq:LVmodel_T0}). We identify three distinct regions. When $\sigma$ is small enough, the species reach a unique fixed point for long times, which corresponds to the single-fixed-point (SFP) phase. The first transition occurs at $\sigma_{SFP}(\mu)$, and is represented with blue points in Fig.~\ref{fig:Langevin_eps_0}. For $\sigma>\sigma_{SFP}(\mu)$, simulations with different initial conditions will not converge to the same fixed point in most interaction graphs; they will either converge to different fixed points or not converge at all. The unbounded growth (UG) transition is located at $\sigma_{UG}(\mu)\geq \sigma_{SFP}(\mu)$, and is represented by the red points in Fig.~\ref{fig:Langevin_eps_0}. Above this line, the abundance of at least one species grows and diverges in most simulations.

Note that the transition at $\sigma_{SFP}(\mu)$ is not purely between a single-fixed-point phase and a multiple-fixed-points phase. Although for large $\mu$ this is indeed the case, for $\mu\leq0$ the system goes directly from reaching a single fixed point to showing unbounded growth (see Appendix \ref{app:unbounded_growth_mutualistic}). In between, for each finite system we have a region in the phase diagram $(\mu, \sigma)$ where all three behaviors coexist. For a fraction $f_{SFP}>0$ of the ecosystems the species will go to a single fixed point, for a fraction $f_{MFP}>0$ they will reach different fixed points depending on the initial conditions, and for a fraction $f_{UG}>0$ some abundance diverges. As we show in Appendix \ref{app:coexistence}, this region of coexistence shrinks when the number of species $N$ grows.

The coexistence poses a problem in predicting the transitions using IBMF. As said before, any fixed point of IBMF is also a stationary solution of the exact dynamics at $T=0$. We need to design a procedure capable of detecting the presence of different fixed points if they exist. We then compare the results when we choose different initial conditions for the average abundances of IBMF. 

\begin{figure}[t]
\centering

\subfloat[]{
\includegraphics[width=0.45\textwidth]{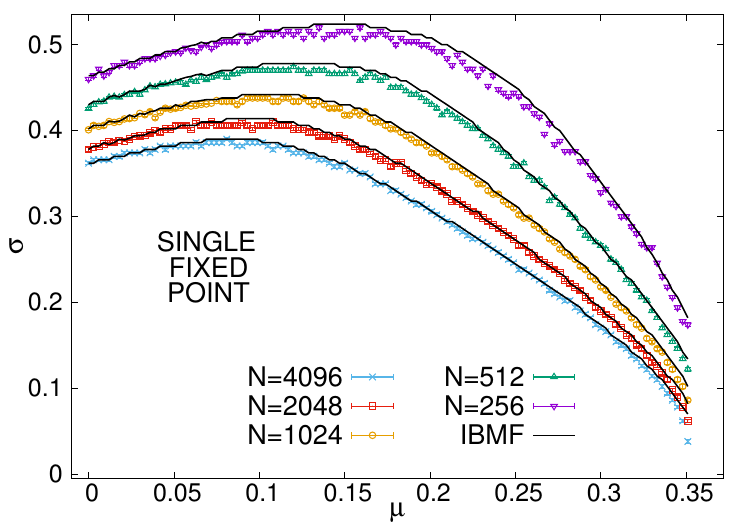} 
\label{fig:IBMFs_Langevin_T0_mult}
}
\subfloat[]{
\includegraphics[width=0.45\textwidth]{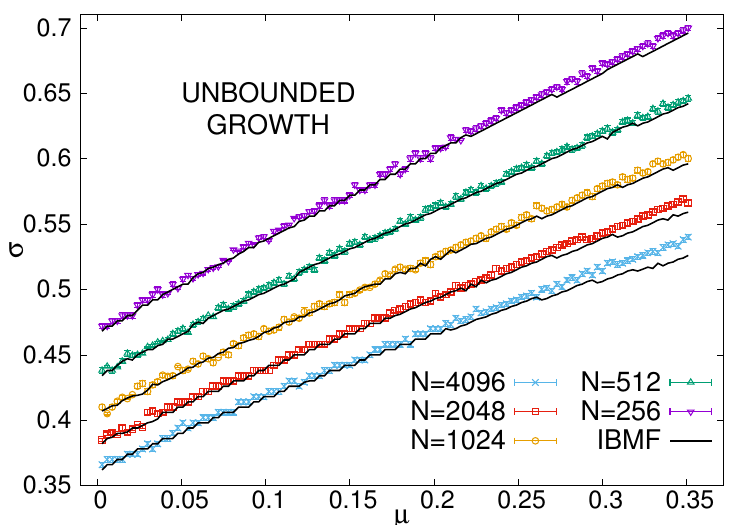} 
\label{fig:IBMFs_Langevin_T0_div}
}

\caption{Transitions of the gLV model for different system sizes at $T=0$. The interactions are asymmetric ($\alpha_{ij}$ is chosen independently of $\alpha_{ji}$) and defined on random regular graphs with connectivity $c=3$. Points represent the results of simulations with immigration rate $\lambda=10^{-6}$, and lines are the predictions made with IBMF for the same sizes. Each transition was determined using $10000$ graphs. Simulations are repeated for $10$ different initial conditions. IBMF was run with damping (see Appendix \ref{app:damping}) for $10$ different random initial conditions. {\bf (a)} For each system size $N$ and average strength $\mu$, points (lines) mark the maximum value of $\sigma$ such that simulations (IBMF) converged to the same fixed point in more than $50\%$ of the interaction graphs. {\bf (b)} Points (lines) mark the minimum value of $\sigma$ such that simulations (IBMF) displayed unbounded growth (not converged) more than $50\%$ of the interaction graphs.}
\label{fig:IBMF_asym}
\end{figure}

In Fig.~\ref{fig:IBMF_asym}, we compare the results of the simulations with the predictions of IBMF. There are three possible outcomes of IBMF, and they are similar to those obtained from the simulations. For $\sigma$ small enough, running IBMF in a specific graph many times with different initial conditions always gives the same fixed point. For $\sigma$ large enough, two things can happen. Either IBMF converges to multiple fixed points for a given graph, provided that we change the initial conditions, or it does not converge at all. Interestingly, for IBMF we also observe a coexistence between these behaviors in a region of the phase diagram. Fig.~\ref{fig:IBMFs_Langevin_T0_mult} shows the transitions between the single-fixed-point phase and the region where we either find multiple fixed points or no convergence. The lines, representing IBMF, accurately reproduce the results from the simulations.

Although we know that if IBMF reaches a fixed point, this is also a fixed point of the exact dynamics, it does not necessarily stop converging when the simulations do. We used IBMF to reproduce the transition to unbounded growth. Fig.~\ref{fig:IBMFs_Langevin_T0_div} shows a very good agreement between simulations and IBMF. This also extends to negative values of $\mu$, which correspond to interactions that are mutualistic on average (see Appendix \ref{app:unbounded_growth_mutualistic}). 

Our results indicate that IBMF is enough to independently describe both the transition to the multiple-fixed-points phase and the transition to unbounded growth. Note that the finite-size effects are relevant in both panels of Fig.~\ref{fig:IBMF_asym}. The transitions obtained with the simulations and with the theory move downward when the number of species $N$ increases. Nevertheless, IBMF is enough to capture these effects correctly, and its description is already accurate for finite systems. In addition, as we show in Appendix \ref{app:mean-abundance-heterogeneous-graphs}, IBMF is also able to correctly predict the global average abundance among all species $\langle n \rangle =\frac{1}{N} \sum_{i=1}^N n_i$, showing a very good agreement with numerical simulations both in the single fixed point and in the multiple fixed points  phases.

As mentioned above, to obtain Fig.~\ref{fig:IBMF_asym} we drew $\alpha_{ij}$ and $\alpha_{ji}$ independently from Gaussian distributions for every pair of interacting species. It is important to mention, however, that the quantitative agreement between IBMF and simulations also holds when $\alpha_{ij}$ and $\alpha_{ji}$ are correlated, as we show in Appendix \ref{app:correlated_alpha_T0}.

\subsection{Directed graphs} \label{subsec:IBMF_directed}
As in Ref.~\cite{StavPRE2024}, we study the emergence of fluctuating abundances $n_i$ in graphs where the degree follows a Poisson distribution with mean $c$, and where most interactions are directed. If the edge $j \to i$ is present, with high probability the edge in the opposite direction does not exist. The graph can be seen as representing a directed flow between the species. We say that species $j$ is \emph{upstream} with respect to species $i$ ($j \to i$). Conversely, species $i$ is said to be \emph{downstream} with respect to species $j$. We then independently draw each associated $\alpha_{ij}$ from the Gaussian distribution $\mathcal{N}(\mu, \sigma)$.

The authors of Ref.~\cite{StavPRE2024} carefully studied the case with homogeneous interactions ($\sigma=0$), and demonstrated that the zero temperature dynamics in this \emph{toy model} can have two distinct outcomes at long times. One possibility is that all abundances converge to a fixed stationary value, with the whole system reaching a fixed point. The second possibility is that not all species converge, resulting in a system with persistent fluctuations. The latter case can also be subdivided into two by taking into account the number of fluctuating species, with one regime with local fluctuations and another with global fluctuations. 

One of the main objects to measure is the probability that we obtain persistent fluctuations $p_{\text{fluc}}$ after running the dynamics in a graph extracted at random for some average connectivity $c$ and interaction strength $\mu$. Note that $p_{\text{fluc}}$ does not distinguish between local and global fluctuations. 

The system undergoes a transition around $\mu=1$ \cite{StavPRE2024}. For $c<\ec$, where $\ec$ is the Euler's constant, and in the limit when the number of species is large ($N \to \infty$), one gets $p_{\text{fluc}}=0$ for $\mu<\mu_c=1$ and $p_{\text{fluc}}>0$ for $\mu>\mu_c=1$. For $c>e$ and also in the limit $N \to \infty$, the same transition occurs but is displaced to smaller $\mu$, and the critical $\mu_c \lesssim 1$ slowly decreases when the connectivity increases.

\begin{figure}[t]
\centering

\subfloat[]{
\includegraphics[width=0.45\textwidth]{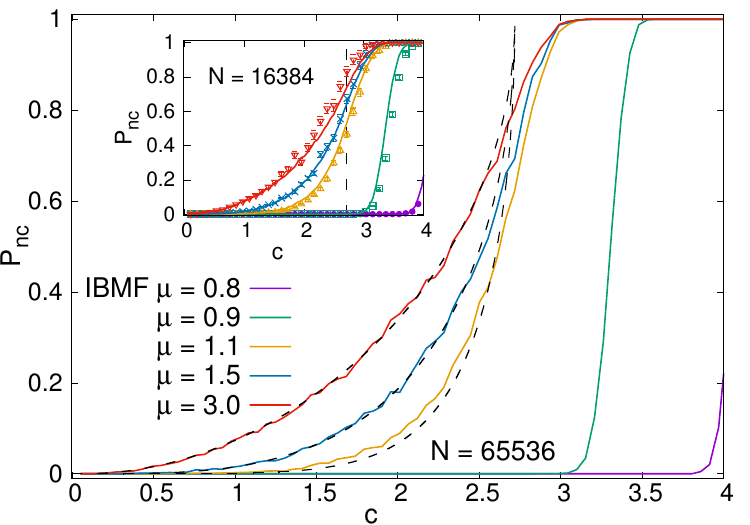}
\label{fig:IBMF_directed_several_mu}
}
\subfloat[]{
\includegraphics[width=0.45\textwidth]{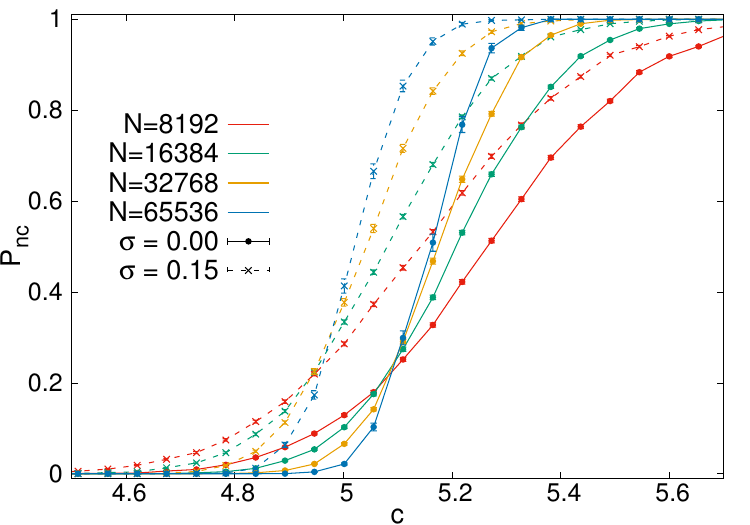}
\label{fig:IBMF_directed_several_sigma}
}

\caption{Probability that IBMF does not converge ($P_{\text{nc}}$) in directed graphs. IBMF is run over different realizations of the interaction graph with a given average connectivity $c$, size $N$, and interaction strength $\mu$. There is no unique function for all $\mu>1$, and dashed lines in the top panel are obtained exactly as in Ref. \cite{StavPRE2024} (see the text for clarification). {\bf (a)} Toy model without noise in the interactions ($\sigma=0$). The colored lines in the main graphic represent the results of IBMF for $N=65536$ and different values of $\mu$. In the inserted graphic, IBMF (lines) is run instead for systems with $N=16384$ species, and the points represent the results of simulations of the dynamics for the same system size. The vertical line marks the value $c=\ec$. The error bars for IBMF predictions are small and are not included in the graphics. {\bf (b)} The interaction strengths are drawn from the distribution $\mathcal{N}(\mu, \sigma)$ with $\mu=0.7$ and two values of $\sigma$. The values of $P_{\text{nc}}$ for different values of $c$ are represented using points with their corresponding error bars. Lines are a guide to the eye.}
\label{fig:IBMF_directed}
\end{figure}

As in the case of asymmetric interactions in a random regular graph (Subsection \ref{subsec:IBMF_asym}), we numerically demonstrate that, for this toy model, the probability of having persistent fluctuations in the simulations can be well approximated by studying the probability that IBMF converges. Fig.~\ref{fig:IBMF_directed_several_mu} presents the results obtained in the toy model. In this case, we observe that adding damping to the iterations is particularly important (see Appendix \ref{app:damping_directed_graphs}). For each graph of interactions generated with the rules described above, we also have two different outcomes. Either the abundances converge to a fixed point, or they continue to exhibit persistent fluctuations. As in simulations, the probability of convergence displays a qualitative change in its behavior around $\mu=1.0$.

The probability $P_{\text{nc}}$ that IBMF does not converge corresponds very well to the predictions made in Ref.~\cite{StavPRE2024} for the probability of having fluctuations, represented with dashed lines in the main graphic of Fig.~\ref{fig:IBMF_directed_several_mu}. When $\mu > 1$, the authors of Ref.~\cite{StavPRE2024} conclude that, in a given graph, the only fluctuating species are located in short cycles of odd length. The species in any cycle of length $n=2k+1$ will fluctuate if two conditions are met: i) all the species that are upstream of the species in the cycle are extinct, and ii) the cycle is unstable, which happens for $\mu > 1 / \cos(\pi / n)$. The value of $\mu$ thus determines the minimum length $n_{\min}$ of the fluctuating cycles. For details on the computation, see Appendix \ref{app:damping_directed_graphs} in this article or directly read Ref.~\cite{StavPRE2024}.

From top to bottom in the figure, the dashed lines correspond to $\mu=3.0$ ($n_{\min}=3$), $\mu=1.5$ ($n_{\min}=5$), and $\mu=1.1$ ($n_{\min}=9$). When $n$ is large, the values of $\mu_c(n)=1/\cos(\pi/n)$ are close to each other and to $\mu=1$, and it is numerically harder to distinguish between two values of $\mu$. However, in Fig.~\ref{fig:IBMF_directed_several_mu}, IBMF results for $\mu=1.1$ (orange points) are not far apart from the corresponding dashed line. 

Below $\mu=1.0$, the results in Fig.~\ref{fig:IBMF_directed_several_mu} are qualitatively different, also in agreement with Ref.~\cite{StavPRE2024}. For $\mu=0.9$, the probability that IBMF does not converge remains close to zero until it abruptly grows towards one around $c\sim3.1$. If we decrease $\mu$ just a bit more to $\mu=0.8$, we get $P_{\text{nc}} \sim 0$ for all $c<4$. In Appendix \ref{app:finite_size_directed}, we study the dependence of the results on the system size to conclude that, effectively, the transition for $\mu < 1$ is qualitatively different from the one for $\mu>1$.

Even for a large system with $N=65536$ species, we observe that the predictions of IBMF deviate from the theory of Ref.~\cite{StavPRE2024}, which is derived in the infinite size limit. In the inserted graphic of the same Fig.~\ref{fig:IBMF_directed_several_mu}, we show that this is not a particular problem of IBMF. With points, we represent the results obtained after simulating the dynamics by integrating Eq.~\eqref{eq:LVmodel_T0}. They are in very good agreement with the predictions of IBMF (lines), also above the critical connectivity $c=\ec$, which is marked with a vertical dashed line. As in Section \ref{subsec:IBMF_asym}, IBMF allows us to capture the finite-size effects in simulations, which is an advantage with respect to previous theoretical predictions. It is important to note that, even when run in single graphs, obtaining results from IBMF is computationally much simpler than performing actual simulations. Remarkably, the average runtime of IBMF for different values of $\mu$ and $c$ is consistently 10 times faster than the runtime of simulations, under equivalent conditions and for the same system size (see Appendix \ref{app:runtimes_directed_ER}).

Moreover, our IBMF equations are not restricted to this toy model, and some variations can also be studied. The authors of Ref.~\cite{StavPRE2024} try a modification to include noise in the interaction strengths. They take $\alpha_{ij}$ from the Gaussian distribution $\mathcal{N}(\mu, \sigma)$, again with probability $c/N$, and zero otherwise. According to the text of that article, for $\sigma=0$ and $\mu=0.7$ the transition occurs around $c\sim5.3$, while for $\sigma=0.15$ and $\mu=0.7$ they get $c\sim 4.9$.

Fig.~\ref{fig:IBMF_directed_several_sigma} shows the probability that IBMF does not converge when the interactions are drawn using this modified toy model. We include results for $\mu=0.7$ and two values of $\sigma$. With $\sigma=0$ (continuous lines), we recover the original toy model and use it as a reference. On the other hand, setting $\sigma=0.15$ (dashed lines) adds noise to the interaction strengths, and the probability that IBMF does not converge increases.

In both cases, we run IBMF for different system sizes. The curves show crossing points at $c \sim 5.1$ and $c \sim 4.95$, for $\sigma=0$ and $\sigma=0.15$, respectively. When the number of species $N$ increases, the probability $P_{\text{nc}}$ has a sharper transition between $P_{\text{nc}} \sim 0$ to the left and $P_{\text{nc}} \sim 1$ to the right of the crossing point. If this trend continues as expected when the number of species is large, the values $c \sim 5.1$ and $c \sim 4.95$ are reliable estimates of the location of the transition between a single equilibrium phase and the phase with global fluctuations. These results are indeed close to the ones mentioned in the text of Ref.~\cite{StavPRE2024}. The small discrepancies should be investigated further by performing the same analysis with data from simulations of the dynamics, which is technically more difficult because simulations take more computational time (see Appendix \ref{app:runtimes_directed_ER}). 

It would be useful to compare the numerical results for $\sigma>0$ with theoretical predictions in the limit of large ecosystems ($N \gg 1$). The very recent contribution of Ref. \cite{Muller_Metz_PRE_2025} could shed light in this direction. There, the authors also consider an IBMF-like closure for the dynamics of the Susceptible-Infectious-Susceptible (SIS) model for epidemics with directed interactions, and derive self-consistent equations in the limit of large systems. If their method can be extended to the gLV model, we expect it to describe the numerical results of our IBMF in the limit of large ecosystems with directed interactions.

\subsection{Including thermal noise}\label{subsec:IBMFsym}

\begin{figure}[]
\centering

\subfloat[]{
\includegraphics[width=0.27\textwidth]{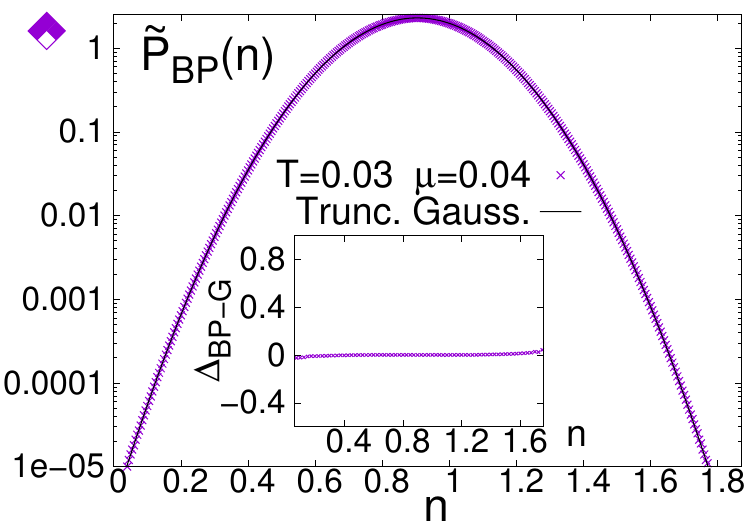} 
\label{fig:BP_Gauss_004}
}
\subfloat[]{
\includegraphics[width=0.27\textwidth]{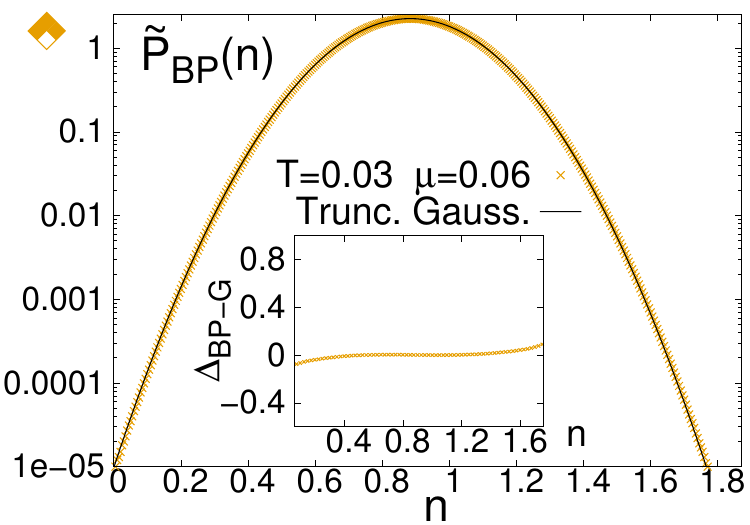} 
\label{fig:BP_Gauss_006}
}
\subfloat[]{
\includegraphics[width=0.27\textwidth]{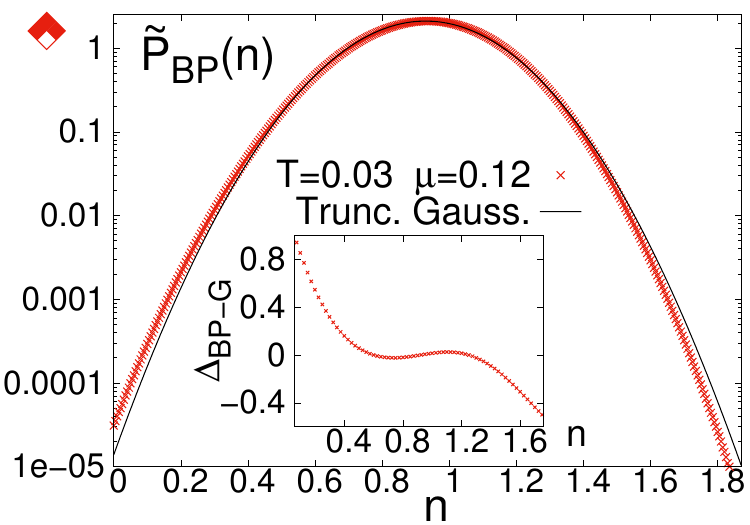} 
\label{fig:BP_Gauss_012}
}

\subfloat[]{
\includegraphics[width=0.45\textwidth]{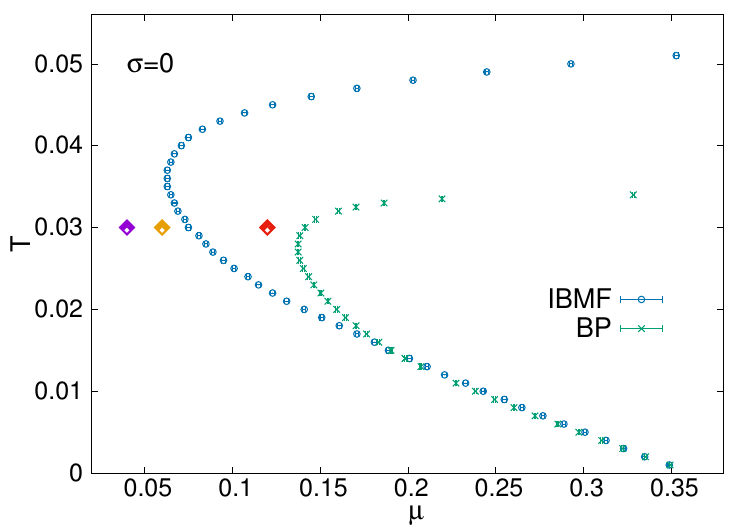}%
\label{fig:IBMFs_BP_phase_diagram}
}
\subfloat[]{
\includegraphics[width=0.45\textwidth]{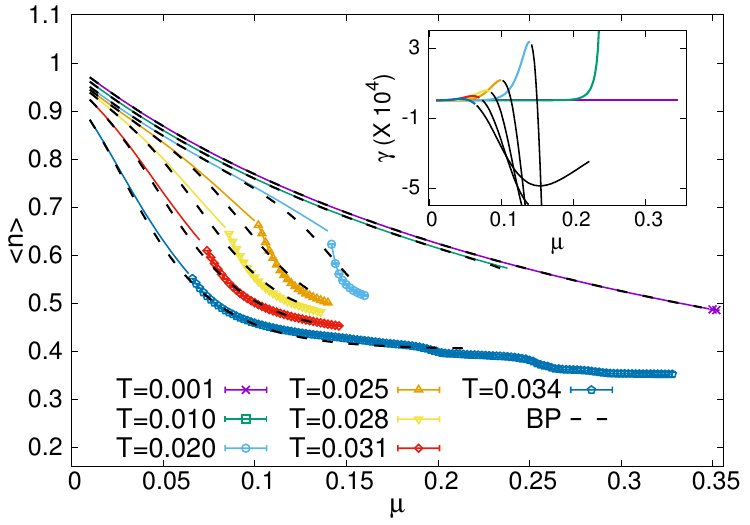}
\label{fig:IBMFs_BP_av_n}
}
\caption{Predictions of IBMF and BP with $T>0$ for symmetric and homogeneous interactions ($\sigma=0$) in random regular graphs with connectivity $c=3$. The immigration rate is $\lambda=10^{-6}$. {\bf (a), (b), and (c)} Distributions $\tilde{P}_{BP}(n)$ obtained with BP (see Eq.~\eqref{eq:PBP_tilde}) for $\mu=0.04, 0.06, 0.12$, at temperature $T=0.03$, and with system size $N=128$. Black continuous lines are fits to the points via truncated Gaussians. The inserted graphics show the relative deviation $\Delta_{BP-G}$ of the points with respect to the fits (Eq.~\eqref{eq:BP_G_dev}). {\bf (d)} For each temperature $T$, we mark the maximum value of $\mu$ where BP converges (green points). We also run IBMF on $10000$ graphs, each with $10$ different initial conditions, and mark the maximum value of $\mu$ where it converges to the same fixed point in at least $50\%$ of the graphs (blue points). System sizes are $N=128$ and $N=1024$ for BP and IBMF, respectively. {\bf (e)} Average abundance as a function of $\mu$. The black dashed lines are obtained with BP where this algorithm converges. The continuous colored lines are obtained with IBMF where it converges to a single fixed point. The colored points represent the average over several IBMF's fixed points, sampled using $10000$ distinct initial conditions. The inserted graphic shows the skewness $\gamma$ of the distribution $\hat{P}_{BP}(n)$. We use colored lines in the region where IBMF finds a single equilibrium point, and continuous black lines elsewhere.}
\label{fig:IBMF_symmetric}
\end{figure}

Subsections \ref{subsec:IBMF_asym} and \ref{subsec:IBMF_directed} show that at $T=0$ the results of IBMF are in good agreement with simulations. When thermal noise is present ($T>0$), the fixed points of IBMF can still be efficiently retrieved using Eqs. \eqref{eq:Z_IBMF}, \eqref{eq:M_IBMF}, and \eqref{eq:m_IBMF}, and taking advantage of the fact that these integrals can be expressed in terms of Kummer's confluent hypergeometric functions (see Section 3 in the SM). However, in this case, IBMF is an approximation that considers the probability distribution of the system to be factorized as $P(\vec{n})=\prod_{i}P(n_i)$. Its predictions, accurate for low temperatures, are expected to fail when $T$ is high enough.

In this section, we study how the results of IBMF depend on the temperature in the gLV model defined over random regular graphs with symmetric interactions. This is a controlled scenario where we have a reliable theoretical technique to compare with, which is BP. We further simplify the setting by eliminating any noise in the interaction strength. Provided that $(i,j)$ is an edge in the random regular graph, we set $\alpha_{ij} = \alpha_{ji}=\mu$, which is equivalent to drawing all $\alpha_{ij}$ from the trivial Gaussian $\mathcal{N}(\mu, 0)$. A first version of the phase diagram $T$ \textit{vs.} $\mu$ is available in Ref.~\cite{Tonolo2026sparse}, where BP is run using discretized abundances. Interestingly, the authors of Ref.~\cite{Tonolo2026sparse} note a re-entrant transition in their phase diagram: for low temperatures, the critical value of $\mu$ decreases when $T$ increases, until it reaches a minimum. Then, it returns and starts increasing as the temperature continues to rise.

To compare BP with IBMF, which works directly with continuous variables, we use our implementation of BP with continuous variables. After obtaining the messages by iterating Eq.~\eqref{eq:BP_up} until convergence, we use Eqs.~\eqref{eq:PiBP} and \eqref{eq:PijBP} to get the \emph{true} marginals. The reader can find details about our implementation in Section 4 of the SM. From Eq.~\eqref{eq:PiBP}, we see that the stationary distribution for a single species has the form $P_{BP}(n_i)=n_i^{\beta \lambda - 1} \tilde{P}_{BP}(n_i) / Z_i$, where:

\begin{eqnarray}
    \tilde{P}_{BP}(n_i)=\frac{1}{\tilde{Z}_{i}} \,  \exp \Big\{-\frac{\beta}{2}(n_i^{2} - 2 n_i) \Big\} \prod_{k \in \partial i^{-} } \int_{0}^{\infty} dn_k \, \eta_{k \to i}(n_k) \, e^{- \beta \alpha_{ik} n_i \, n_k} \label{eq:PBP_tilde}
\end{eqnarray}
can be interpreted as an auxiliary probability distribution if $\tilde{Z}_{i}$ is taken as the proper normalization factor. The messages $\eta_{k \to i}(n_k)$, necessary to compute $\tilde{P}_{BP}(n_i)$, are the fixed point solution of Eq.~\eqref{eq:BP_up}. 

Since the abundance $n_i$ must be positive, the distribution in Eq.~\eqref{eq:PBP_tilde} is defined only for $n_i\geq0$. Whenever interactions are absent ($\alpha_{ik}=\mu=0$ for all $i$ and $k$), $\tilde{P}_{BP}(n_i)$ becomes a truncated Gaussian centered at $n_i=1$. Letting $\mu$ increase away from zero, one gets a distribution $\tilde{P}_{BP}(n_i)$ that is no longer strictly a truncated Gaussian. Moreover, when the interactions are homogeneous (all $\alpha_{ik}=\mu$), BP converges to the same $\tilde{P}_{BP}(n_i)$ for all sites $i$. This index can be dropped, and the average distribution $\tilde{P}_{BP}(n)=\sum_i \tilde{P}_{BP}(n_i) /N$ is equal to $\tilde{P}_{BP}(n_i)$ itself.

Figs.~\ref{fig:BP_Gauss_004}, \ref{fig:BP_Gauss_006}, and \ref{fig:BP_Gauss_012} show that, even with non-negligible thermal noise and non-zero values of $\mu$, the distribution $\tilde{P}_{BP}(n)$ obtained with BP is not far from Gaussian. These three distributions are obtained for the same temperature $T=0.03$, using $\mu=0.04$, $\mu=0.06$, and $\mu=0.12$, respectively. While $\tilde{P}_{BP}(n)$ is represented with colored points in the main graphics, the continuous black lines are the result of fitting truncated Gaussians $\tilde{P}_{G}(n)$ to the data. The inserted graphics show the relative deviation $\Delta_{BP-G}$ of the points with respect to the fits:

\begin{equation}
    \Delta_{BP-G} = \frac{\tilde{P}_{BP}(n)-\tilde{P}_{G}(n)}{\tilde{P}_{G}(n)} \label{eq:BP_G_dev} \:\:\:.
\end{equation}

For $\mu=0.04$ and $\mu=0.06$ (Figs.~\ref{fig:BP_Gauss_004} and \ref{fig:BP_Gauss_006}, respectively), the relative deviation is small for all the values of the abundance. In the corresponding main graphics, the points are indeed very close to the fits. The most significant difference occurs at the tails of the distribution. Close to $n=0$, we get that $\tilde{P}_{BP}(n)$ is below the truncated Gaussian, while for $n$ large it is above. In other words, the presence of thermal noise and interactions tilts the true distribution and gives slightly more weight to large abundances. The ecosystem is a bit more favorable for the species to thrive.

The difference with respect to a Gaussian increases with $\mu$ and is more evident at $\mu=0.12$ (Fig.~\ref{fig:BP_Gauss_012}). There, the scenario has been reversed. The main and inserted graphics show that now the true distribution is above the Gaussian for $n \sim 0$, and is below for large $n$. The non-Gaussianity of the distribution gives more weight to species that are close to extinction. In terms of the model, this corresponds to an ecosystem that can support fewer species. Eventually, the situation is no longer compatible with the existence of a fixed point of BP, and this algorithm stops converging around $\mu =0.1405(5)$. In summary, we observe two distinct types of non-Gaussianity in BP at $T=0.03$: one that is tilted towards larger abundances, which occurs at small values of $\mu$, and another one tilted towards extinctions, which occurs close to the point where BP stops converging.

In the phase diagram in Fig.~\ref{fig:IBMFs_BP_phase_diagram}, the points where BP stops converging at each temperature are marked with green crosses. The results confirm the re-entrant transition detected in Ref.~\cite{Tonolo2026sparse}. When the temperature goes to zero, one recovers the exact result $\mu \approx 0.354$ of Ref.~\cite{StavPLOS2022} (see Appendix \ref{app:BP_T0}) for the transition from the single equilibrium phase to the phase with multiple equilibria. The blue circles, on the other hand, represent the prediction made with IBMF for the same transition. We run this approximation for several random regular graphs, in each case using different initial conditions for the average abundances. To the left of the circles, different runs of IBMF converge to the same average abundances in at least $50 \%$ of the graphs. To the right, we find instead distinct stationary values of the average abundances just by changing the initial conditions in at least $50 \%$ of the graphs. IBMF reproduces very closely the results of BP for low temperatures and, as expected, deviates from it for high temperatures. However, it maintains qualitative agreement with BP,  also displaying a re-entrant transition. 

It is also important to mention that the computational cost of running IBMF is considerably lower than the cost of BP. When an extensive use of computational resources is required, as in Fig.~\ref{fig:IBMFs_BP_phase_diagram}, this advantage of IBMF plays an important role and one can use it to study larger systems. Nevertheless, we checked that the transition points presented in this figure do not change when the system size is increased, neither for IBMF nor for BP.

Since IBMF is a factorized \emph{ansatz} for the stationary distribution $P_{\infty}(\vec{n}) = \prod_i P_{\infty}(n_i)$, and each of the factors $P_{\infty}(n_i)$ is a Gaussian multiplied by the factor $n_i^{\beta \lambda-1}$, it can be used to shed light on the discussion about the non-Gaussianity of BP's solution. The latter, illustrated here in Figs.~\ref{fig:BP_Gauss_004}, \ref{fig:BP_Gauss_006}, and \ref{fig:BP_Gauss_012}, has already been noticed in Ref.~\cite{Tonolo2026sparse}. To study it in more detail, we computed the stationary average abundance $\langle n \rangle$ with both techniques, IBMF and BP, for different temperatures. The black dashed lines in Fig.~\ref{fig:IBMFs_BP_av_n} are obtained with BP in the range of values of $\mu$ where this algorithm converges. In turn, the colored continuous lines represent the prediction of IBMF when it converges to a single fixed point. However, this is not the only way to estimate the average abundance with IBMF. Even in the region where the fixed point is not unique, we compute $\langle n \rangle$ by averaging over the different fixed points of IBMF (colored points in the figure). Interestingly, the predictions of IBMF closely follow the dashed lines of BP, also in the region where the fixed point of IBMF is not unique. In other words, this approximation describes the average abundance well at all temperatures under consideration. In Appendix \ref{app:mean-abundance-heterogeneous-graphs}, we show that IBMF's predictions for the average abundance are also in very good agreement with simulations in the case of heterogeneous, asymmetric interactions at $T=0$.

Furthermore, IBMF's transition, marked with blue circles in Fig.~\ref{fig:IBMFs_BP_phase_diagram}, is related to the type of non-Gaussianity displayed by BP. In the inserted graphic of Fig.~\ref{fig:IBMFs_BP_av_n}, we present the skewness $\gamma$ of the distribution $\tilde{P}_{BP}(n)$ (see Eq.~\eqref{eq:PBP_tilde}) to quantify this non-Gaussianity. When the distribution is tilted towards large abundances, we get $\gamma > 0$. On the other hand, when $\tilde{P}_{BP}(n)$ is tilted towards extinctions, we get $\gamma < 0$.

We use colored lines in the inserted graphic to represent the skewness of $\tilde{P}_{BP}(n)$ in the region where IBMF converges to a single fixed point. Black continuous lines are used, in turn, in the region where IBMF converges to different fixed points. The results indicate that around the same value of $\mu$ where BP starts developing a distribution $\tilde{P}_{BP}(n)$ that is tilted towards extinctions, IBMF stops converging to a single fixed point.

\section{Conclusions} \label{sec:discussion}

In conclusion, our local closures for the global Fokker-Planck equations, and in particular the Individual Based Mean Field (IBMF) method, provide a powerful and versatile tool for analyzing the stationary states of the generalized Lotka-Volterra model on sparse graphs. We have demonstrated its efficacy across a range of scenarios, from asymmetric interactions on undirected graphs to directed networks and systems with thermal noise. IBMF faces its greatest challenge in the case of symmetric interactions with thermal noise, where its assumption of species independence breaks down due to correlations. These correlations are precisely what the Belief Propagation (BP) method captures, but BP is restricted to symmetric interactions. This highlights a key trade-off: BP offers higher accuracy for symmetric networks, while IBMF provides a versatile and effective solution for the more general and common case of asymmetric couplings.

It is important to note the qualitative differences between the output of IBMF and that of simulations, which are particularly relevant in the presence of thermal noise ($T>0$). From simulations, one obtains realizations of the temporal evolution of individual abundances. Since $T>0$, the stationary abundances do not stop oscillating around their averages, and the data must then be post-processed to compute, for example, the stationary average abundances. From IBMF, instead, we directly obtain the average abundances in the stationary point after the iteration process converges. There is no need for post-processing.

More importantly, when the fixed point is not unique and $T>0$, the analysis of the simulations is technically more complex. For finite systems, it is hard to numerically assess whether each abundance is oscillating around a single stationary average, or the whole dynamics is fluctuating between different distinct fixed points. In the same scenario, running IBMF directly gives a set of stationary average abundances, and we are able to identify regions with a single fixed point and with multiple fixed points without much effort (see Fig. \ref{fig:IBMF_symmetric}). This constitutes a considerable practical advantage of IBMF over simulations to be exploited in the future.

Our analysis reveals that the phase diagram for sparse, asymmetric interactions (Fig.~\ref{fig:Langevin_eps_0}) is qualitatively distinct from its symmetric counterpart; notably, we observe a transition towards a multiple-equilibria phase at a positive $\sigma$ even for $\mu\gtrsim0$, whereas symmetric interactions always lead to a single fixed point at small $\mu$ \cite{Tonolo2026sparse}. This sparse topology also induces a different stability landscape compared to fully-connected systems \cite{bunin2017ecological}, with a transition from single fixed point to multiple fixed points occurring at a positive $\mu$ for $\sigma=0$, a phenomenon linked to the intrinsic instability of sparse competitive loops identified in \cite{StavPLOS2022}. 

Moving from fully-connected to sparse interactions also has a relevant impact on the stability of the fixed points. In the phase with multiple fixed points, the fully-connected gLV model with asymmetric interactions is well-known to display signatures of chaotic dynamics \cite{Roy_2019}. Recent contributions \cite{Arrola_arxiv_2026, ros2023generalized} show how the right amount of asymmetry in the interactions destabilizes the exponentially many fixed points of the model. On the contrary, for the gLV model with sparse, asymmetric interactions in the multiple fixed point phase, we find that both the simulations and IBMF converge to distinct stable fixed points and stay there. The study of how these fixed points destabilize when the connectivity increases is left for future work.

The observed finite-size effects in the transition lines are correctly captured by IBMF. Notably, the transition lines to both multiple fixed points and unbounded growth progressively shift toward lower heterogeneity $\sigma$ as the system size $N$ increases. This suggests that the sparse ecological models under consideration may be intrinsically unstable in the infinite-species limit whenever heterogeneity is finite. This observation is consistent with previous RMT results~\cite{valigi2024localsignstability, valigi2025spectraltails, mambuca2022dynamical}, which show that, in general, sparse random matrices remain stable in the large size limit only if the interactions are purely antagonistic or unidirectional. However, for finite sizes we do observe a region where there is a single stable fixed point, a fact that is accurately predicted by our method.

Looking forward to new applications, the computational efficiency and general applicability of IBMF make it a promising candidate for predicting stable states in real ecological networks, when direct data on interaction strengths is available \cite{ruiter1995energetics, stein2013ecological, jacquet2016no, Barbier2021Fingerprints} or, in its absence, when one has access to the relevant parameters from which the interaction strengths can be drawn \cite{barbier2025gettingmore, Pasqualini_2025}. Furthermore, the methodological framework is not restricted to ecology and could be fruitfully generalized to analyze a wide class of models in economics, evolutionary game theory, and other fields defined on complex, sparse, and even asymmetric interaction networks.

\section{Acknowledgments}
We thank Giacomo Gradenigo and Chiara Cammarota for early discussions on this problem. Our special gratitude goes to Mattia Tarabolo, Roberto Mulet, and Luca Dall'Asta, with whom we maintained a productive scientific dialogue throughout the writing process, and who generously contributed their own novel insights. This project has been supported by the FIS 1 funding scheme (SMaC - Statistical Mechanics and Complexity) from Italian MUR (Ministry of University and Research) and from the project MIUR-PRIN2022, “Emergent Dynamical Patterns of Disordered Systems with Applications to Natural Communities”, code 2022WPHMXK, funded by European Union — Next Generation EU, Mission 4 Component 1 CUP: 
B53D23005310001. This study was conducted using the DARIAH HPC
cluster at CNR-NANOTEC in Lecce, funded by the ”MUR PON
Ricerca e Innovazione 2014-2020” project, code PIR01 00022.

 % Bibliography
%\bibliographystyle{unsrt}
\bibliography{ref_ecology_2025}

\appendix

\section{From Ito's rule to the Fokker-Planck equation}\label{app:derivation_gen_FP}

Let $A(\vec{n})$ be a generic observable that depends on the whole system $\vec{n}=(n_1, \ldots, n_N)$ at time $t$, but does not explicitly depend on time. For example, $A(\vec{n})$ could be the average abundance $A(\vec{n})=\sum_{i} n_i(t) / N$. Following Ito's rule:

\begin{equation}
\frac{d}{dt} \mathbb{E}[A(\vec{n})] = \mathbb{E}\Big[ \sum_{i=1}^{N} \frac{\partial A(\vec{n})}{\partial n_i} \, \frac{dn_i}{dt}\Big] + T \, \mathbb{E}\Big[ \sum_{i=1}^{N} \frac{\partial^{2} A(\vec{n})}{\partial n_i^{2}} \, n_i\Big]
 \label{eq:Ito} \: ,
\end{equation}
where $\mathbb E[\, \cdot \,]$ is the average over the probability density $P_{t}(\vec{n})$ of having abundances $\vec{n}$ at time $t$, which is defined in a stochastic process where each trajectory is given by a specific realization of the thermal noise and a specific choice for the initial conditions. In other words, $\mathbb E[\, \cdot \,]$ is an average over the thermal noise and the initial conditions. Using this definition and Eq.~\eqref{eq:LVmodel} for $dn_i/dt$, one gets:

\begin{eqnarray}
\int_0^{\infty} d\vec{n} A(\vec{n}) \frac{\partial}{\partial t} P_{t}(\vec{n})  \!\!\!\! &=& \!\!\!\! \sum_{i=1}^{N} \int_0^{\infty} d\vec{n} \, P_t(\vec{n}) \, \big[ n_i(1 - n_i - \sum_{j \in \partial i^{-}} \alpha_{ij} n_j) + \lambda \big] \frac{\partial A(\vec{n})}{\partial n_i} + \nonumber  \\
& & +  T \, \sum_{i=1}^{N} \int_0^{\infty} d\vec{n}\, P_t(\vec{n}) \, n_i \frac{\partial^{2} A(\vec{n})} {\partial n_i^{2}} \: . \:\:\:\:\:\:\:\:
 \label{eq:Ito_2}
\end{eqnarray}

To obtain Eq.~\eqref{eq:Ito_2} one needs to use the fact that $\langle \xi_i(t) \rangle = 0$, where $\xi_i(t)$ is the Gaussian noise that appears in Eq. \eqref{eq:LVmodel}. Integrating by parts and using that, to have finite moments, $P_t(\vec{n}) \to 0$ faster than $n_i^{-2}$ when $n_i \to \infty$, leads to:

\begin{eqnarray}
\int_0^{\infty} d\vec{n} A(\vec{n}) \frac{\partial}{\partial t} P_t(\vec{n}) \!\!\!\! &=& \!\!\!\! - \sum_{i=1}^{N} \int_0^{\infty} d\vec{n} A(\vec{n})  \, \frac{\partial }{\partial n_i} \Big\{ \big[n_i(1 - n_i - \sum_{j \in \partial i^{-}} \alpha_{ij} n_j) + \lambda \big] P_t(\vec{n})  \Big\} + \nonumber \\
& & \!\!\!\!\!\!\!\!\!\!\!\!\!\!\!\!\!\!\!\!\!\!\!\!\!\!\!\!\!\!\!\!\!\!\!\!\!\!\!\!\!\!\!\!\!\!\!\!\!\!\!\!\!\!\!\!\!\!\!\!\!\!\!\!\!\!\!\!\!\!\!\!\!\!\!\!\! + \: T \, \sum_{i=1}^{N} \int_0^{\infty} d\vec{n} A(\vec{n}) \frac{\partial^{2}}{\partial n_i^{2}} \big\{ n_i P_t(\vec{n}) \big\} + \sum_{i=1}^{N} (T-\lambda) \int_0^{\infty} \Big[ \prod_{k \neq i} dn_k \Big] \lim_{n_i \to 0^+} \big[ A(\vec{n}) \, P_t(\vec{n}) \big] \: .
 \label{eq:Ito_3}
\end{eqnarray}

In the usual diffusion problems where one follows this procedure, the variables are defined in the whole open space $x \in (-\infty, \infty)$. The property $\lim_{x\to\pm \infty}P(x) =0$ kills all the terms that come from evaluating the integrands in $x \to \pm \infty$. However, now one has a variable $n_i$ defined in $[0, +\infty)$, and $\lim_{n_i \to 0^+}P_t(\vec{n}) \neq 0$ in general. The last term in Eq.~\eqref{eq:Ito_3} highlights the role of the conditions at the border $n_i=0$, and this is relevant to find the right local closures. 

To continue from here, however, we should impose the proper boundary conditions for this Fokker-Planck equation. To guarantee that $P_t(\vec{n})$ keeps properly normalized, the current of probability density must be zero at every border $n_i=0$, using what is called a reflecting boundary condition \cite{Gardiner2009stochastic}. We must then enforce the relations $(T-\lambda) \lim_{n_i \to 0^+} \, P_t(n_i) = 0$ for all species $i$, where $P_t(n_i) = \int_0^{\infty} [\,\prod_{k\neq i} dn_k] P_t(\vec{n})$ is the single-site probability for the abundance of species $i$. Therefore, the last term in Eq. \eqref{eq:Ito_3} vanishes. In other words, we can neglect the surface terms that arise after integrating by parts.

Since $A(\vec{n})$ is a generic function, the only way to fulfill this relation is to have

\begin{eqnarray}
 \frac{\partial P_t(\vec{n})}{\partial t} =  T \sum_{i=1}^{N} \frac{\partial^{2}}{\partial n_i^{2}} \big\{ n_i P_t(\vec{n}) \big\}  -\sum_{i=1}^{N}  \frac{\partial }{\partial n_i} \Big\{ \big[n_i(1 - n_i - \sum_{j \in \partial i^{-}} \alpha_{ij} n_j) + \lambda \big] P_t(\vec{n})  \Big\} , \label{eq:FokkerPlanck_app} 
\end{eqnarray}
which is the right Fokker-Planck equation, valid for any graph $G(V, E)$.

\section{Solution for the isolated variable} \label{app:FokkerPlanck_single}

For the rest of this article, it will be useful to obtain the stationary solution of Eq.~\eqref{eq:FokkerPlanck} in the particular case where there is only one variable $n$. The equation is then:

 \begin{eqnarray}
      \frac{\partial}{\partial t} P_t(n) = T  \frac{\partial^{2}}{\partial n^{2}} \big\{ n P_t(n) \big\} - \frac{\partial }{\partial n} \Big\{ \big[n(1 - n) + \lambda \big] P_t(n)  \Big\} \label{eq:FokkerPlanck_one} \: . 
 \end{eqnarray}

In the steady state:

 \begin{eqnarray}
0 = T  \frac{\partial^{2}  \big\{ n P_\infty(n) \big\} }{\partial n^{2}} \! - \frac{\partial }{\partial n} \Big\{ \big[n(1 - n) + \lambda \big] P_\infty(n)  \Big\} \,. \:\:\:\: \label{eq:FokkerPlanck_one_stat}
\end{eqnarray}

Integrating over $n$ and remembering that $\lim_{n \to \infty} n^{2} P_{\infty}(n) = 0$, the integration constant goes away. We get:

\begin{eqnarray}
 \frac{d }{d n} P_{\infty}(n) =  \frac{1 - n}{T} \, P_{\infty}(n) + \Big(\frac{\lambda}{T} - 1 \Big) \, \frac{P_{\infty}(n)}{n} \: . 
\end{eqnarray}

Solving this differential equation with separable variables is simple. The result is:

\begin{equation}
P_{\infty}(n) = \frac{1}{Z} \, n^{\beta \lambda - 1} \, \exp \Big\{ -\frac{\beta}{2} \,(n - 1)^{2} \Big\} \: ,
 \label{eq:sol_single_var_app}
\end{equation}
where $Z$ is a normalization constant and $\beta \equiv 1/T$.

Once one has the solution (Eq.~\eqref{eq:sol_single_var_app}) to the Fokker-Planck equation for a single species (Eq.~\eqref{eq:FokkerPlanck_one}), it is not hard to see what will be the solution for IBMF in the stationary regime. In the open space $n_i\in(0, +\infty)$, the equation to fulfill is:

\begin{eqnarray}
 0 = -\frac{\partial }{\partial n_i} \Big\{ \big[n_i(1 - n_i - \sum_{j \in \partial i^{-}} \alpha_{ij}\, m_{j}(\infty)) + \lambda \big] P_{\infty}(n_i) \Big\} + T \frac{\partial^{2}}{\partial n_i^{2}} \big\{ n_i P_{\infty}(n_i) \big\} \label{eq:IBMF_stat} \: .
\end{eqnarray}

This is essentially the same Eq.~\eqref{eq:FokkerPlanck_one_stat}, where one substitutes $1 - n_i$ by the mean-field expression $1 - n_i - \sum_{j \in \partial i^{-}} \alpha_{ij}\, m_{j}(\infty)$. Thus, if the solution to the single variable was Eq.~\eqref{eq:sol_single_var_app}, the solution to Eq.~\eqref{eq:IBMF_stat} is:

\begin{equation}
P_{\infty}(n_i) = \frac{1}{Z_i} \, n_i^{\beta \lambda - 1} \, \exp \Big\{ -\frac{\beta}{2} \,(n_i - M_i)^{2} \Big\}
 \label{eq:sol_IBMF_app_B} \: ,
\end{equation}
where $M_i = 1 - \sum_{j \in \partial i^{-}} \alpha_{ij}\, m_{j}(\infty)$. 

The stationary solution in Eqs. \eqref{eq:sol_single_var_app} and \eqref{eq:sol_IBMF_app_B} are normalizable functions for any $\lambda > 0$. Therefore, they give valid distributions $P_{\infty}(n_i)$ that solves the Fokker-Planck equation when $t \to \infty$ for any $\lambda > 0$. Nevertheless, we identify two distinct qualitative behaviors of $P_{\infty}(n_i)$, depending on the value of the immigration rate $\lambda$. When $\lambda > T$, we get that the probability density is zero at the border ($\lim_{n_i \to 0^+} P_{\infty}(n_i) = 0$), which is consistent with the boundary conditions imposed by us while deriving the Fokker-Planck equation. In this case, the immigration effectively counteracts the thermic noise and the species are strongly repelled from extinction ($n_i=0$). When $0 < \lambda < T$, the stationary distribution $P_{\infty}(n_i)$ diverges at $n_i=0$. This contradicts the boundary condition $(T-\lambda) \lim_{n_i \to 0^+} \, P_t(n_i) = 0$. 

However, the mathematical inconsistency in the definition of the Fokker-Planck problem for $0 < \lambda < T$ does not impede $P_{\infty}(n_i)$ from being a valid solution of Eq.~\eqref{eq:FokkerPlanck_one_stat} also in this interval. We can still give a physical interpretation to this case: when the effect of the immigration $\lambda$ is small, a finite fraction of the species goes nearly extinct. Indeed, our simulations in Fig.~\ref{fig:individual_abundances}, for a system of $N$ interacting species in a random graph, remain stable even for $\lambda \ll T$. We observe, both from simulations and from IBMF, how a fraction of species spend long times close to extinction.

\section{Limits of IBMF with parallel updates at zero temperature}\label{app:parallel_up}

Eq. \eqref{eq:IBMFT0}, which is the zero-temperature limit of IBMF, can be straightforwardly recast to matrix form as:

\begin{equation}
 \vec{n} = \vec{1} - \hat{J} \, \cdot \, \vec{n} \: ,\label{eq:IBMFT0_matrix}
\end{equation}
where $\vec{n}=(n_1, \ldots n_N)$ is the vector of species abundances, $\vec{1}$ is a vector full of ones, and $\hat{J}$ is the interaction matrix, whose elements are $J_{ij}=\alpha_{ij}$. In the case with homogeneous interactions, we have that $\hat{J}$ can be expressed in a simple way in terms of the adjacency matrix $\hat{A}$ associated with the interaction graph. By its definition, the element $A_{ij}$ of this matrix is zero if $\alpha_{ij}=0$, and is one otherwise. Thus, when all nonzero $\alpha_{ij}$ are equal to the same number $\mu$, Eq. \eqref{eq:IBMFT0_matrix} transforms into $\vec{n} = \vec{1} - \mu \, \hat{A} \, \cdot \, \vec{n}$.

Similarly to Ref.~\cite{StavPLOS2022}, we can use the properties of $\hat{A}$ to derive the exact single-to-multiple-equilibria transition in the case of random regular graphs with homogeneous interactions. The same equation can be rewritten as $(\mathbb{I} + \mu \, \hat{A}) \cdot  \vec{n} = \vec{1}$, where $\mathbb{I}$ is the identity matrix. Then, the solution is obtained after a matrix inversion by making $\vec{n}^{\ast} = (\mathbb{I} + \mu \, \hat{A})^{-1} \cdot \vec{1}$. For the matrix $\mathbb{I} + \mu \, \hat{A}$ to be invertible, all its eigenvalues must be nonzero. When $A$ is the adjacency matrix of a large random regular graph, we can use the fact that its smallest eigenvalue must be close to $\lambda_{\text{min}}^{A} =-2\sqrt{c-1}$ \cite{McKay_1981}. Then, the smallest eigenvalue of $\mathbb{I} + \mu \, \hat{A}$ is $\lambda_{\text{min}}=1-2\mu \sqrt{c-1}$. To compute the fixed point $\vec{n}^{\ast}$, or equivalently, for the matrix $\mathbb{I} + \mu \, \hat{A}$ to be invertible, one should verify the relation $\lambda_{\text{min}} >0$. In other words, the interaction strength $\mu$ must satisfy the relation:

\begin{equation}
 \mu < \mu^{\ast} \equiv \frac{1}{2 \sqrt{c-1}} \: ,\label{eq:transition_RRG_T0}
\end{equation}
which is the same result presented in Ref.~\cite{StavPLOS2022}. Since the eigenvalues of $\hat{A}$ cover the interval $\lambda^{A} \in [-2\sqrt{c-1}, 2\sqrt{c-1}]$, for $\mu \geq \mu^{\ast}$ one could always find an eigenvalue in this bulk of the distribution such that $1 + \lambda_{\text{min}}^{A} = 0$, and the matrix will not be invertible for $\mu \geq \mu^{\ast}$. Furthermore, given the homogeneity of the interactions, Eq.~\eqref{eq:IBMFT0} must admit the solution:

\begin{equation}
n = \frac{1}{1+c \, \mu} \: ,
 \label{eq:abundance_RRG_T0}
\end{equation}
which is also presented in Ref.~\cite{StavPLOS2022} as the unique fixed point for $\mu < \mu^{\ast}$.

It is also interesting to understand what would happen if one uses Eq.~\eqref{eq:IBMFT0} in an iterative way by setting $\vec{n}_{k+1} = \vec{1} - \hat{J} \cdot \vec{n}_{k}$, choosing a given initial $\vec{n}_0$. This, simply put, is a parallel update that gives a full vector $\vec{n}_{k+1}$ using information about the previous vector $\vec{n}_{k}$. The result of iterating $k$ times can be expressed as:

\begin{align}
 \vec{n}_{k}  &= \Big(\vec{1} - \hat{J} \cdot \big( \vec{1} - \hat{J} \cdot (\vec{1} - \hat{J} \cdot  \ldots \cdot (1 - \hat{J} \cdot \vec{n}_0) ) \big) \Big)\\
  \vec{n}_{k}  &= (-\hat{J})^{k} \cdot \, \vec{n}_0 + \sum_{i=0}^{k-1} (-\hat{J})^{i} \cdot \vec{1} \: \:\: .\label{eq:parallel_IBMF_1}
\end{align}

In Eq. \eqref{eq:parallel_IBMF_1}, the power $(-\hat{J})^{i}$ must be interpreted as the product of the matrix $\hat{J}$ with itself $i$ times, which returns a matrix. Now, we can use the expression for the geometric sum of matrices to explicitly write the result of the sum on the right-hand side and get:

\begin{align}
 \vec{n}_{k} &= (-\hat{J})^{k} \cdot \, \vec{n}_0 + (\mathbb{I} + \hat{J})^{-1} \cdot \, \big( \, (\mathbb{I} - (-\hat{J})^{k}) \cdot \vec{1} \big) \nonumber \\
 \vec{n}_{k} &=  (\mathbb{I} + \hat{J})^{-1} \cdot \vec{1} + (-\hat{J})^{k} \cdot \Big(\vec{n}_0 - (\mathbb{I} + \hat{J})^{-1} \cdot \vec{1} \Big) \:\:\: ,
\end{align}
where we exploited the fact that $(-\hat{J})^{k}$ and $(\mathbb{I} + \hat{J})^{-1}$ commute.

This expression will converge to the right solution if $\lim_{k \to \infty} (-\hat{J})^{k}=\hat{0}$, or, equivalently, if the eigenvalue of $\hat{J}$ with the maximum absolute value $|\lambda^{J}|_{\max}$ is smaller than one. At this point, it is important to note that this conclusion is independent on the specific interaction graph. The convergence of IBMF with parallel updates at zero temperature can be determined by computing the eigenvalue of $\hat{J}$ with the largest absolute value. 

Unfortunately, the maximum eigenvalue associated to the adjacency matrix of a random regular graph is not inside the bulk $[-2\sqrt{c-1}, \,2 \sqrt{c-1}]$. There is an eigenvalue outside of the bulk and its value is $\lambda_{\max}^{\hat{A}}=c$. Therefore, $|\lambda^{\hat{J}}|_{\max} = c \mu$ and the iterations will not converge for any $\mu \geq \mu^{\text{par}} = 1 / c $. As $\mu^{\text{par}}$ is smaller than $\mu^{\ast}=1/(2 \sqrt{c-1})$ for any $c > 2$, in these cases there will be an interval $\mu \in [\mu^{\text{par}}, \mu^{\ast})$ where the parallel iterations will not converge to the right solution, even if that solution exists. 

To overcome this problem, the results in the main text are obtained using a sequential update. In practice, this means that each $n_i^{k}$ is updated to $n_i^{k+1}$ asynchronously. We choose a random order of the species, and one by one we apply Eq.~\eqref{eq:IBMFT0}. When $n_i$ is recomputed, the new value is ready to be used in the next update. It is important to mention that the problem with parallel updates is already known and has been solved using sequential updates in other contexts \cite{Caltagirone_BPconv_2014}.

\section{Finite size effects on undirected graphs}\label{app:FSundirected}

In Fig.~\ref{fig:individual_abundances} of the main text, we compare the stationary abundances obtained from IBMF with those obtained from dynamical simulations, for each species on a given graph. We consider a random regular graph with connectivity $c=3$ and asymmetric interactions and we observe perfect agreement for species with high average abundances (that correspond to dominant species), while IBMF underestimates the stationary abundances of species with very low abundances. Simulations reveal that these low-abundance species intermittently switch between two states: one characterized by a finite (not small) abundance and another with very low abundance. The analysis presented in the main text was performed for a system of size $N=1024$. Here, we show that these results are robust and do not depend on the system size.

\begin{figure}[t]
\centering

\subfloat[]{
\includegraphics[width=0.48\textwidth]{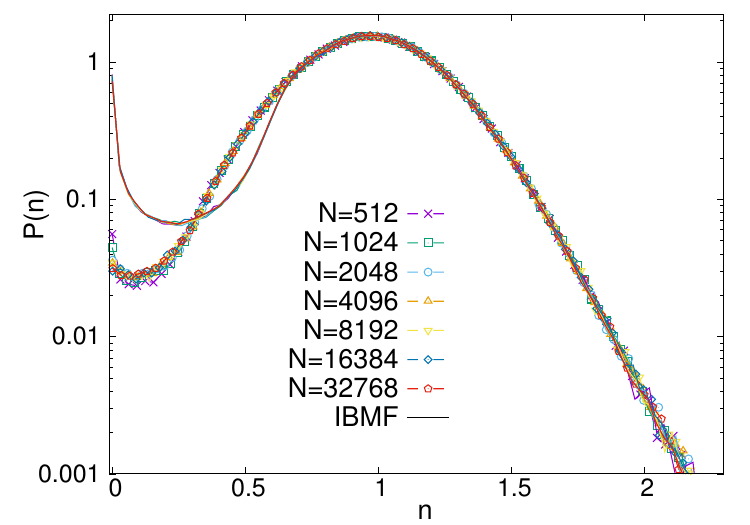}
\label{fig:prob_av_individual_abund}
}
\subfloat[]{
\includegraphics[width=0.48\textwidth]{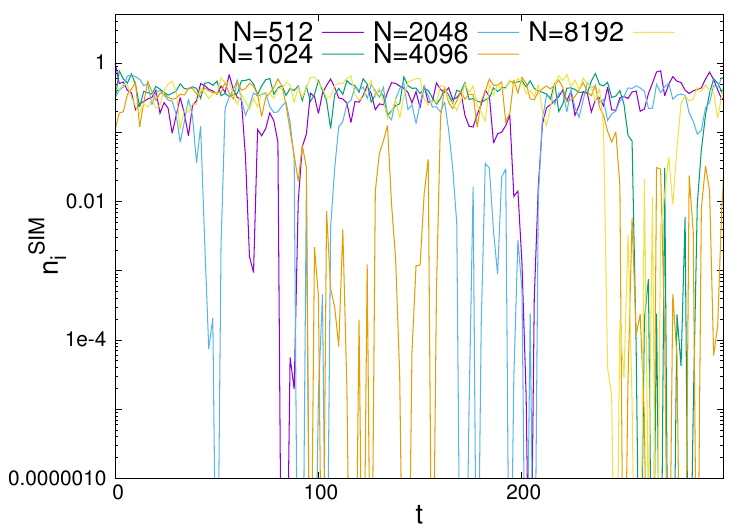} 
\label{fig:low-n_dynamics}
}

\caption{Species abundance distribution {\bf (a)} and low abundance species dynamics {\bf (b)} in a random regular graph with connectivity $c=3$, asymmetric interactions with mean $\mu=0$ and standard deviation $\sigma=0.15$, at finite temperature $T=0.015$. The immigration rate is set to $\lambda=10^{-6}$. Different system sizes are explored. {\bf (a)} Probability density of stationary average abundances, computed using both IBMF (solid lines) and dynamical simulations (symbols), for different values of $N$. For each system size, $S$ independent graphs are generated. For each graph, the stationary average abundances of all $N$ species are computed, and the probability density is constructed from the histogram of the resulting $N\times S$ abundance values. The number of graph realizations $S$ depends on $N$, varying from $S=1000$ for smaller systems to $S=50$ for larger ones. Probability densities are independent of system sizes. {\bf (b)} Dynamics of a single species with low average abundance for various sizes $N$. The switching between two different states persists independently of system size.}
\label{fig:FSundirected}
\end{figure}

Fig.~\ref{fig:prob_av_individual_abund} shows the probability density of stationary average abundances of the different species in the ecosystem, computed using both IBMF and simulations, for system sizes ranging from $N=512$ to $N=32768$. This density, also known as species abundance distribution, reaches a maximum close to $n=1$ and decays exponentially to zero for large abundances. For low abundances, the distribution has a second distinctive maximum, which we associate with a finite subset of species that are close to extinction. As detailed in the caption of Fig.~\ref{fig:FSundirected}, the densities are constructed by averaging over many graph realizations. Both methods give distributions that are essentially independent of $N$. All the curves coincide for $n > 0.6$, which means that IBMF correctly predicts the species abundance distribution for a finite fraction of the species with large average abundances, and that fraction does not change when the number of species increases. For small abundances, IBMF deviates from simulations, but again, the respective distributions are independent of $N$. %In particular, \david{this means that} 
Accordingly, the fraction of stationary abundances for which IBMF deviates from simulations remains unchanged.

In Fig.~\ref{fig:low-n_dynamics}, we show that the switching dynamics of low-abundance species persists across system sizes ranging from $N=512$ and $N=8192$. For all the system sizes considered, we select a single species with small average abundance and plot its abundance in time. We conclude that the switching dynamics is characteristic of at least some species whose average abundance is underestimated by IBMF.

\section{Unbounded growth for asymmetric mutualistic interactions} \label{app:unbounded_growth_mutualistic}

Fig. \ref{fig:Langevin_eps_0} in Subsection \ref{subsec:IBMF_asym} shows that, in random regular graphs with Gaussian asymmetric interactions, the phase with multiple equilibria exists only for positive values of the average interaction strength $\mu$. A positive value of $\mu$ corresponds to ecosystems in which most species develop competitive interactions. On the other hand, when $\mu$ is negative the interactions are mostly mutualistic (species abundances grow together). In this case, we have only one transition line $\sigma_c(\mu)$ that separates two phases. For $\sigma<\sigma_c(\mu)$ the abundances converge to a single equilibrium state, while for $\sigma>\sigma_c(\mu)$ at least one abundance grows indefinitely and diverges for long times. The latter is called unbounded growth.

\begin{figure}[t]
\centering
\includegraphics[width=0.48\textwidth]{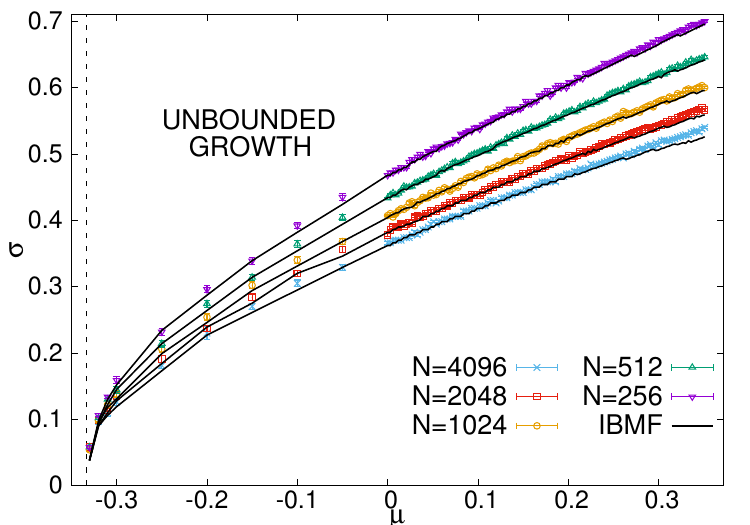}

\caption{Transitions of the Generalized Lotka-Volterra model for different system sizes at $T=0$. The interactions are asymmetric ($\alpha_{ij}$ is chosen independently of $\alpha_{ji}$) and defined on random regular graphs with connectivity $c=3$. Points represent the results of simulations with immigration rate $\lambda=10^{-6}$, and lines are the predictions made with IBMF for the same sizes. Each transition was determined using $10000$ graphs. Simulations are repeated for $10$ different initial conditions. IBMF was run with sequential updates for $10$ different random initial conditions. Points (lines) mark the minimum value of $\sigma$ such that simulations (IBMF) displayed unbounded growth (not converged) in more than $50\%$ of the interaction graphs. The vertical line marks the limit value $\mu=-1/c\approx-0.333$}
\label{fig:IBMF_Langevin_mutualistic}
\end{figure}

The results in the right panel of Fig. \ref{fig:IBMF_asym} are easily extended to $\mu<0$. Fig. \ref{fig:IBMF_Langevin_mutualistic} shows that IBMF (lines) maintains a good agreement with the results of the simulations (points) for random regular graphs with connectivity $c=3$. For $\mu < -1/c\approx-0.333$ the abundances diverge for any value of $\sigma$, and the transition line goes to $\sigma(-1/3)=0$. This is consistent with the fact that at $\sigma=0$ the interaction strengths $\alpha_{ij}$ are homogeneous and all equal to $\mu$. The solution in the single equilibria phase is then $n_i=1/(1+c \, \mu)$ for all species (for $i=1,\ldots,N)$. Given that $n_i$ must be non-negative, when $\mu<-1/c$ we do not have a feasible solution anymore and the abundances diverge in any simulation.

\section{Transitions and coexisting phases for asymmetric interactions}\label{app:coexistence}

We identified two distinct types of transitions in the phase diagram of Fig. \ref{fig:Langevin_eps_0}, for a system with undirected, asymmetric interactions. First, a transition at $\sigma_{SFP}(\mu)$ from a region ($\sigma < \sigma_{SFP}(\mu)$) where the dynamics reaches a single fixed point to a region ($\sigma > \sigma_{SFP}(\mu)$) where one of two things happens: either the abundances reach multiple fixed points depending on the initial conditions, or at least one of them diverges. Second, a transition at $\sigma_{UG}(\mu)$ such that at least one species abundance diverges in most ecosystems generated with $\sigma>\sigma_{UG}(\mu)$.

To define these transitions quantitatively, we can introduce the fraction $f_{SFP}$ of ecosystems for which the species reach a single equilibrium, the fraction $f_{MFP}$ of ecosystems for which they reach multiple fixed points depending on the initial conditions, and the fraction $f_{UG}$ of ecosystems for which at least one species abundance diverges. These three, of course, are functions of the parameters $(\mu, \sigma)$ used to generate the network of interactions. Then, the transition at $\sigma_{SFP}(\mu)$ can be defined on the curve where $f_{SFP}\big(\mu, \sigma(\mu)\big)=1/2$, {\it i.e.}, where 50\% of the generated ecosystems have a single fixed point. Analogously, the transition at $\sigma_{UG}(\mu)$ can be defined on the curve where $f_{SFP}\big(\mu, \sigma(\mu)\big)=1/2$, , {\it i.e.}, where 50\% of the generated ecosystems display unbounded growth.

\begin{figure}[t]
\centering

\subfloat[]{
\includegraphics[width=0.45\textwidth]{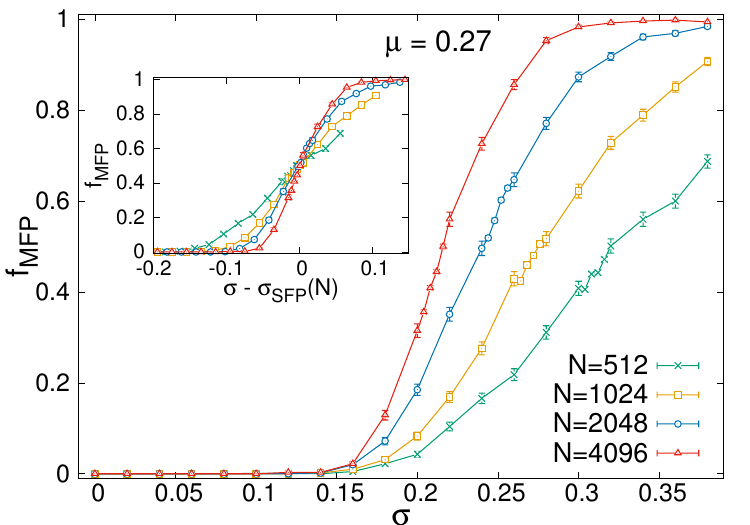} 
\label{fig:simulations_T0_fraction_mult}
}
\subfloat[]{
\includegraphics[width=0.45\textwidth]{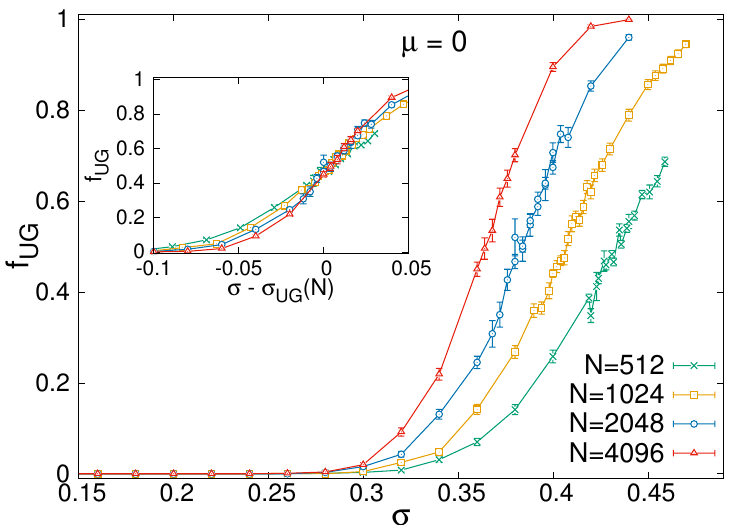}
\label{fig:simulations_T0_fraction_div}
}

\caption{Fraction of ecosystems with multiple fixed points ($f_{MFP}$, {\bf panel (a)}) and fraction of ecosystems displaying unbounded growth ($f_{UG}$, {\bf panel (b)}) for systems with asymmetric interactions, where $\alpha_{ij}$ is drawn independently from $\alpha_{ij}$ from the Gaussian distribution $\mathcal{N}(\mu, \sigma)$. The fractions are shown as a function of $\sigma$ for fixed values of $\mu$ ($\mu=0.27$ and $\mu=0$ for panel (a) and panel (b), respectively). Different colors correspond to different system sizes $N$. In the inserted graphics, the values of $\sigma$ are rescaled so that all transitions are localized at the point with coordinates $(0,1/2)$. In panel (a), we subtract $\sigma_{SFP}(N)$ from $\sigma$, while in panel (b) we subtract $\sigma_{UG}(N)$ from $\sigma$.}
\label{fig:simulations_T0_fractions}
\end{figure}

In Fig. \ref{fig:simulations_T0_fraction_mult}, we show the fraction of ecosystems with multiple fixed points $f_{MFP}$ for $\mu=0.27$ and different values of $\sigma$ for which we do not observe unbounded growth in any sample. In this case, we simply have that $f_{SFP}=1-f_{MFP}$ and in this situation the transition, as defined above, takes place at the value $\sigma_{SFP}(\mu=0.27)$ where $f_{SFP} =f_{MFP}=1/2$. The figure indicates that the transition is moving towards smaller values of $\sigma$ when the system size $N$ increases. In other words, $\sigma_{SFP} \equiv \sigma_{SFP}(\mu, N)$ is also a function of $N$.

With a standard statistical mechanics procedure, in the insert of fig. \ref{fig:simulations_T0_fraction_mult} we plot $f_{MFP}$ as a function of $ \sigma-\sigma_{SFP}(\mu=0.27,N)$. In this way the transition point corresponds to the point $(0,1/2)$ for all the curves at different $N$ and it is easier to visualize that the transition becomes more abrupt when the number of species $N$ increases. Although for any finite $N$ we observe an interval in $\sigma$ where both $f_{SFP}$ and $f_{MFP}$ are positive, this region of coexistence shrinks when $N$ increases.

Fig. \ref{fig:simulations_T0_fraction_div} shows, instead, the fraction of samples displaying unbounded growth for different values of $\sigma$ at $\mu=0$. For this value of $\mu$, we observe very few samples with multiple fixed points. As before, the transition $\sigma_{UG}\equiv\sigma_{UG}(\mu, N)$, located at the value of $\sigma$ where $f_{UG}=1/2$, is a function of both $\mu$ and the number of species $N$. The inserted graphic also indicates that this transition is more abrupt when $N$ increases. Therefore, the interval in $\sigma$ for which we observe coexistence between unbounded growth and other behaviors shrinks when the ecosystem becomes larger.

\begin{figure}[t]
\centering

\includegraphics[width=0.48\textwidth]{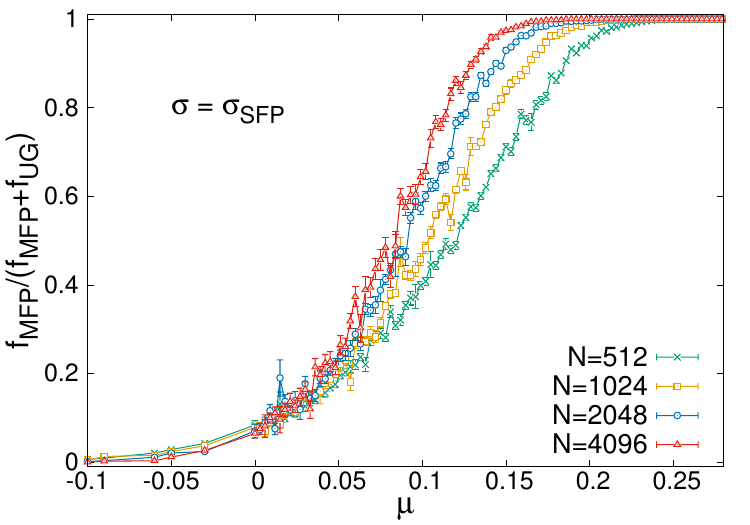} 
\caption{Measuring the dominance of the multiple fixed points over the unbounded growth for ecosystems with asymmetric interactions, where $\alpha_{ij}$ is drawn independently from $\alpha_{ij}$ from the Gaussian distribution $\mathcal{N}(\mu, \sigma)$. We report the ratio $f_{MFP} / (f_{MFP}+f_{UG})$ for different values of $\mu$ for systems generated over the transition line $\sigma=\sigma_{SFP}(\mu,N)$. Different colors represent different ecosystem sizes $N$. Curves at different values of $N$ cross around $\mu=0.$}
\label{fig:simulations_T0_over_transitions}
\end{figure}

Finally, in Fig. \ref{fig:simulations_T0_over_transitions} we also show that the region in $\mu$ where we have coexisting phases shrinks when $N$ increases. In order to do so, we measure the dominance of the ecosystems with multiple fixed points along the transition curve $\sigma_{SFP}(\mu, N)$. For different values of $\mu$ and $N$, we first determine the location of $\sigma_{SFP}(\mu, N)$ by analyzing data like the one in Fig. \ref{fig:simulations_T0_fraction_mult}. We then measure $f_{MFP}$ and $f_{UG}$ precisely at $\sigma=\sigma_{SFP}(\mu, N)$, and compute the ratio $f_{MFP} / (f_{MFP}+f_{UG})$. When this ratio is close to zero, the transition is dominated by samples displaying unbounded growth. When it is close to one, the transition is dominated by samples where the dynamics reach different fixed points depending on the initial conditions. In between, we have coexistence of all three phases, including also the one with single fixed points since $f_{SFP}=1/2$ at $\sigma=\sigma_{SFP}(\mu, N)$ by definition.

We again observe that the curves become steeper when the number of species $N$ increases. Moreover, they seem to cross around $\mu \sim 0$. To the left (for $\mu<0$), we expect that the unbounded growth dominates for any $\sigma>\sigma_{SFP}(\mu, N)$, provided that $N$ is large enough. More precisely, $f_{UG}$ should be close to one in that region. The reader should note that, for $\mu <0$, we have $\sigma_{SFP}(\mu, N)=\sigma_{UG}(\mu, N)$ (see Fig. \ref{fig:Langevin_eps_0}).

To the right (for $\mu > 0$), we expect the samples with multiple fixed points to dominate immediately above $\sigma_{SFP}(\mu, N)$. Unbounded growth will appear again if we keep increasing $\sigma$. In other words, $f_{MFP}$ will be close to one in the interval $\sigma_{SFP}(\mu, N)<\sigma<\sigma_{UG}(\mu, N)$ for any fixed $\mu>0$ and for $N$ large enough. Furthermore, the interval in $\mu$ for which we observe coexistence between samples displaying a single fixed point, multiple fixed points, and unbounded growth also shrinks when the number of species increases.

\section{Use of damping to improve convergence} \label{app:damping}

\begin{figure}[t]
\centering

\subfloat[]{
\includegraphics[width=0.45\textwidth]{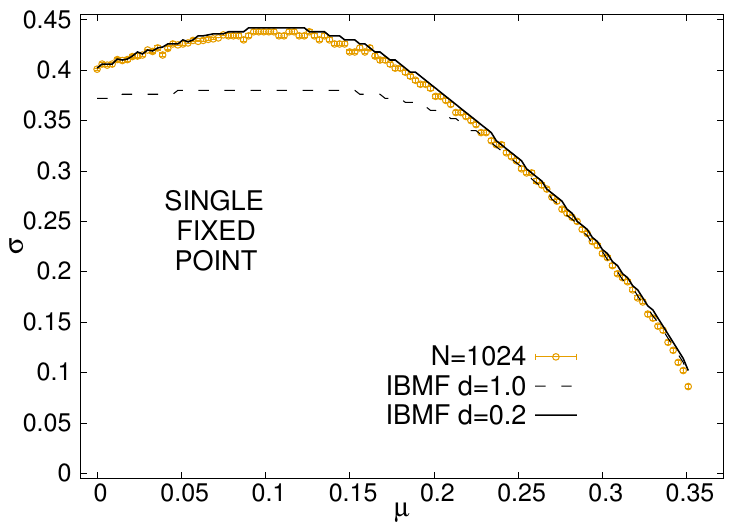} 
\label{fig:IBMF_T0_different_damping}
}
\subfloat[]{
\includegraphics[width=0.45\textwidth]{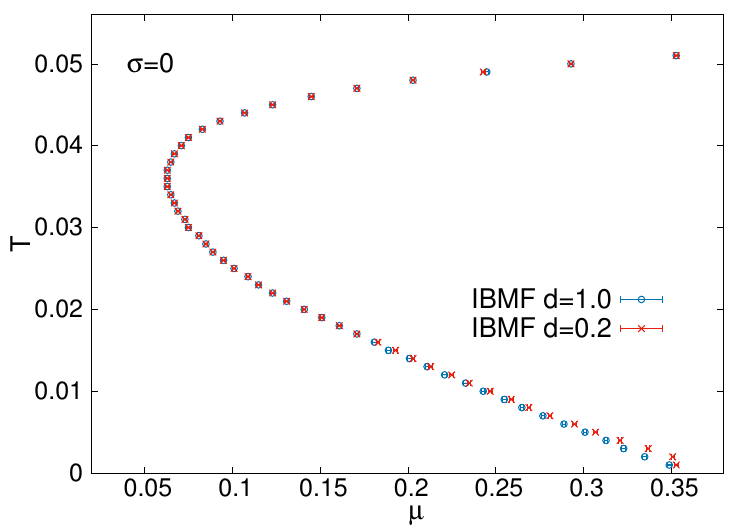}
\label{fig:IBMF_sigma0_different_damping}
}

\caption{Effects of damping in the results of IBMF for the phase diagrams of the gLV model. The equations are run on random regular graphs with size $N=1024$ and connectivity $c=3$. Each transition was determined using $10000$ graphs. In panel {\bf (a)} (panel {\bf (b)}), we mark the maximum value of $\sigma$ (resp. $\mu$) such that simulations or IBMF converged to the same fixed point in more than $50\%$ of the interaction graphs. IBMF was run on each graph with damping ($d=0.2$) and without damping ($d=1$) for $10$ different random initial conditions. {\bf (a)} Phase diagram at $T=0$. The interactions are asymmetric ($\alpha_{ij}$ is chosen independently of $\alpha_{ji}$). Points represent the results of simulations with immigration rate $\lambda=10^{-6}$, and lines are obtained with IBMF. Simulations are repeated for $10$ different initial conditions. {\bf (b)} Results of IBMF in the presence of thermal noise for graphs with symmetric and homogeneous interactions (drawn using $\sigma=0$). The immigration rate is $\lambda=10^{-6}$.}
\label{fig:IBMF_different_damping}
\end{figure}

As usual in the scenario of iterating equations until the quantities reach a fixed point, some standard tricks can be used to help IBMF converge. Perhaps the most common mechanism is to add damping to the iterations. Given the update rule $m_i=f_i(\{m_k\}_{k\in \partial i^{-}})$, one chooses a parameter $d \in [0, 1]$, and updates the vector $\{m_i^{k}\}_{i=1,\ldots, N}$ of the average abundances by doing:

\begin{equation}
 m_i^{k+1} =d \cdot f\big( \{m_k\}_{k\in \partial i^{-}} \big) + \:(1-d) \cdot  m_i^{k} \label{eq:IBMFT0_damping} \:\:\: . \:
\end{equation}

The value $d=1$ corresponds to the original case, where IBMF is iterated without damping. When $d=0$ nothing happens to $m_i^{k}$. A quick study shows that, for IBMF, the probability of convergence is maximized for some intermediate $d$ around $d=0.2$. The results in Subsections \ref{subsec:IBMF_asym} and \ref{subsec:IBMF_directed} are obtained using precisely this value ($d=0.2$). In the latter case, the impact of damping is explained in detail in Appendix \ref{app:damping_directed_graphs}. In the first case, achieving convergence with IBMF is an important issue due to the coexistence of the phases of multiple fixed points and unbounded growth. 

For $T=0$ any fixed point of IBMF is also a fixed point for simulations, and we can be sure that whenever we find different fixed points with IBMF this has implications also for simulations. However, the phase of unbounded growth is determined by the divergence of the abundances, and, as said in the main text, it could be that the iteration process of IBMF does not converge while the simulations do. Fig. \ref{fig:IBMF_T0_different_damping} shows that the differences between IBMF without damping and simulations are indeed noticeable only for $\mu \leq 0.2$, where unbounded growth starts to dominate. Therefore, one needs to add damping to overcome the convergence problems that are not physical and are only related to the algorithmic dynamics of the iterations.

On the other hand, the phase diagram of Fig. \ref{fig:IBMFs_BP_phase_diagram} is produced without damping (using $d=1$). Here, we do not find any problems in achieving convergence with IBMF. In fact, in that phase diagram the unbounded growth phase is not present. The iteration process reaches a fixed point for any temperature $T$ and average interaction strength $\mu$. The only relevant question is, at a given temperature, what is the smallest value of $\mu$ where we can find two different fixed points. Fig. \ref{fig:IBMF_sigma0_different_damping} shows that the answer is approximately the same in most of the phase diagram, except at very low temperatures. As expected, the effect of adding damping, if any, is to move the transition to larger values of $\mu$. Intuitively, the damping could stabilize one fixed point more than the others, preventing the algorithm from sampling them with the right probability. When one wants to correctly locate the transition between the single and the multiple attractor phases, the correct physical results are obtained by using IBMF without damping ($d=1$).

\section{Global averages from IBMF and simulations in heterogeneous graphs}\label{app:mean-abundance-heterogeneous-graphs}

To confirm the fact that IBMF can be used to predict global statistical properties, such as the average abundance of all the species in the ecosystem, here we present results for a random regular graph with heterogeneous and asymmetric interactions. In this case, we do not consider thermal noise ($T=0$), and the scenario coincides with the one of Subsection \ref{subsec:IBMF_asym} in the main text. 

\begin{figure}[t]
\centering
\includegraphics[width=0.48\textwidth]{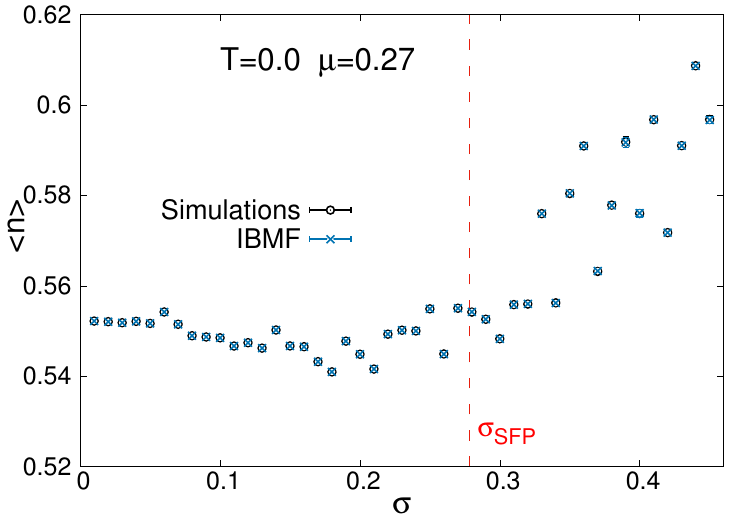}
\caption{Average abundance in a single ecosystem as a function of $\sigma$, with $\mu=0.27$ and $T=0$. Black dots correspond to simulations of the dynamics, while blue crosses correspond to IBMF. Both are computed on a random regular graph with connectivity $c=3$, size $N=1024$, and asymmetric interactions. For each value of $\sigma$, both methods are run on the same interaction graph from 1000 different initial conditions. Averages are first computed over the $N$ stationary abundances obtained in each run and then over initial conditions. The red dashed vertical line marks the value of $\sigma$ at which the transition from a single fixed point to multiple fixed points occurs, as reported in the main text. Error bars are of the size of the standard deviation of the average abundance, computed over different initial conditions.}
\label{fig:IBMF_simulations_av-n_heterogeneous}
\end{figure}

Fig.~\ref{fig:IBMF_simulations_av-n_heterogeneous} shows the average abundances for a single graph with connectivity $c=3$ and size $N=1024$. For fixed $\mu=0.27$, averages are computed using both IBMF and simulations, for different values of the interaction heterogeneity $\sigma$. For each value of $\sigma$, the averages are obtained by first averaging over the $N$ species and then over multiple initial conditions. As shown in Fig.~\ref{fig:IBMF_simulations_av-n_heterogeneous}, the results from IBMF and simulations are in perfect agreement across the entire range of $\sigma$, both in the single fixed point phase and in the multiple fixed points one. The transition between the two phases is indicated by the red dashed line. This agreement in global observables further supports the validity of the IBMF approach.

\section{IBMF for graphs with correlated couplings} \label{app:correlated_alpha_T0}

In Subsection \ref{subsec:IBMF_asym}, we present the results of IBMF and simulations of the gLV model on random regular graphs with asymmetric interactions. In that case, the interaction graphs were built drawing $\alpha_{ij}$ and $\alpha_{ji}$ independently for every pair of interacting species. This choice automatically sets the connected correlation $\langle \alpha_{ij} \alpha_{ji} \rangle - \langle \alpha_{ij} \rangle \langle \alpha_{ji} \rangle$ to zero. 

It is important, however, to verify that the accuracy of IBMF's predictions extends to cases where the connected correlation is not zero. Fortunately, we can use a simple procedure to build correlated couplings. For each pair $i \to j$ and $j \to i$ of interacting species (let $i < j$ just to fix ideas), we do one of two things: i) with probability $\epsilon$ we choose $\alpha_{ij}$ from the Gaussian $\mathcal{N}(\mu, \sigma)$ and then we set $\alpha_{ji}=\alpha_{ij}$, or ii) with probability $1-\epsilon$ we independently draw $\alpha_{ij}$ and $\alpha_{ji}$ from the same Gaussian. Evidently, the setting used in Subsection \ref{subsec:IBMF_asym} corresponds to $\epsilon=0$.

\begin{figure}[t]
\centering

\subfloat[]{
\includegraphics[width=0.45\textwidth]{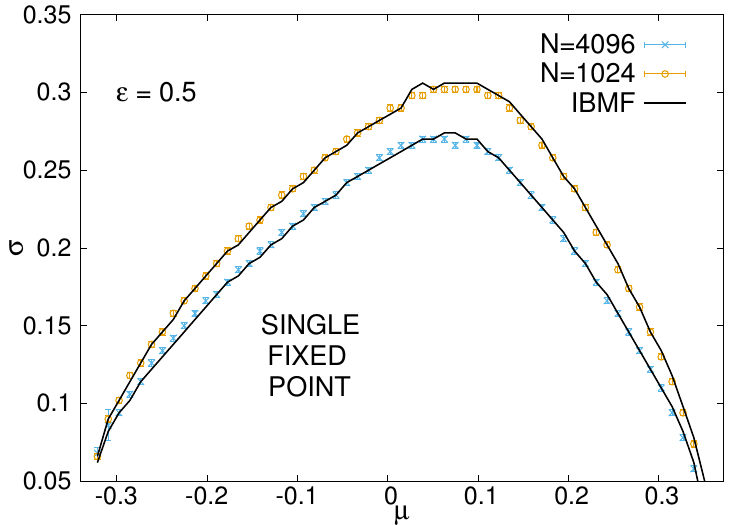} 
\label{fig:IBMFs_Langevin_T0_mult_eps05}
}
\subfloat[]{
\includegraphics[width=0.45\textwidth]{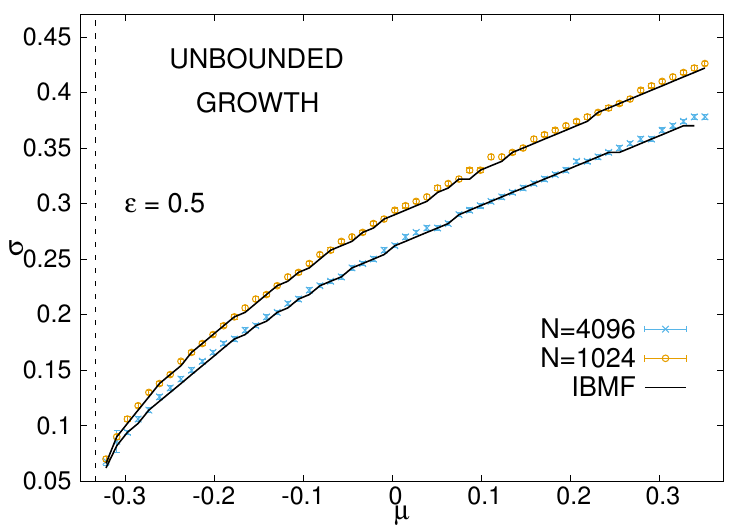} 
\label{fig:IBMFs_Langevin_T0_div_eps05}
}

\caption{Transitions of the gLV model for different system sizes at $T=0$. The interactions are asymmetric and defined on random regular graphs with connectivity $c=3$. The couplings are chosen such that with probability $\epsilon=0.5$ we have $\alpha_{ij}=\alpha_{ji}$. Points represent the results of simulations with immigration rate $\lambda=10^{-6}$, and lines are the predictions made with IBMF for the same sizes. Each transition was determined using $10000$ graphs. Simulations are repeated for $10$ different initial conditions. IBMF was run with damping (see Appendix \ref{app:damping}) for $10$ different random initial conditions. {\bf (a)} For each system size $N$ and average strength $\mu$, points (lines) mark the maximum value of $\sigma$ such that simulations (IBMF) converged to the same fixed point in more than $50\%$ of the interaction graphs. {\bf (b)} Points (lines) mark the minimum value of $\sigma$ such that simulations (IBMF) displayed unbounded growth (not converged) in more than $50\%$ of the interaction graphs.}
\label{fig:IBMF_asym_eps05}
\end{figure}

Fig. \ref{fig:IBMF_asym_eps05} shows the results for $\epsilon=0.5$, chosen such that the interactions are still asymmetric, but correlated. As in Subsection \ref{subsec:IBMF_asym}, the predictions obtained with IBMF for the transitions of the gLV model are in very good agreement with simulations in this case. IBMF provides a precise description of the model's fixed points at $T=0$, valid for different values of $\epsilon$.

\section{Convergence of IBMF in directed graphs} \label{app:damping_directed_graphs}

The convergence of IBMF is sensitive to the addition of damping. In the toy model described in Subsection \ref{subsec:IBMF_directed}, using no damping ($d=1$ in Eq. \eqref{eq:IBMFT0_damping}) has negative implications on the convergence, and the results no longer coincide with the predictions in Ref. \cite{StavPRE2024}. Specifically, as we will discuss in this appendix, numerical simulations only exhibit fluctuations in the presence of unstable isolated odd cycles, while the undamped IBMF algorithm fails to converge also in the presence of isolated even cycles; the inclusion of damping reconciles the IBMF results with the simulations.

\begin{figure}[t]
\centering

\subfloat[]{
\includegraphics[width=0.45\textwidth]{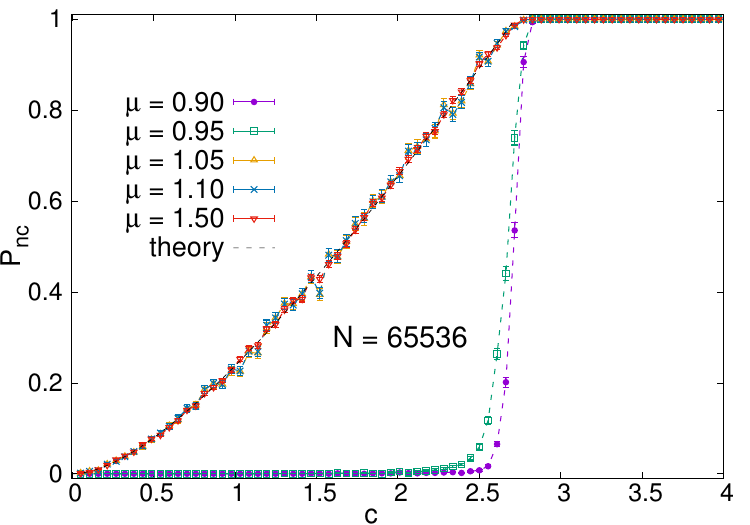} 
\label{fig:IBMF_without_damping_several_mu}
}
\subfloat[]{
\includegraphics[width=0.45\textwidth]{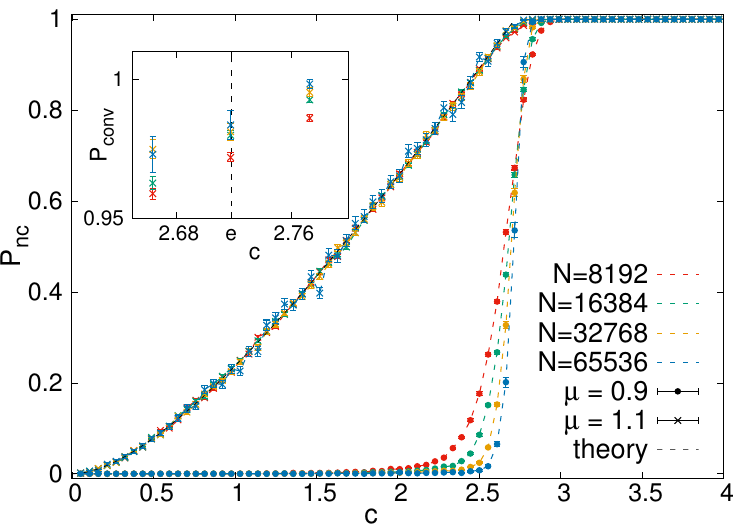}
\label{fig:IBMF_without_damping_several_N}
}

\caption{Probability that IBMF does not converge ($P_{\text{nc}}$) in instances of the toy model (directed graphs with $\sigma=0$). Each point is obtained by running IBMF for $10000$ different realizations of the interaction graph with a given average connectivity $c$, size $N$, and interaction strength $\mu$. In both panels, IBMF is run without damping, and for $\mu>1$ its results follow a unique function represented using a dashed line. This function is the analytical result (see the computation in the text) for the probability of having at least one isolated cycle in the graph formed by species with completely polarized abundances ($n_i=0$ or $n_i=1$). It is close to zero for small $c$ and goes to one exactly at $c=e$. {\bf (a)} IBMF convergence without damping for $N=65536$ and different values of $\mu$. {\bf (b)} IBMF convergence without damping for two values of the interaction strength around the transition ($\mu=0.9$ and $\mu=1.1$) obtained with several system sizes. The inserted graphic shows an enlargement of the zone where $P_{\text{conv}}\sim 1$.}
\label{fig:IBMF_directed_damping_1}
\end{figure}

In Fig.~\ref{fig:IBMF_without_damping_several_mu}, we show the probability that IBMF, without damping, does not converge ($P_{\text{nc}}$) for large graphs with different average connectivities and interaction strengths. In this case, $P_{\text{nc}}$ is independent of $\mu$ for all $\mu>1$. It follows a slowly increasing function that goes from zero at $c=0$ to one at $c=e$. This function is represented with a dashed line in the figure, and we give its precise mathematical form below. This behavior is also nearly independent of the size $N$, as can be seen in Fig.~\ref{fig:IBMF_without_damping_several_N}. Only close to $P_{\text{nc}}\sim 1$, for $c\sim e$, finite-size effects make IBMF have a small deviation from the dashed line. The empirical $P_{\text{nc}}$ is not exactly equal to one at $c=e$ for finite sizes, but the inserted graphic shows that $P_{\text{nc}}$ increases when the number of species $N$ increases.

To compute the function followed by IBMF for $d=1$ and $\mu>1$, we can do something analogous to what the authors of Ref. \cite{StavPRE2024} did to predict the probability of fluctuations in simulations. Their results indicate that, for any $\mu>1$ and in the stationary state, all species in the graph are completely polarized, \textit{i.e.}, either they are extinct ($n_i=0$) or they reach their carrying capacity ($n_i=1$). First, we write the probability $\phi$ that a species is isolated and therefore can have $n_i=1$. In tree-like graphs, the neighbors belong to nearly independent subgraphs, and we get $\phi=\sum_{k=0}^{\infty} p(k) (1-\phi)^{k}$, where $k$ is the connectivity of the species and $p(k)$ is its distribution. When $p(k)$ is Poisson, the authors of Ref. \cite{StavPRE2024} show that $\phi=W(c)/c$, where $c$ is the average connectivity and $W(x)$ is the Lambert $W$ function. Second, we compute the probability that a directed cycle of length $n$ is isolated. Indeed, a species in a directed cycle has one incoming edge from another species inside the cycle. For the cycle to be isolated, we need all the other incoming edges to correspond to extinct species. Since in Poisson graphs the number of these other incoming edges is also distributed as Poisson with the same mean, and the neighbors of different species in the cycle are independent, the probability that the cycle of length $n$ is isolated is simply $\phi^{n}$.

The number of directed cycles of length $n$ in a Poisson graph with mean connectivity $c$ is also Poisson distributed with mean $c^{n}/n$ \cite{Bollobas}. Thus, the number of directed isolated cycles is Poisson distributed with mean $(c \,  \phi)^{n}/n=[W(c)]^{n}/n$. The probability of having at least one cycle is then:

\begin{eqnarray}
    f(c) \!\!\!\! &=& \!\!\!\!  1 - \exp \Big\{-\sum_{n=2}^{\infty} \frac{[W(c)]^{n}}{n} \Big\} \nonumber \\
    f(c) \!\!\!\!  &=&  \!\!\!\! 1 - \frac{c}{W(c)} \, \big(\, 1-W(c) \, \big) \label{eq:prob_any_cycle} \: ,
\end{eqnarray}
where we used that $W(c) \, e^{W(c)}=c$. This function $f(c)$ is the one represented with dashed lines in Fig. \ref{fig:IBMF_directed_damping_1}.

On the other hand, for $\mu<1$, the probability that IBMF does not converge without damping is very close to zero for all $c<e$. In this regime, the results for different system sizes $N$ have a nice crossing point at $c>e$, as can be seen in Fig. \ref{fig:IBMF_without_damping_several_mu} for $\mu=0.9$. This is a familiar feature of a first-order phase transition that happens exactly at that crossing point. To the left of the crossing, the probability that IBMF does not converge goes to zero when the number of species $N$ goes to infinity. To the right, the probability approaches one as $N \to \infty$. 

Adding damping considerably helps IBMF to converge at any $\mu$. For the interesting case $\mu>1$, setting $d=0.2$ avoids the problems caused by a subset of the directed isolated cycles. Now, the probability $P_{\text{nc}}$ does not follow a unique function $f(c)$ for all $\mu>0$ (see Fig. \ref{fig:IBMF_directed}). They coincide, instead, with the predictions made by the authors of Ref. \cite{StavPRE2024}. Their computation is analogous, but with a key difference. They concluded that, for the simulations, the cycles with even length do not cause fluctuations. If we exclude the even values of $n$, Eq.~\eqref{eq:prob_any_cycle} changes to:

\begin{eqnarray}
    f_{3}(c) \!\!\!\! &=& \!\!\!\! 1 - \exp \Big\{-\sum_{k=1}^{\infty} \frac{[W(c)]^{2k+1}}{2k+1} \Big\} \nonumber \\
    f_{3}(c) \!\!\!\! &=& \!\!\!\! 1  - \frac{c}{W(c)} \, \sqrt{\frac{1-W(c)}{1+W(c)}} \label{eq:prob_odd_cycle} \:\: .
\end{eqnarray}

Eq. \eqref{eq:prob_odd_cycle} gives the probability $f_3(c)$ of having at least one directed isolated cycle with odd length. Furthermore, a cycle of odd length $n$ will be unstable, and thus will fluctuate, for all $\mu > \mu_c(n) = 1 / \cos(\pi / n) > 1$ \cite{StavPRE2024}. Therefore, Eq. \eqref{eq:prob_odd_cycle} gives the probability of fluctuations for any $\mu>1/\cos(\pi/3)=2$. When $\mu < 2$ the cycles of length $n=3$ are stable, but the ones with $n=5$ are still unstable for any $\mu>1/\cos(\pi/5)\approx1.24$. Thus, to compute the line that corresponds to the blue points (done for $\mu=1.5$) in Fig. \ref{fig:IBMF_directed_several_mu}, we simply need to subtract the number of cycles with length $n=3$ from the sum in Eq. \eqref{eq:prob_odd_cycle}. We get the probability:

\begin{eqnarray}
    f_{5}(c) \!\!\!\! &=& \!\!\!\! 1 - \exp \Big\{-\sum_{k=2}^{\infty} \frac{[W(c)]^{2k+1}}{2k+1} \Big\} \label{eq:prob_odd_cycle_greater_5} \\
    f_{5}(c) \!\!\!\! &=& \!\!\!\! 1-\exp \Big\{\frac{1}{2} \ln\Big(\, \frac{1+W(c)}{1-W(c)} \, \Big) \! + W(c) + \! \frac{[W(c)]^{2}}{3} \Big\} \nonumber \: ,
\end{eqnarray}
that we also plot using a dashed line in Fig. \ref{fig:IBMF_directed_several_mu}, showing that it indeed coincides very well with the results of IBMF obtained at $\mu=1.5$.

This is the procedure to follow for any $\mu$. Fig. \ref{fig:IBMF_directed_several_mu} indicates that, once we use damping, the convergence of IBMF stops being affected by the cycles of even length and coincides with the theoretical predictions for the probability of fluctuations as presented in Ref. \cite{StavPRE2024}.

\section{Finite size effects of IBMF on directed graphs}  \label{app:finite_size_directed}

\begin{figure}[t]
\centering
\includegraphics[width=0.48\textwidth]{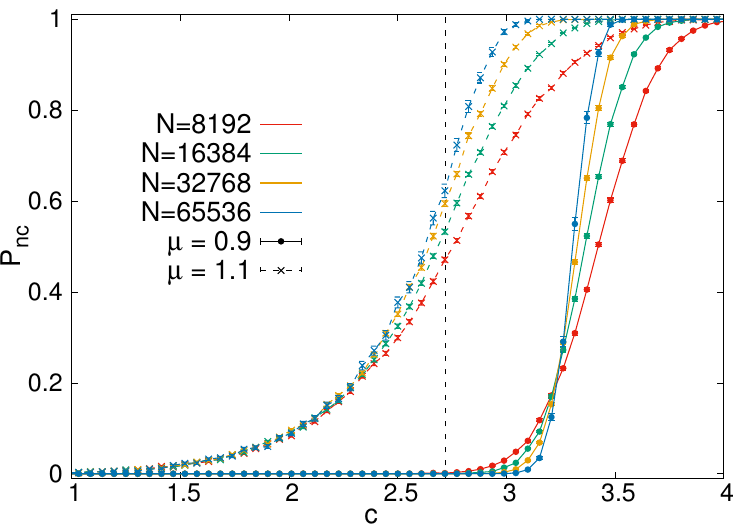}

\caption{Probability that IBMF does not converge ($P_{\text{nc}}$) in instances of the modified toy model (directed graphs with $\sigma=0$). Each point is obtained by running IBMF for $10000$ different realizations of the interaction graph with a given average connectivity $c$ and several sizes $N$. IBMF is run with damping for two values of the interaction strength around the transition ($\mu=0.9$ and $\mu=1.1$). The vertical line marks the value $c=e$.}
\label{fig:IBMF_directed_damping_02_several_N}
\end{figure}

Fig. \ref{fig:IBMF_directed_damping_02_several_N} shows the finite size effects for two values of $\mu$ around $\mu=1.0$ when we run IBMF for the toy model defined in Subsection \ref{subsec:IBMF_directed}. We observe two distinct types of transitions in the probability that IBMF does not converge ($P_{\text{nc}}$). As with the simulations in Ref. \cite{StavPRE2024}, when $\mu>1$ the results for finite systems do not reach $P_{\text{nc}}=1$ exactly at $c=e$. We present numerical evidence that, for $\mu=1.1$, the probability $P_{\text{nc}}$ increases with the system size and the points move to the left towards the line $c=e$.

The curves for $\mu=0.9$, instead, have a clear crossing point at $c \sim 3.28$. When the number of species $N$ increases, the probability $P_{\text{nc}}$ has a sharper transition between $P_{\text{nc}} \sim 0$ to the left and $P_{\text{nc}} \sim 1$ to the right of the crossing point. Therefore, the value $c \sim 3.28$ is a good estimate for the location of the transition between the single equilibrium phase and the phase with global fluctuations. Indeed, it is compatible with the results in Fig. 1 of Ref. \cite{StavPRE2024}.

\section{Comparing runtimes of IBMF and simulations in directed graphs}\label{app:runtimes_directed_ER}

One of the advantages of IBMF is that it can be implemented efficiently (see Section 3 of the SM). Here, we show that running IBMF in single graphs is considerably faster than running simulations. Although we present data for the toy model on directed graphs of Subsection \ref{subsec:IBMF_directed}, this conclusion is generally applicable to all the scenarios discussed in this article.

\begin{figure}[t]
\centering
\includegraphics[width=0.48\textwidth]{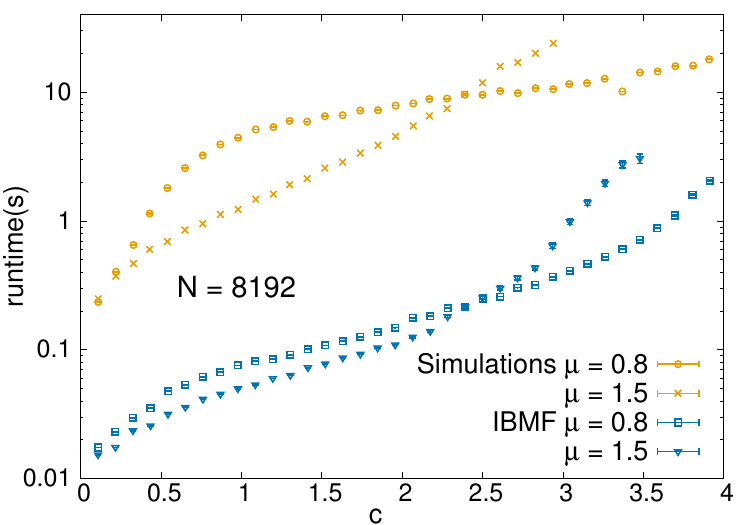}

\caption{Average runtime (in seconds) required to obtain IBMF predictions and to run simulations for the toy model (directed graphs with $\sigma=0$, see Subsection \ref{subsec:IBMF_directed}). Runs that show persistent fluctuations are not included. System size is $N=8192$ in both cases. The points, with error bars, are averages over different graphs for which IBMF and the simulations converge to a fixed point. We consider $10000$ graphs in total, and discarded the cases where fewer than $200$ lead to convergence.}
\label{fig:directed_ER_runtimes}
\end{figure}

In Fig. \ref{fig:directed_ER_runtimes}, we compare the average wall-clock time required for IBMF to converge to a fixed point with the corresponding simulation wall-clock time. The average discards all the samples that lead to persistent fluctuations for long times. For all connectivities and for both values of $\mu$ in the figure, IBMF is consistently around 10 times faster to reach convergence. Both algorithms, available at Ref. \cite{GenLotkaVolterra_SparseGraphs}, were run on a single CPU Intel Xeon Gold 6248 2.5G.

It is important to note that the iterations required by IBMF to converge do not possess a physical meaning. The process is discrete, and at each step we update one of the average abundances $m_i$ using Eq.~\eqref{eq:IBMFT0}. On the other hand, the simulation involves integrating a differential equation (see Eq.~\eqref{eq:LVmodel_T0}) whose time $t$ does have a physical meaning and is a continuous variable. Therefore, the wall-clock time is sensitive to the precision of the integration in time. In this case, we use an adaptive step size to optimize the number of steps needed to reach convergence.

Our results indicate that, even at zero temperature, where the dynamics is simpler to simulate, it is advantageous to run IBMF instead. It gives fast and accurate predictions, as can be seen in Subsections \ref{subsec:IBMF_asym} and \ref{subsec:IBMF_directed}.

\section{Zero-temperature limit of BP}\label{app:BP_T0}

We discuss here the zero temperature limit of BP for random regular graphs with symmetric and homogeneous interactions ($\sigma=0$, see Section \ref{subsec:IBMFsym}). We show that the exact results for the single-to-multiple equilibria \cite{StavPLOS2022} can be easily retrieved after properly taking the limit $T \to 0$. This has already been suggested by the numerical results in Ref. \cite{Tonolo2026sparse}, obtained with the discretized version of BP, and by the results presented here in Fig. \ref{fig:IBMFs_BP_phase_diagram}.

When the temperature is small, the probability densities concentrate around the mean values. We can then assume that the message $\eta_{i \to j}(n_i)$ can be written as the multiplication of a Gaussian factor and a power-law factor $n_i^{\beta \lambda - 1}$ as follows:

\begin{equation}
\eta_{i \to j}(n_i) = \frac{1}{z_{i \to j}} n_i^{\beta \lambda - 1} \exp \Big\{ -\frac{\beta}{2 Q_{i \to j}^{2}} \big(n_i - m_{i \to j} \big)^{2} \Big\} \: .
 \label{eq:Gauss_eta}
\end{equation}

This means that every message $\eta_{i \to j}(n_i)$ can be parameterized using the mean $m_{j \to i}$ and the variance $Q_{j \to i}^{2}$. The update rule (Eq. \eqref{eq:BP_up}) becomes:

\begin{eqnarray}
 \eta_{i \to j}(n_i) \sim  n_i^{\beta \lambda - 1} \, e^{-\beta(n_i^{2} - 2 n_i)/2}  \prod_{k \in \partial i^{-} \setminus j}  \int_{0}^{\infty} \!\!\!\! dn_k \, n_k^{\beta \lambda - 1}   e^{-\beta \, \alpha_{ik} \, n_i \, n_k  -\beta (n_k^{2}-2n_k m_{k\to i})\, / \, 2 \, Q_{k \to i}^{2}} \:. \label{eq:BP_up_Gauss_2}
\end{eqnarray}

When $\beta \to \infty$, with $\beta \lambda$ finite, the integral is dominated by the value $n_k=n_k^{\ast}$ such that the argument of the exponential is maximum. Finding this maximum is equivalent to compute the minimum of $f(n_k) = n_k^{2} - 2 \, n_k \, m_{k \to i} + 2 \, \alpha_{ik} \, n_i \, n_k \, Q_{k \to i}^{2}$. The result is $n_k^{\ast} = m_{k \to i} - \alpha_{ik} \, n_i \, Q_{k \to i}^{2}$. Therefore:

\begin{eqnarray}
 \eta_{i \to j}(n_i) \!\!\!\! &\sim& \!\!\!\!  n_i^{\beta \lambda - 1} \, e^{-\beta(n_i^{2} - 2 n_i)/2} \: \Big[\prod_{k \in \partial i^{-} \setminus j} \big( m_{k \to i} - \alpha_{ik} \, n_i \, Q_{k \to i}^{2}\big)^{\beta \lambda - 1} \Big] \times \nonumber \\
 & & \!\!\!\! \times \exp \Big\{ \sum_{k \in \partial i^{-} \setminus j} \frac{\beta}{2 Q_{k \to i}^{2}} \big(m_{k \to i} - \alpha_{ik} \, n_i \, Q_{k \to i}^{2} \big)^{2} \Big\} \label{eq:BP_up_Gauss_3}  \: .
\end{eqnarray}

Using again that, when $\beta \to \infty$ with $\beta \lambda$ finite, the distribution $\eta_{i \to j}(n_i)$ will concentrate around $n_i = n_i^{\ast}$ such that the argument of the exponential is maximum, we get:

\begin{eqnarray}
 \eta_{i \to j}(n_i) = \frac{n_i^{\beta \lambda - 1} }{z_{i \to j}}  \, \exp \Big\{ \!\!-\frac{\beta}{2}  n_i^{2} \, \big(1 - \sum_{k \in \partial i^{-} \setminus j} \alpha_{ik}^{2} Q_{k \to i}^{2} \big) +  \beta \, n_i \, \big( 1 - \sum_{k \in \partial i^{-} \setminus j} \alpha_{ik} \, m_{k \to i} \big) \Big]  \Big\} \: .
\end{eqnarray}

Comparing with Eq. \eqref{eq:Gauss_eta}, we can easily identify that:

\begin{eqnarray}
 Q_{i \to j}^{2} \!\!\!\! &=& \!\!\!\! \frac{1}{1 - \sum_{k \in \partial i^{-} \setminus j} \alpha_{ik}^{2} Q_{k \to i}^{2}} \\
 m_{i \to j} \!\!\!\! &=& \!\!\!\! \frac{1 - \sum_{k \in \partial i^{-} \setminus j} \alpha_{ik} \, m_{k \to i}}{1 - \sum_{k \in \partial i^{-} \setminus j} \alpha_{ik}^{2} Q_{k \to i}^{2}} \:\:\: .
\end{eqnarray}

These equations are known as relaxed Belief Propagation \cite{Kabashima_2003}, and in this case correspond to the zero temperature expansion of BP. When we have a random regular graph with homogeneous interactions ($\alpha_{ij} \equiv \mu$ for all edges in the graph), all sites become equivalent and

\begin{eqnarray}
 Q_{\to}^{2} \!\!\!\! &=& \!\!\!\! \frac{1}{1 - (c-1) \mu^{2} Q_{\to}^{2}} \label{eq:BP_Q_RRG}\\
 m_{\to} \!\!\!\! &=& \!\!\!\! \frac{1 - (c - 1) \mu \, m_{\to}}{1 - (c - 1) \mu^{2} Q_{\to}^{2}} \label{eq:BP_m_RRG} \:\: .
\end{eqnarray}

From Eq. \eqref{eq:BP_Q_RRG}, we can obtain a closed expression for the variance $Q_{\to}^{2}$:

\begin{equation}
 Q_{\to}^{2} = \frac{1}{2(c-1)\mu^{2}} \Big(1 \pm \sqrt{1-4(c-1)\mu^{2}} \Big) \:.
\end{equation}

From where it follows that, in order to have $Q_{\to}^{2} \in \mathbb{R}$, the strength of the interactions must fulfill the relation

\begin{equation}
 \mu \leq \mu^{\ast} \equiv \frac{1}{2 \sqrt{c-1}} \:\:\: .\label{eq:RRG_exact_transition}
\end{equation}

This result, again, coincides with the exact relation obtained in Ref. \cite{StavPLOS2022} and we already presented it in Eq. \eqref{eq:transition_RRG_T0}. Finally, we can also use that

\begin{equation}
 m_i = \frac{1 - \sum_{k \in \partial i^{-}} \alpha_{ik} \, m_{k \to i}}{1 - \sum_{k \in \partial i^{-}} \alpha_{ik}^{2} Q_{k \to i}^{2}} \: ,
\end{equation}

In the case with homogeneous interactions we have $\alpha_{ik}^{2}=\mu^2$ and $\alpha_{ik}^{2}=\mu^2$ for all edges in the graph. Thus: 

\begin{equation}
 m \equiv \frac{1 - c \mu \, m_{\to}}{1 - c \mu^{2} Q_{\to}^{2}} \: ,
\end{equation}

together with Eqs. \eqref{eq:BP_Q_RRG} and \eqref{eq:BP_m_RRG} to get another exact result:

\begin{equation}
 m = \frac{1}{1 + c \mu} \label{eq:RRG_exact_m} \:\:\: .
\end{equation}

With Eqs. \eqref{eq:RRG_exact_transition} and \eqref{eq:RRG_exact_m}, we recover two known results for the single-to-multiple-equilibria transition for $\beta \to \infty$ \cite{StavPLOS2022}. In terms of BP, this transition is simply a boundary $\mu^{\ast}=1 / (2 \sqrt{c-1})$ such that, for $\mu > \mu^{\ast}$ and at low temperature ($\beta \gg 1$), it is impossible to have a ``Gaussian" stationary point like the one in Eq. \eqref{eq:Gauss_eta} (more precisely, a Gaussian multiplied by the power $n^{\beta \lambda - 1}$).

\end{document}

\maketitle

\section{Local closures} \label{sec:dynclos}

To write the local closures, one needs to marginalize Eq. 2 in the main text to obtain the differential equations for the local probabilities. This will be done first with $P_t(n_i) = \int_0^{\infty} [\,\prod_{k\neq i} dn_k] P_t(\vec{n})$. Marginalizing over all the abundances except $n_i$, one gets:

\begin{eqnarray}
 \frac{\partial}{\partial t} P_t(n_i) \!\!\!\! &=& \!\!\!\! -\frac{\partial }{\partial n_i} \Big\{ \big[n_i(1 - n_i) + \lambda \big] P_t(n_i) \Big\}  + T \frac{\partial^{2}}{\partial n_i^{2}} \big\{ n_i P_t(n_i) \big\} \nonumber \\
 & & + \frac{\partial }{\partial n_i} \Big\{ n_i \sum_{j \in \partial i^{-}} \alpha_{ij} \int_{0}^{\infty} dn_j \,  n_j P_t(n_i, n_j) \Big\} \nonumber \\
 & & -\sum_{k\neq i} \int_0^{\infty} dn_k  \frac{\partial }{\partial n_k} \Big\{ \int_0^{\infty} \big( \prod_{r \neq k, i} dn_r \big) \big[n_k(1 - n_k - \sum_{l \in \partial k^{-}} \alpha_{kl} n_l) + \lambda \big] P_t(\vec{n})  \Big\} +\nonumber \\
 & &  + T \sum_{k\neq i} \int_0^{\infty} dn_k \frac{\partial^{2}}{\partial n_k^{2}} \big\{ n_k P_t(n_i, n_k) \big\} \: .
\end{eqnarray}

Assuming that $\lim_{n_k \to \infty} n_k^{2} \, P_t(\vec{n}) = 0$, this equation can be simplified to:

\begin{eqnarray}
 \frac{\partial}{\partial t} P_t(n_i) \!\!\!\! &=& \!\!\!\! -\frac{\partial }{\partial n_i} \Big\{ \big[n_i(1 - n_i) + \lambda \big] P_t(n_i) \Big\}  + T \frac{\partial^{2}}{\partial n_i^{2}} \big\{ n_i P_t(n_i) \big\} \nonumber \\
 & & + \frac{\partial }{\partial n_i} \Big\{ n_i \sum_{j \in \partial i^{-}} \alpha_{ij} \int_{0}^{\infty} dn_j \,  n_j P_t(n_i, n_j) \Big\} \nonumber \\
 & & + \sum_{k\neq i} \lambda \lim_{n_k \to 0^{+}} P_t(n_i, n_k) - T \sum_{k\neq i} \lim_{n_k \to 0^{+}} P_t(n_i, n_k) \: . \label{eq:der_Pi_2}
\end{eqnarray}

Now, the boundary conditions that we imposed on the Fokker-Planck equation become useful. We should enforce that, for each species $k$, the relation $(T - \lambda) \lim_{n_k \to 0^{+}} P_{\infty}(n_k) =0$. Because of this, the last two terms in Eq. \eqref{eq:der_Pi_2} cancel, leaving:

\begin{eqnarray}
 \frac{\partial}{\partial t} P_t(n_i) \!\!\!\! &=& \!\!\!\! -\frac{\partial }{\partial n_i} \Big\{ \big[n_i(1 - n_i) + \lambda \big] P_t(n_i) \Big\}  + T \frac{\partial^{2}}{\partial n_i^{2}} \big\{ n_i P_t(n_i) \big\} \nonumber \\
 & & + \frac{\partial }{\partial n_i} \Big\{ n_i \sum_{j \in \partial i^{-}} \alpha_{ij} \int_{0}^{\infty} dn_j \,  n_j P_t(n_i, n_j) \Big\} \: . \label{eq:der_Pi_3}
\end{eqnarray}

Defining the conditional average 

\begin{equation}
 m_{j \to i}(n_i, t) \equiv \int_{0}^{\infty} dn_j \,  n_j P_t(n_j \mid n_i) \label{eq:def_mji_supp},
\end{equation}
it is possible to write:

\begin{equation}
 \int_{0}^{\infty} dn_j \,  n_j P_t(n_i, n_j) = m_{j \to i}(n_i, t) \, P_t(n_i)
\end{equation}
and
\begin{eqnarray}
 \frac{\partial}{\partial t} P_t(n_i) = -\frac{\partial }{\partial n_i} \Big\{ \big[n_i(1 - n_i - \sum_{j \in \partial i^{-}} \alpha_{ij} \, m_{j \to i}(n_i, t)) + \lambda \big] P_t(n_i) \Big\} + T \frac{\partial^{2}}{\partial n_i^{2}} \big\{ n_i P_t(n_i) \big\} \: .\label{eq:der_Pi_final_supp}
\end{eqnarray}

Since one does not know the shape of $m_{j \to i}(n_i, t)$, Eq. \eqref{eq:der_Pi_final_supp} cannot be solved directly. Its stationary solutions must fulfill the equation:

\begin{eqnarray}
 0 = -\frac{d }{d n_i} \Big\{ \big[n_i(1 - n_i - \sum_{j \in \partial i^{-}} \alpha_{ij} \, m_{j \to i}(n_i)) + \lambda \big] P_{\infty}(n_i) \Big\} + T \frac{d^{2}}{d n_i^{2}} \big\{ n_i P_{\infty}(n_i) \big\} \: ,\label{eq:Pi_stat_exact}
\end{eqnarray}
where $m_{j \to i}(n_i)$ is the stationary value of the conditional average $m_{j \to i}(n_i, t)$ and $P_{\infty}(n_i)$ is the stationary distribution of the $i$-th species abundance. Integrating once over $n_i$, gives:

\begin{eqnarray}
 \frac{d}{dn_i}P_{\infty}(n_i) = \Big[\beta(1 - n_i - \sum_{j \in \partial i^{-}} \alpha_{ij} \, m_{j \to i}(n_i)) + \frac{\beta \lambda - 1}{n_i} \Big] P_{\infty}(n_i) \: ,\label{eq:Pi_stat_exact_2}
\end{eqnarray}
where $\beta=1/T$. This equation has the following formal solution:

\begin{equation}
 P_{\infty}(n_i) = \frac{1}{Z_{i}} \, n_i^{\beta \lambda - 1} \, \exp \Big\{-\frac{\beta}{2}(n_i^{2} - 2 n_i) \Big\} \exp \Big\{-\beta \sum_{j \in \partial i^{-}} \alpha_{ij} \, \int_{0}^{n_i} dx \, m_{j \to i}(x) \Big\} \: .
 \label{eq:Pi_formal}
\end{equation}

Although Eq. \eqref{eq:Pi_formal} gives a simple expression that will be useful in the future, it is not a solution to the problem because we still do not have the functions $m_{j \to i}(n_i)$. But to obtain $m_{j \to i}(n_i)$ one needs to compute the conditional probabilities, or equivalently, the pair probabilities $P_t(n_i, n_j)$. The corresponding differential equation can be obtained following a similar procedure.

\begin{eqnarray}
 \frac{\partial}{\partial t} P_t(n_i, n_j) \!\!\!\! &=& \!\!\!\! -\frac{\partial }{\partial n_i} \Big\{ \big[n_i(1 - n_i - \alpha_{ij} n_j) + \lambda \big] P_t(n_i, n_j) \Big\}  + T \frac{\partial^{2}}{\partial n_i^{2}} \big\{ n_i P_t(n_i, n_j) \big\} \nonumber \\
 & &\!\!\!\! -\frac{\partial }{\partial n_j} \Big\{ \big[n_j(1 - n_j - \alpha_{ji} n_i) + \lambda \big] P_t(n_i, n_j) \Big\}  + T \frac{\partial^{2}}{\partial n_j^{2}} \big\{ n_j P_t(n_i, n_j) \big\} \nonumber \\
 & &\!\!\!\! + \frac{\partial }{\partial n_i} \Big\{ n_i \sum_{k \in \partial i^{-} \setminus j} \alpha_{ik} \int_{0}^{\infty} dn_k \,  n_k P_t(n_k, n_i, n_j) \Big\} \nonumber \\
 & &\!\!\!\! + \frac{\partial }{\partial n_j} \Big\{ n_j \sum_{l \in \partial j^{-} \setminus i} \alpha_{jl} \int_{0}^{\infty} dn_l \,  n_l P_t(n_i, n_j, n_l) \Big\}  \label{eq:der_Pij_1} \: .
\end{eqnarray}

Similarly as before, let us define the conditional average $m_{k \to i,j}(n_i, n_j, t) \equiv \int_{0}^{\infty} dn_k P_t(n_k, t \mid n_i, n_j)$. Eq. \eqref{eq:der_Pij_1} reduces to:

\begin{eqnarray}
 \frac{\partial}{\partial t} P_t(n_i, n_j) \!\!\!\! &=& \!\!\!\! -\frac{\partial }{\partial n_i} \Big\{ \big[n_i(1 - n_i - \alpha_{ij} n_j - \sum_{k \in \partial i^{-} \setminus j} \alpha_{ik} \, m_{k \to i,j}(n_i, n_j)) + \lambda \big] P_t(n_i, n_j) \Big\} \nonumber \\
 & &\!\!\!\! -\frac{\partial }{\partial n_j} \Big\{ \big[n_j(1 - n_j - \alpha_{ji} n_i - \sum_{l \in \partial j^{-} \setminus i} \alpha_{jl} \, m_{l \to i,j}(n_i, n_j)) + \lambda \big] P_t(n_i, n_j) \Big\}    \nonumber \\
 & & + T \frac{\partial^{2}}{\partial n_i^{2}} \big\{ n_i P_t(n_i, n_j) \big\} + T \frac{\partial^{2}}{\partial n_j^{2}} \big\{ n_j P_t(n_i, n_j) \big\} \label{eq:der_Pij_final} \: .
\end{eqnarray}

Again, one does not know the shape of $m_{k \to i,j}(n_i, n_j, t)$, but to obtain it, one needs to solve the differential equation for $P_t(n_k, n_i, n_j)$. This process builds a hierarchy of differential equations that, in the end, goes back to the Fokker-Planck equation for the full system (Eq. (2) in the main text). This is not surprising, since no approximations have been done, the solution of these local Fokker-Planck equations must be as difficult as the solution of the original Fokker-Planck equation for the whole system.

To close the hierarchy at some point, one can make some factorization of the joint probability densities so that they are expressed in terms of probability densities of a lower level in the hierarchy. The first approximation one could do is known in the studies of epidemic spreading as Individual Based Mean Field (IBMF) \cite{WangEpid2003, pastor2015epidemic} and it is simply:

\begin{equation}
P_t(n_i, n_j) \approx P_t(n_i) \, P_t(n_j) \: ,
 \label{eq:IBMF_app}
\end{equation}
which implies that:

\begin{eqnarray}
m_{j \to i}(n_i, t) \!\!\!\! &=& \!\!\!\! \int_{0}^{\infty} dn_j \, n_j P_t(n_j \mid n_i) = \int_{0}^{\infty} dn_j \, n_j \frac{P_t(n_i, n_j)}{P_t(n_i)} \nonumber \\
m_{j \to i}(n_i, t) \!\!\!\! &\approx& \!\!\!\!  \int_{0}^{\infty} dn_j \, n_j \,  \frac{P_t(n_i) \, P_t(n_j)}{P_t(n_i)}  = \int_{0}^{\infty} dn_j \, n_j P_t(n_j) \equiv m_j(t) \: ,
 \label{eq:red_m_IBMF}
\end{eqnarray}
where $m_{j}(t)$ is the expected value of the abundance $n_j$ at time $t$.

The IBMF differential equation is then:

\begin{eqnarray}
 \frac{\partial}{\partial t} P_t(n_i) \!\!\!\! &=& \!\!\!\! -\frac{\partial }{\partial n_i} \Big\{ \big[n_i(1 - n_i - \sum_{j \in \partial i^{-}} \alpha_{ij}\, m_{j}(t)) + \lambda \big] P_t(n_i) \Big\}  + T \frac{\partial^{2}}{\partial n_i^{2}} \big\{ n_i P_t(n_i) \big\} \label{eq:IBMF} \: ,
\end{eqnarray}
which must be completed with the definition $m_{j}(t) \equiv \int_{0}^{\infty} dn_j \, n_j \, P_t(n_j)$.

The IBMF is the first local closure one can provide. However, it is possible to go ahead and propose closures that stop at higher levels of the hierarchy. To express $P_t(n_k, n_i, n_j)$ in terms of pair probability densities like $P_t(n_i, n_j)$, one could use information about the actual graph of interactions. In the third line of Eq. \eqref{eq:der_Pij_1}, the probability density $P_t(n_k, n_i, n_j)$ is actually defined in a graph $G(V, E)$ that contains the edges $k \to i$ and $i \to  j$, but not necessarily the edges $k \to j$ or $j \to k$. In tree-like graphs, with high probability when the number of species is large, given that the edges $k \to i$ and $i \to  j$ are present, we will have that $k$ and $j$ are not directly connected by any edge. Moreover, the length of the cycles diverges with the system size $N$, so the only short path that connects $k$ and $j$ necessarily passes through $i$. So, expecting that it works especially well in tree-like graphs, one could propose:

\begin{equation}
P_t(n_k, n_i, n_j) \approx \frac{P_t(n_k, n_i) \, P_t(n_i, n_j)}{P_t(n_i)} \equiv P_t(n_k \mid n_i) \, P_t(n_i) \, P_t(n_j \mid n_i)  \label{eq:PBMF_app} \: ,
\end{equation}

This is known as Pair Based Mean Field (PBMF) in the above-mentioned context of epidemic spreading \cite{cator2012second, mata2013pair} and is a factorization of the conditional measure. Given $n_i$ at time $t$, one assumes $n_j$ and $n_k$ are independent. This is not strange for those who know the replica symmetric cavity method, whose algorithmic counterpart is Belief Propagation (BP). However, even in the cases where BP is the exact solution in equilibrium (provided also that equilibrium exists), this approximation is not necessarily valid for the dynamics, since temporal correlations can still forbid the factorization, and $n_j$ and $n_k$ are not independent given only the value of $n_i$ at time $t$.

Anyways, one can assume Eq. \eqref{eq:PBMF_app} to be valid and continue:

\begin{eqnarray}
m_{k \to i, j}(n_i, n_j, t) \!\!\!\! &=& \!\!\!\! \int_{0}^{\infty} dn_k \, n_k P_t(n_k \mid n_i, n_j) = \int_{0}^{\infty} dn_k \, n_k \frac{P_t(n_k, n_i, n_j)}{P_t(n_i, n_j)} \nonumber \\
m_{k \to i, j}(n_i, n_j, t) \!\!\!\! &\approx& \!\!\!\!  \int_{0}^{\infty} dn_k \, n_k \frac{P_t(n_k, n_i)}{P_t(n_i)} \equiv m_{k \to i}(n_i, t) \: .
 \label{eq:red_m_PBMF}
\end{eqnarray}

Thus, the PBMF differential equation is:

\begin{eqnarray}
 \frac{\partial}{\partial t} P_t(n_i, n_j) \!\!\!\! &=& \!\!\!\! -\frac{\partial }{\partial n_i} \Big\{ \big[n_i(1 - n_i - \alpha_{ij} n_j - \sum_{k \in \partial i^{-} \setminus j} \alpha_{ik} \, m_{k \to i}(n_i, t)) + \lambda \big] P_t(n_i, n_j) \Big\} \nonumber \\
 & &\!\!\!\! -\frac{\partial }{\partial n_j} \Big\{ \big[n_j(1 - n_j - \alpha_{ji} n_i - \sum_{l \in \partial j^{-} \setminus i} \alpha_{jl} \, m_{l \to j}(n_j, t)) + \lambda \big] P_t(n_i, n_j) \Big\}    \nonumber \\
 & & + T \frac{\partial^{2}}{\partial n_i^{2}} \big\{ n_i P_t(n_i, n_j) \big\} + T \frac{\partial^{2}}{\partial n_j^{2}} \big\{ n_j P_t(n_i, n_j) \big\} \, \label{eq:PBMF} ,
\end{eqnarray}
which must be completed with the definition $m_{j \to i}(n_i, t) \equiv \int_{0}^{\infty} dn_j \,  n_j P_t(n_j \mid n_i)$.

Both closures, IBMF and PBMF, are in principle solvable, at least numerically. However, when the system is large this is a difficult task.

\section{Connections with known results}\label{sec:connections}

\subsection{Belief Propagation as stationary solution of Pair Based Mean Field} \label{subsec:BP}

As mentioned before, the approximation in Eq. \eqref{eq:PBMF_app} is valid in equilibrium in all the cases where BP is also valid, since BP respects this factorization of the conditional measure. It is no surprise then that the BP equations, introduced in Ref. \cite{Tonolo2026sparse} for symmetric interactions ($\alpha_{ij} = \alpha_{ji}$), are a stationary solution of PBMF equations for symmetric interactions. Nevertheless, proving it could be a useful exercise for the future.

BP's update rule, as presented in Ref. \cite{Tonolo2026sparse}, is:

\begin{equation}
 \eta_{i \to j}(N_i) = \frac{1}{z_{i \to j}} \, \frac{1}{N_i + \Delta} \, \exp \Big\{-\frac{\beta}{2}(N_i^{2} - 2 N_i) \Big\} \prod_{k \in \partial i^{-} \setminus j} \sum_{N_k=0}^{\infty} \eta_{k \to i}(N_k) \exp \Big\{ - \beta \alpha_{ik} N_i \, N_k \Big\} \label{eq:BP_up_original} \: ,
\end{equation}
where the constants $r$ and $K$ are set to one for simplicity. The constant $z_{i \to j}$ is a normalization factor, $\Delta>0$ is a small parameter and $\eta_{i \to j}(N_i)$ are BP's messages. One must remember that in this case $\alpha_{ik} = \alpha_{ki}$, so one could write one or the other and Eq. \eqref{eq:BP_up_original} remains valid.

First of all, one must notice that Eq. \eqref{eq:BP_up_original} is written in a discretized space, so $N_i$ is actually taking discrete values. That is the reason why in Eq. \eqref{eq:BP_up_original} one sees a sum over $N_k$ and not an integral. It is always possible to recast this expression in terms of the continuous variables $n_i$. Besides substituting sums by integrals, one has to pay particular attention to the factor $(N_i + \Delta)^{-1}$. As $\lambda$ in the continuous case, here $\Delta$ ensures that the divergence when $N_i$ goes to zero is not so critical that it makes the probability distribution non-normalizable. Any positive $\Delta$ will remove this divergence at $N_i = 0$. Thus, in the continuous version, one must simply substitute the factor $(N_i + \Delta)^{-1}$ by the analogous factor $n_i^{\beta \lambda - 1}$, present in Eqs. 3 and 8 in the main text.

The resulting update rule is:

\begin{equation}
 \eta_{i \to j}(n_i) = \frac{1}{z_{i \to j}} \, n_i^{\beta \lambda - 1} \, \exp \Big\{-\frac{\beta}{2}(n_i^{2} - 2 n_i) \Big\} \prod_{k \in \partial i^{-} \setminus j} \int_{0}^{\infty} dn_k \, \eta_{k \to i}(n_k) \exp \big\{ - \beta \alpha_{ik} n_i \, n_k \big\} \label{eq:BP_up_supp} \: ,
\end{equation}

It will be helpful to look closely at the integral

\begin{equation}
 \mathcal{Z}_{k \to i}(n_i) = \int_{0}^{\infty} dn_k \, \eta_{k \to i}(n_k) \exp \big\{ - \beta \alpha_{ik} n_i \, n_k \big\} \: .
\end{equation}

In understanding its meaning, it is useful to write the stationary pair probability density in terms of BP's messages:

\begin{equation}
 P_{\infty}(n_i, n_k) =  \eta_{i \to k}(n_i) \, \eta_{k \to i}(n_k) \, e^{- \beta \alpha_{ik} n_i \, n_k} \label{eq:Pij_BP} \:\: .
\end{equation}

Then, the conditional probability density is

\begin{eqnarray}
 P_{\infty}(n_k \mid n_i) \!\!\!\! &=& \!\!\!\! \frac{\eta_{i \to k}(n_i) \, \eta_{k \to i}(n_k) \, e^{- \beta \alpha_{ik} n_i \, n_k}}{\int_{0}^{\infty} dn_k \, \eta_{i \to k}(n_i) \, \eta_{k \to i}(n_k) e^{- \beta \alpha_{ik} n_i \, n_k}} = \frac{ \eta_{k \to i}(n_k) \, e^{- \beta \alpha_{ik} n_i \, n_k}}{\int_{0}^{\infty} dn_k \, \eta_{k \to i}(n_k) e^{- \beta \alpha_{ik} n_i \, n_k}} \nonumber \\
 P_{\infty}(n_k \mid n_i) \!\!\!\! &=& \!\!\!\! \frac{1}{\mathcal{Z}_{k \to i}(n_i)} \, \eta_{k \to i}(n_k) \, e^{- \beta \alpha_{ik} n_i \, n_k} \: \: .\label{eq:symmetric_conditional_P}
\end{eqnarray}

So $\mathcal{Z}_{k \to i}(n_i)$ is the normalization factor of the conditional probability density. Furthermore, its derivative with respect to $n_i$ is:

\begin{eqnarray}
 \frac{\partial}{\partial n_i} \mathcal{Z}_{k \to i}(n_i)  \!\!\!\! &=& \!\!\!\! -\beta \, \alpha_{ik} \, \int_{0}^{\infty} dn_k \, n_k \, \eta_{k \to i}(n_k) e^{- \beta \alpha_{ik} n_i \, n_k} \nonumber \\
 \frac{\partial}{\partial n_i} \mathcal{Z}_{k \to i}(n_i)  \!\!\!\! &=& \!\!\!\!-\beta \, \alpha_{ik} \mathcal{Z}_{k \to i}(n_i) \, \frac{\int_{0}^{\infty} dn_k \, n_k \, \eta_{k \to i}(n_k) e^{- \beta \alpha_{ik} n_i \, n_k}}{\mathcal{Z}_{k \to i}(n_i)} \:\: .
\end{eqnarray}

Remembering the definition of $m_{k \to i}(n_i)$ and Eq. \eqref{eq:Pij_BP}, one gets

\begin{equation}
 \frac{\partial}{\partial n_i} \mathcal{Z}_{k \to i}(n_i) = -\beta \, \alpha_{ik} \, \mathcal{Z}_{k \to i}(n_i) \, m_{k \to i}(n_i) \: ,
\end{equation}
from where it is easy to see that

\begin{equation}
 \mathcal{Z}_{k \to i}(n_i) \propto \exp \Big\{ -\beta \, \alpha_{ik} \int_{0}^{n_i} dx \, m_{k \to i}(x) \Big\} \: .
\end{equation}

Returning to Eq. \eqref{eq:BP_up_supp}, the expression can be rewritten in a form that is very similar to the formal solution for $P_{\infty}(n_i)$ in Eq. \eqref{eq:Pi_formal}, and that is connected with the differential equation for PBMF (Eq. \eqref{eq:PBMF}):

\begin{equation}
 \eta_{i \to j}(n_i) = \frac{1}{z_{i \to j}} \, n_i^{\beta \lambda - 1} \, \exp \Big\{-\frac{\beta}{2}(n_i^{2} - 2 n_i) \Big\} \exp \Big\{-\beta \sum_{k \in \partial i^{-} \setminus j} \alpha_{ik} \, \int_{0}^{n_i} dx \, m_{k \to i}(x) \Big\} \label{eq:BP_up_expform} \: .
\end{equation}

Thus, the pair probability can be written as:

\begin{eqnarray}
 & & \!\!\!\!\!\!\!\!\!\!\!\! P_{\infty}(n_i, n_j) = \frac{1}{Z_{ij}}  (n_i \, n_j)^{\beta \lambda - 1} \, e^{-\frac{\beta}{2}(n_i^{2} - 2 n_i)} \, e^{-\frac{\beta}{2}(n_j^{2} - 2 n_j)} \, e^{- \beta \alpha_{ij} n_i \, n_j} \times \nonumber \\
 & & \!\!\! \times \exp \Big\{-\beta \sum_{k \in \partial i^{-} \setminus j} \alpha_{ik} \, \int_{0}^{n_i} dx \, m_{k \to i}(x) \Big\} \exp \Big\{-\beta \sum_{l \in \partial j^{-} \setminus i} \alpha_{jl} \, \int_{0}^{n_j} dx \, m_{l \to j}(x) \Big\} \label{eq:Pij_BP_expform} \: .
\end{eqnarray}

Everything is now ready to use the update rule and obtain an expression for the derivative of the probability density $P_{\infty}(n_i, n_j)$ with respect to $n_i$:

\begin{equation}
 \frac{\partial}{\partial n_i} P_{\infty}(n_i, n_j) = P_{\infty}(n_i, n_j) \Big[ \frac{\beta \lambda - 1}{n_i} - \beta(n_i - 1) - \beta \, \alpha_{ij} \, n_j - \beta \, \sum_{k \in \partial i^{-} \setminus j} \alpha_{ik} \, m_{k \to i}(n_i) \Big] \label{eq:Pij_BP_der} \: .
\end{equation}

Therefore

\begin{eqnarray}
   T \frac{\partial}{\partial n_i} \big\{ n_i P_t(n_i, n_j) \big\} -\big[n_i(1 - n_i - \alpha_{ij} n_j - \sum_{k \in \partial i^{-} \setminus j} \alpha_{ik} \, m_{k \to i}(n_i)) + \lambda \big] P_t(n_i, n_j) \nonumber \\
 = P_t(n_i, n_j) \, \Big\{T + T \, n_i \, \Big[ \frac{\beta \lambda - 1}{n_i} - \beta(n_i - 1) - \beta \, \alpha_{ij} \, n_j - \beta \, \sum_{k \in \partial i^{-} \setminus j} \alpha_{ik} \, m_{k \to i}(n_i) \Big] - \nonumber \\
 - \big[n_i(1 - n_i - \alpha_{ij} n_j - \sum_{k \in \partial i^{-} \setminus j} \alpha_{ik} \, m_{k \to i}(n_i)) + \lambda \big] \Big\} {\bf = 0} \: .
\end{eqnarray}

Something analogous happens with the derivative with respect to $n_j$. This means that, in the open region $(n_i, n_j) \in (0, +\infty) \times (0, +\infty)$, the right hand side of Eq. \eqref{eq:PBMF} is equal to zero. In other words, BP is a stationary solution for PBMF when the interactions are symmetric.

\subsection{Zero temperature and large connectivity limit of IBMF} \label{subsec:T0}

So far, we have presented two stationary distributions that are approximate solutions to the problem. First, IBMF, which is the simplest local closure one can do, but at the same time is of a general purpose. Regardless of the degree of symmetry in the interactions, the stationary solution of IBMF is:

\begin{equation}
P_{\infty}(n_i) = \frac{1}{Z_i} \, n_i^{\beta \lambda - 1} \, \exp \Big\{ -\frac{\beta}{2} \,(n_i - h_i)^{2} \Big\} \: ,
 \label{eq:sol_IBMF_repeat}
\end{equation}
with $h_i=1- \sum_{j \in \partial i^{-}} \alpha_{ij} \, m_{j}$ and $m_j=\int_{0}^{\infty} dn_j \, n_j \, P_{\infty}(n_j)$.

When taking the zero temperature limit (equivalently $\beta \to \infty$), it is crucial to know how to treat the parameter $\lambda$. One must remember that $\lambda$ has the role of ensuring that the distribution is normalizable, by avoiding the extinction of all the species. It is usually interpreted as an immigration rate that must be taken positive, but small, to study the phenomenology of the model in its purest version possible. Therefore, one should take $\beta \to \infty$ but keep $\beta \lambda$ finite, since $\lambda \sim 0$.

We will compare the final result of taking this limit with Eq. (13) in Ref. \cite{Roy_2019}, which gives an expression for the steady state abundance of the fully connected model at $T=0$:

\begin{equation}
 n = \max \Big(0, \, \frac{1-\hat{\mu} \, m_{\infty} - \hat{\sigma} \, \zeta_{\infty}}{1 - \epsilon \, \hat{\sigma}^{2} \, \chi_{\text{int}}} \Big) \label{eq:DMFT_result} \: .
\end{equation}

%The reader should note
It is worth noticing that a very similar expression already appeared in the pioneering contribution in Ref. \cite{Opper_PRL_1992}. Its authors derive the Dynamical Mean-Field Theory (DMFT) equations for a similar setting, which can be mapped to the gLV model, and obtain a stationary solution of the form \eqref{eq:DMFT_result}.

Remember that $\mu$ is the average value of the couplings $\alpha_{ij}$, and $\sigma^{2}$ is its variance. When the system is fully connected, it is necessary to rescale these parameters and send them to zero when $N\to \infty$, but keeping finite  $\hat{\mu} = N \, \mu$ and $\hat{\sigma}^{2} = N \, \sigma^{2}$. The number $\epsilon \in [-1, 1]$ is the correlation between $\alpha_{ij}$ and $\alpha_{ji}$, or, in other words, the level of symmetry in the couplings. Besides these parameters, in Eq. \eqref{eq:DMFT_result} one has: $m_{\infty}$, which is the expected value of $n$; $\zeta_{\infty}$, which is a random Gaussian variable; and $\chi_{\text{int}}$, which is the integrated response of $n$ to the effect of a small external field.

To recover this result, one can start by taking the limit $T \to 0$ in Eq. \eqref{eq:sol_IBMF_repeat}. As we discussed in Section IV of the main text, the probability density then concentrates at one point:

\begin{equation}
n_i = \max \Big( 0, 1 - \sum_{j \in \partial i^{-}} \alpha_{ij}\, m_{j} \Big) \: .
 \label{eq:sol_IBMF_T0}
\end{equation}
This way of writing it ensures that $n_i$ must always be non-negative. When the connectivity is large, the sum $\theta = \sum_{j \in \partial i^{-}} \alpha_{ij}\, m_{j}$ is distributed as a Gaussian with the following first and second moments:

\begin{eqnarray}
 \langle \theta \rangle \!\!\!\! &=& \!\!\!\! \sum_{j \in \partial i^{-}} \langle \alpha_{ij} \, m_{j} \rangle  \label{eq:av_theta_1} \\
 \langle \theta^{2} \rangle \!\!\!\! &=& \!\!\!\! \sum_{j \in \partial i^{-}} \sum_{k \in \partial i^{-}} \langle \alpha_{ij} \, \alpha_{ik} \, m_{j} \, m_k  \rangle \label{eq:av_theta_sqr_1} \:\: .
\end{eqnarray}

The averages in Eqs. \eqref{eq:av_theta_1} and \eqref{eq:av_theta_sqr_1} are taken over the disorder in the interactions, \textit{i.e.}, over the possible values of the whole matrix of couplings $\overleftrightarrow{\alpha}$ with elements $\alpha_{ij}$. Then we must realize that $m_{j}$ is actually a function of $\overleftrightarrow{\alpha}$, and that it is not possible to compute the averages in Eqs. \eqref{eq:av_theta_1} and \eqref{eq:av_theta_sqr_1} as if $m_{j}$ were independent of the specific realization of the disorder $\overleftrightarrow{\alpha}$. However, one can explicitly average Eqs. \eqref{eq:av_theta_1} and \eqref{eq:av_theta_sqr_1} over part of the disorder since most of the elements of $\overleftrightarrow{\alpha}$ are indeed independent of $\alpha_{ij}$:

\begin{eqnarray}
 \langle \theta \rangle  \!\!\!\! &=& \!\!\!\! \sum_{j \in \partial i^{-}} \langle \alpha_{ij} \, m_{j}(\overleftrightarrow{\alpha}) \rangle = \sum_{j \in \partial i^{-}} \langle \, \alpha_{ij} \, m_{j}(\alpha_{ij}, \alpha_{ji}) \, \rangle \label{eq:av_theta_2} \\
 \langle \theta^{2} \rangle \!\!\!\! &=& \!\!\!\! \sum_{j \in \partial i^{-}} \sum_{k \in \partial i^{-}} \langle \, \alpha_{ij} \, \alpha_{ik} \, m_{j}(\alpha_{ij}, \alpha_{ji}, \alpha_{ik}, \alpha_{ki}) \, m_k(\alpha_{ij}, \alpha_{ji}, \alpha_{ik}, \alpha_{ki}) \, \rangle \label{eq:av_theta_sqr_2} \: .
\end{eqnarray}

The dependence of $m_{j}$ on the disorder has been reduced to the only elements that are not independent of $\alpha_{ij}$, which are $\alpha_{ij}$ itself and possibly $\alpha_{ji}$ (if $\epsilon > 0$). The new $m_{j}(\alpha_{ij}, \alpha_{ji})$ is already an average over almost all the couplings, except for the ones associated with the edges $i \to j$ and $j \to i$. Therefore, the right-hand sides of Eqs. \eqref{eq:av_theta_2} and \eqref{eq:av_theta_sqr_2} are no longer averages over all the couplings that appear in the arguments of the functions. More specifically, the average $\langle \, \alpha_{ij} \, m_{j}(\alpha_{ij}, \alpha_{ji}) \, \rangle$ in Eq. \eqref{eq:av_theta_2} is taken over the values of $\alpha_{ij}$ and $\alpha_{ji}$, considering that they can be correlated. In the same way, the average in Eq. \eqref{eq:av_theta_sqr_2} is taken over the values of $\alpha_{ij}$, $\alpha_{ji}$, $\alpha_{ik}$, and $\alpha_{ki}$. To compute those averages, it will be useful to write the couplings in terms of standardized Gaussians $x_{ij} \sim \mathcal{N}(0, 1)$ as:

\begin{eqnarray}
 & &\alpha_{ij} = \mu + \sigma \, x_{ij}, \:\:\:\: \text{with} \nonumber \\
 & & \langle x_{ij} \rangle = 0; \:\:\:\:\:\: \langle x_{ij} x_{kl} \rangle = \delta_{ij, kl} + \epsilon \delta_{ij, lk} \label{eq:introd_x} \:\:\: .
\end{eqnarray}
  
Then $m_{j}(\alpha_{ij}, \alpha_{ji})$ becomes a function of $x_{ij}$ and $x_{ji}$. However, when the connectivity is large, it is safe to assume that the dependence of $m_{j}(x_{ij}, x_{ji})$ on each $x_{ij}$, associated to only one of the many edges that contain the node $j$, is very weak. Taking an expansion of $m_{j}(x_{ij}, x_{ji})$ in powers of $x_{ij}$ and $x_{ji}$ up to the first order leads to:

\begin{equation}
 m_{j}(x_{ij}, x_{ji}) \approx m_{j}(0, 0) + x_{ij} \Big\{ \frac{\partial }{\partial x_{ij}} m_{j}(x_{ij}, 0) \Big\} \Big|_{x_{ij}=0} + x_{ji} \Big\{ \frac{\partial }{\partial x_{ji}} m_{j}(0, x_{ji}) \Big\} \Big|_{x_{ji}=0} \:\: .
\end{equation}

Remembering that $x_{ij}$ actually does not appear in the equation for $dn_j /dt$, but in the equation for $dn_i /dt$, one realizes that its effect in $m_{j}(x_{ij}, x_{ji})$ must be even weaker than the effect of $x_{ji}$. Finally:

\begin{equation}
 m_{j}(x_{ij}, x_{ji}) \approx m_{j}(x_{ji}) \approx m_{j}(0) + x_{ji} \Big\{ \frac{\partial }{\partial x_{ji}} m_{j}(x_{ji}) \Big\} \Big|_{x_{ji}=0} \label{eq:expansion_m} \:\: .
\end{equation}

Inserting Eqs. \eqref{eq:introd_x} and \eqref{eq:expansion_m} back into Eq. \eqref{eq:av_theta_2} gives:

\begin{eqnarray}
\langle \theta \rangle \!\!\!\! &=& \!\!\!\! \sum_{j \in \partial i^{-}} \Big\langle \, (\mu + \sigma \, x_{ij}) \, \Big( m_{j}(0) + x_{ji} \Big\{ \frac{\partial }{\partial x_{ji}} m_{j}(x_{ji}) \Big\} \Big|_{x_{ji}=0} \Big) \, \Big\rangle \nonumber \\
\langle \theta \rangle \!\!\!\! &=& \!\!\!\! \mu \sum_{j \in \partial i^{-}} \langle m_{j}(0) \rangle  + \sigma \sum_{j \in \partial i^{-}} \Big\langle x_{ij} x_{ji} \Big\{ \frac{\partial }{\partial x_{ji}} m_{j}(x_{ji}) \Big\} \Big|_{x_{ji}=0}  \, \Big\rangle \: \: .
\end{eqnarray}

Let the connectivity be equal to $c$. After identifying that $\langle m_{j}(0) \rangle \equiv m_{\infty}$, where $m_{\infty}$ is the parameter in Eq. \eqref{eq:DMFT_result} above, and by using the expression for the second moment in Eq. \eqref{eq:introd_x}, one gets:

\begin{eqnarray}
\langle \theta \rangle \!\!\!\! &=& \!\!\!\! \mu \, c \, m_{\infty}  + \sigma \, c \, \epsilon \, \Big\langle \Big\{ \frac{\partial }{\partial x_{ji}} m_{j}(x_{ji}) \Big\} \Big|_{x_{ji}=0} \, \Big\rangle \label{eq:av_theta_3} \: \: .
\end{eqnarray}

The derivative in Eq. \eqref{eq:av_theta_3} gives the average response of $m_{j}$ to a small perturbation in the couplings. This will be related to the response to a small external field, and to see the relation more explicitly, it is convenient to recast the differential equation for $n_j$ (see Eq. 1 in the main text) in terms of $x_{ji}$:

\begin{equation}
 \frac{dn_j}{dt} = -n_j(1 - n_j - \mu \,  \sum_{l \in \partial j^{-}} n_{l} - \sigma \, \sum_{l \in \partial j^{-}} x_{jl} \, n_l) + \xi_j(t) + \lambda \:\:\: .
\end{equation}

Taking $i$ outside the sum over the neighbors:

\begin{equation}
 \frac{dn_j}{dt} = -n_j(1 - n_j - \mu \,  \sum_{l \in \partial j^{-}} n_{l} - \sigma \, \sum_{l \in \partial j^{-} \setminus i} x_{jl} \, n_l - \sigma \, x_{ji} \, n_i) + \xi_j(t) + \lambda  \:\:\: .
\end{equation}

The effect of a small perturbation $x_{ji}$ is the same as the one provoked by a small external field with value $h_j(t) = - \sigma \, x_{ji} \, n_i(t)$. The two responses must be related by:

\begin{equation}
\Big\langle \Big\{ \frac{\partial }{\partial x_{ji}} m_{j}(x_{ji}) \Big\} \Big|_{x_{ji}=0} \, \Big\rangle = \Big\langle \Big\{ \frac{\partial h_j}{\partial x_{ji}} \, \frac{\partial }{\partial h_{j}} m_{j}(h_j) \Big\} \Big|_{h_j=0}  \, \Big\rangle  = -\sigma \, n_i \, \chi_{\text{int}}
 \label{eq:responses} \:\:\: .
\end{equation}

Inserting Eq. \eqref{eq:responses} back into Eq. \eqref{eq:av_theta_3}, one finally obtains an expression for $\langle \theta \rangle$ in terms of $m_{\infty}$ and $\chi_{\text{int}}$, which are two of the parameters in the known result from DMFT (see Eq. \eqref{eq:DMFT_result}):

\begin{eqnarray}
\langle \theta \rangle \!\!\!\! &=& \!\!\!\! \mu \, c \, m_{\infty}  - \sigma^{2} \, c \, \epsilon \, n_i \, \chi_{\text{int}} \label{eq:av_theta_final} \:\: .
\end{eqnarray}

The same needs to be done with the second moments:

\begin{eqnarray}
 \langle \theta^{2} \rangle \!\!\!\! &=& \!\!\!\! \sum_{j \in \partial i^{-}} \sum_{k \in \partial i^{-}} \big\langle \, (\mu + \sigma \, x_{ij}) \, (\mu + \sigma \, x_{ik}) \, ( m_{\infty} - \sigma \, n_i \, \chi_{\text{int}}  \, x_{ji} ) \, ( m_{\infty} - \sigma \, n_i \, \chi_{\text{int}}  \, x_{ki} ) \, \big\rangle \nonumber \\
 \langle \theta^{2} \rangle \!\!\!\! &=& \!\!\!\! \sum_{j \in \partial i^{-}} \sum_{k \in \partial i^{-}} \big\langle \, \big[\mu^{2} + \mu \, \sigma \, (x_{ij} + x_{ik}) + \sigma^{2} \, x_{ij} \, x_{ik} \big] \times \nonumber \\ 
 & & \qquad\qquad\qquad\qquad \times \big[ m_{\infty}^{2} - m_{\infty} \sigma \, n_i \, \chi_{\text{int}}   \, (x_{ji} + x_{ki}) + \sigma^{2} \, n_i^{2} \, \chi_{\text{int}}^{2}  \, x_{ji} \, x_{ki} \big] \, \big\rangle \nonumber \\
 \langle \theta^{2} \rangle \!\!\!\! &=& \!\!\!\! c^{2} \, \mu^{2} \, m_{\infty}^{2} + c \, \sigma^{2} \, m_{\infty}^{2} - \mu \, \sigma^{2} \, m_{\infty} \, n_i \, \chi_{\text{int}} \sum_{j \in \partial i^{-}} \sum_{k \in \partial i^{-}} \big\langle \, (x_{ij} + x_{ik}) \, (x_{ji} + x_{ki})  \, \big\rangle + \nonumber  \\
 & &  + c \, \mu^{2} \sigma^{2} \, n_i^{2} \chi_{\text{int}}^{2} +  \sigma^{4} \, n_i^{2} \, \chi_{\text{int}}^{2} \sum_{j \in \partial i^{-}} \langle x_{ij}^{2} \, x_{ji}^{2} \rangle   +  \sigma^{4} \, n_i^{2} \, \chi_{\text{int}}^{2} \sum_{j \in \partial i^{-}} \sum_{k \in \partial i^{-} \setminus j} \langle x_{ij} \, x_{ji} \, x_{ik} \, x_{ki} \rangle \nonumber \\
 \langle \theta^{2} \rangle \!\!\!\! &=& \!\!\!\! c^{2} \, \mu^{2} \, m_{\infty}^{2} + c \, \sigma^{2} \, m_{\infty}^{2} - \mu \, \sigma^{2} \, m_{\infty} \, n_i \, \epsilon \, \chi_{\text{int}} \, \big(4 c + 2 c(c - 1) \big) + \nonumber  \\
 & & \qquad\qquad + c \, \mu^{2} \sigma^{2} \, n_i^{2} \chi_{\text{int}}^{2} +  \sigma^{4} \, n_i^{2} \, \chi_{\text{int}}^{2} \, \big(c + c(c - 1) \epsilon^{2} \big) \nonumber \\  \langle \theta^{2} \rangle \!\!\!\! &=& \!\!\!\! c^{2} \, \mu^{2} \, m_{\infty}^{2} + c \, \sigma^{2} \, m_{\infty}^{2} - 2 \, \mu \, \sigma^{2} \, m_{\infty} \, n_i \, \epsilon \, \chi_{\text{int}} \, c \, (c + 1)  + \nonumber  \\
 & & \qquad\qquad + c \, \mu^{2} \sigma^{2} \, n_i^{2} \chi_{\text{int}}^{2} +  \sigma^{4} \, n_i^{2} \, \chi_{\text{int}}^{2} \, \big(c + c(c - 1) \epsilon^{2} \big) \label{eq:av_theta_sqr_3} \:\:\: .
\end{eqnarray}
where we assumed that $\langle x_{ij}^{2} \, x_{ji}^{2} \rangle = 1$ and that $\langle x_{ij} \, x_{ji} \, x_{ik} \, x_{ki} \rangle = \epsilon^{2}$, if $i \neq k$.

Then, the variance of $\theta$ is:

\begin{eqnarray}
 s_{\theta}^{2} \!\!\!\! &=& \!\!\!\! \langle \theta^{2} \rangle - \langle \theta \rangle^{2} \nonumber \\
 s_{\theta}^{2} \!\!\!\! &=& \!\!\!\! c \, \sigma^{2} \, m_{\infty}^{2} - 2 \, \mu \, \sigma^{2} \, m_{\infty} \, n_i \, \epsilon \, \chi_{\text{int}} \, c + c \mu^{2} \sigma^{2} n_i^{2} \chi_{\text{int}} + \sigma^{4} \, n_i^{2} \, \chi_{\text{int}}^{2} \, c \, (1 - \epsilon^{2})  \:\:\: .\label{eq:variance_sigma_final}
\end{eqnarray}

Now, to complete the procedure, one needs to rescale the average $\mu$ and the variance $\sigma^{2}$ of the couplings for $\langle \theta \rangle$ and $\langle \sigma^{2} \rangle$ to remain finite. The natural choice is

\begin{equation}
\hat{\mu} = c \, \mu \qquad , \qquad\qquad\qquad\qquad\qquad \hat{\sigma}^{2} = c \, \sigma^{2} \: \: \: .
 \label{eq:rescale_mu_sigma}
\end{equation}

After neglecting the terms that go to zero when $c \to \infty$, we finally obtain:

\begin{eqnarray}
 \langle \theta \rangle \!\!\!\! &=& \!\!\!\! \hat{\mu} \, m_{\infty}  - \hat{\sigma}^{2}  \, \epsilon \, n_i \, \chi_{\text{int}} \label{eq:rescaled_av_theta}  \\
 s_{\theta}^{2} \!\!\!\! &=& \!\!\!\! \hat{\sigma}^{2} \, m_{\infty}^{2} \label{eq:rescaled_av_theta_sqr} \:\:\: .
\end{eqnarray}

Remembering Eq. \eqref{eq:sol_IBMF_T0}, the stationary abundance for large connectivity and at $T=0$ was:

\begin{equation}
n_i = \max \Big( 0, 1 - \sum_{j \in \partial i^{-}} \alpha_{ij}\, m_{j}(\infty) \Big)  \equiv \max (0, \, 1 - \theta) = \max(0, \, 1 - \langle \theta \rangle - s_{\theta} \, \nu) \:\:\: .
\end{equation}
where $\nu \sim \mathcal{N}(0, 1)$ is a standardized Gaussian. Substituting Eqs. \eqref{eq:rescaled_av_theta} and \eqref{eq:rescaled_av_theta_sqr} into this expression, we get:

\begin{equation}
n_i = \max(0, \, 1 - \hat{\mu} \, m_{\infty}  + \hat{\sigma}^{2}  \, \epsilon \, n_i \, \chi_{\text{int}} - \hat{\sigma} \, m_{\infty} \, \nu ) \:\:\: .
\end{equation}

Rearranging terms, we finally recover Eq. (13) of \cite{Roy_2019}:

\begin{equation}
 n_i = \max \Big( 0, \, \frac{1 - \hat{\mu} \, m_{\infty} - \sigma \, \zeta_{\infty}}{1 - \epsilon \, \hat{\sigma}^{2} \, \chi_{\text{int}}} \Big) \: ,
\end{equation}
where $\zeta_{\infty}$ is a Gaussian with zero mean $\langle \zeta_{\infty} \rangle = 0$ and variance $\langle \zeta_{\infty}^{2} \rangle = m_{\infty}$. The latter is also consistent with the third line of Eq. (12) in Ref. \cite{Roy_2019}.

\section{Efficient implementation of IBMF} \label{sec:eff_IBMF}

To use IBMF to obtain the actual values of the averages $m_{i}$, one needs to numerically compute integrals of the form

\begin{equation}
I_{k}(\beta, \lambda, M) = \int_{0}^{\infty} \, dn \, n^{\beta \lambda - 1 + k} \, \exp \Big\{ -\frac{\beta}{2} \,(n - M)^{2} \Big\}
 \label{eq:integral_k_supp} \:\: ,
\end{equation}
with the parameter $k$ taking the value $k=0, 1$ in our case.

Luckily, the integral \eqref{eq:integral_k_supp} can be expressed in terms of known special functions, called parabolic cylinder functions. They have the integral representation (see 9.241 in Ref. \cite{Gradshteyn7ed}):

\begin{equation}
D_{-p}(z) = \frac{e^{-z^{2}/4}}{\Gamma(p)} \int_{0}^{\infty} dx \, e^{-xz - x^{2}/2} x^{p-1} \:\: ,
 \label{eq:def_D}
\end{equation}
which is valid for $\text{Re}[p]>0$ and where $\Gamma(-p)$ is the Euler's Gamma function. Looking back at Eq. \eqref{eq:integral_k_supp}, we can change variables making $x=n \sqrt{\beta}$ to get

\begin{equation}
I_{k} = \Big[ \frac{1}{\beta} \Big]^{\frac{\beta \lambda + k}{2}} \, e^{-\frac{\beta M^{2}}{2}} \int_{0}^{\infty} \, dx \, x^{\beta \lambda - 1 + k} \, \exp \Big\{ -\frac{x^{2}}{2} + \sqrt{\beta \, M^{2}} \, x \, M \Big\} 
 \label{eq:integral_k_cv} \:\:.
\end{equation}

After comparing Eqs. \eqref{eq:def_D} and \eqref{eq:integral_k_cv}, we identify $p=\beta \lambda + k$ and $z = -\sqrt{\beta \, M^{2}}$, and we can write the integral $I_k$ in terms of the parabolic cylinder functions as:

\begin{equation}
I_k(\beta, \lambda, M) = \Big[ \frac{1}{\beta} \Big]^{\frac{\beta \lambda + k}{2}} \exp\Big\{- \frac{\beta M^{2}}{4 } \Big\} \: \Gamma(\beta \lambda + k) \: D_{-\beta \lambda - k} \Big(- \sqrt{\beta M^{2}} \, \Big)
 \label{eq:integral_k_in_D} \:\: .
\end{equation}

Eq. \eqref{eq:integral_k_in_D} is very convenient since we can express the first moments of the distribution $P_{\infty} (n_j)$ in terms of these integrals. Indeed, one has:

\begin{equation}
 m_j  =  \frac{I_1(\beta, \lambda, M_{j})}{I_0(\beta, \lambda, M_{j})} = \sqrt{\beta } \, \lambda \, \frac{D_{-\beta \lambda - 1} \big(- \sqrt{\beta M_j^{2}} \, \big)}{D_{-\beta \lambda} \big(- \sqrt{\beta M_j^{2}} \, \big)} \label{eq:mji_D} \:\: ,
\end{equation}
where $M_{j}=1-\sum_{k \in \partial j^{-}} \alpha_{jk} \, m_k$.

Using again Ref. \cite{Gradshteyn7ed} (9.240), we can write the parabolic cylinder functions in terms of the more practical Kummer's confluent hypergeometric function $\Phi(a, b; z)$, which can be found already tabulated in different programming languages. The relation is:

\begin{eqnarray}
 & &D_{-\beta \lambda - k} \big(- \sqrt{\beta M^{2}} \, \big) = \frac{2^{-(\beta \lambda + k)/2} e^{-\beta M^{2}/4} \sqrt{\pi}}{\Gamma\Big(\frac{\beta \lambda + 1 + k}{2} \Big) \, \Gamma\Big(\frac{\beta \lambda + k}{2} \Big)} \times  \\
 & & \!\!\!\!\!\! \times \Big\{ \Gamma\Big(\frac{\beta \lambda + k}{2} \Big) \, \Phi \Big[\frac{\beta \lambda + k}{2}, \frac{1}{2}, \frac{\beta M^{2}}{2} \Big] + \sqrt{2 \beta M^{2}} \: \Gamma\Big(\frac{\beta \lambda + 1 + k}{2} \Big) \, \Phi \Big[\frac{\beta \lambda + 1 + k}{2}, \frac{3}{2}, \frac{\beta M^{2}}{2} \Big] \Big\} \nonumber
\end{eqnarray}

Then, the first moment can be written as:

\begin{equation}
 m_{j} = \frac{\Gamma\Big(\frac{\beta \lambda + 1}{2} \Big) \, \Phi \Big[\frac{\beta \lambda + 1}{2}, \frac{1}{2}, \frac{\beta M_{j}^{2}}{2} \Big] + \sqrt{\frac{\beta}{2}} \, M_{j} \, \beta \, \lambda \, \Gamma\Big(\frac{\beta \lambda}{2} \Big) \, \Phi \Big[1+\frac{\beta \lambda}{2}, \frac{3}{2}, \frac{\beta M_{j}^{2}}{2} \Big]}{\sqrt{\frac{\beta}{2}} \, \Gamma\Big(\frac{\beta \lambda}{2} \Big) \, \Phi \Big[\frac{\beta \lambda}{2}, \frac{1}{2}, \frac{\beta M_{j}^{2}}{2} \Big] + \beta M_{j} \, \Gamma\Big(\frac{\beta \lambda + 1}{2} \Big) \, \Phi \Big[\frac{\beta \lambda + 1}{2}, \frac{3}{2}, \frac{\beta M_{j}^{2}}{2} \Big]} \label{eq:up_m_hyper}
\end{equation}

We can then set an initial condition for all the $m_{i}$, with $i=1, \ldots, N$, and iterate Eq. \eqref{eq:up_m_hyper}. This is computationally fast, and the result of the iteration process depends on the relevant parameters of the model $(\mu, \sigma, \beta)$ and on the interaction graph. The code is available at Ref. \cite{GenLotkaVolterra_SparseGraphs}.

\section{Continuous BP equations}

The continuous BP equations for the Generalized Lotka-Volterra (gLV) model, as introduced in Eq. (13) in the main text, read:

\begin{equation}\label{eq:suppl-BP-continuous}
 \eta_{i \to j}(n_i) = \frac{1}{z_{i \to j}} \, n_i^{\beta \lambda - 1} \, \exp \Big\{-\frac{\beta}{2}(n_i^{2} - 2 n_i) \Big\} \prod_{k \in \partial i^{-} \setminus j} \left(\int_{0}^{\infty} dn_k \, \eta_{k \to i}(n_k) \exp \big\{ - \beta \alpha_{ik} n_i \, n_k \big\}\right) \:\: .
\end{equation}

Eq.~\eqref{eq:suppl-BP-continuous} is the update rule for the cavity marginals, or messages, $\eta_{i\to j}(n_i)$, which are the probability distributions for the abundance of species (node) $i$ once we cut the edges with species $j$. As usual, we call $\eta_{i\to j}(n_i)$ \textit{node-to-link} messages. What we immediately notice from Eq.~\eqref{eq:suppl-BP-continuous} is that $\eta_{i\to j}(n_i)$ contains the factor $n_i^{\beta\lambda-1}$, which diverges in $n_i=0$, for $\beta\lambda<1$. To avoid such divergence, we introduce the \textit{link-to-node} cavity messages, which we denote $\hat{\eta}_{i\to j}(n_j)$ and which represent the distribution of the abundance of species $j$ once all its links, except the one with $i$, are cut. These cavity marginals are defined as:
\begin{align}\label{eq:suppl-cav-factor-to-node}
   \hat\eta_{i \to j}(n_j) &\propto \int_{0}^{\infty} dn_i \, \eta_{i \to j}(n_i) \exp \big\{ - \beta \alpha_{ij} n_i \, n_j \big\} \:\:.
\end{align}

Inserting Eq.~\eqref{eq:suppl-BP-continuous} into Eq.~\eqref{eq:suppl-cav-factor-to-node}, we get:
\begin{align}\label{eq:suppl-hat-eta2}
 \hat\eta_{i \to j}(n_j)\propto \int_{0}^{\infty} dn_i \, n_i^{\beta \lambda - 1} \, \exp \Big\{-\frac{\beta}{2}(n_i^{2} - 2 n_i) \Big\} \exp \big\{ - \beta \alpha_{ij} n_i \, n_j \big\} \prod_{k \in \partial i^{-} \setminus j} \hat\eta_{k\rightarrow i}(n_i) \:,
\end{align}
which is now the update rule for the link-to-node cavity marginals $\hat\eta_{i \to j}(n_j)$.
Importantly, once the messages $\hat\eta_{i\rightarrow j}(n_j)$ are determined, the cavity marginals $\eta_{i\rightarrow j}(n_i)$ follow directly, as we can see by rewriting Eq.~\eqref{eq:suppl-BP-continuous} as:

\begin{equation}\label{eq:suppl-BP-continuous-with-hateta}
 \eta_{i \to j}(n_i) \propto n_i^{\beta \lambda - 1} \, \exp \Big\{-\frac{\beta}{2}(n_i^{2} - 2 n_i) \Big\} \prod_{k \in \partial i^{-} \setminus j} \hat\eta_{k\to i}(n_i).
\end{equation}
We can then analyze the BP convergence directly at the level of the link-to-node cavity messages $\hat\eta_{i\rightarrow j}(n_j)$. As said before, the advantage is that $\hat\eta_{i\rightarrow j}(n_j)$ is a well-behaved function that allows us to avoid the divergence of $\eta_{i\rightarrow j}(n_i)$ at $n_i=0$.

If we initialize all the messages $\hat\eta_{i \to j}(n_j)$ to the uniform distribution over $n_j$, the update rule~\eqref{eq:suppl-hat-eta2} in the first iteration step becomes:

\begin{align}\label{eq:suppl-1step}
    \hat\eta_{i\rightarrow j}(n_j)\propto\int_{0}^{\infty} dn_i \, n_i^{\beta \lambda - 1} \, \exp \Big\{-\frac{\beta}{2}(n_i^{2} - 2 n_i) \Big\} \exp \big\{ - \beta \alpha_{ij} n_i \, n_j \big\}.
\end{align}
The advantage of this initialization is that we can rewrite Eq.~\eqref{eq:suppl-1step} as
\begin{align}
    \hat\eta_{i\rightarrow j}(n_j)&\propto\int_0^\infty dn_i n_i^{\beta\lambda-1}\exp \Big\{-\frac{\beta}{2}n_i^2+\beta n_i(1-\alpha_{ij}n_j)\Big\}\label{eq:suppl-etahat-rewritten1}\\
    &=\frac{1}{\beta^{\frac{\beta\lambda}{2}}}\int_0^\infty dx\, x^{-p-1}\exp\Big\{-\frac{x^2}{2}-zx\Big\}\label{eq:suppl-etahat-rewritten2},
\end{align}
where in the last passage we introduced the change of variable $x=\sqrt{\beta}n_i$, with $p=-\beta\lambda$, $z=\sqrt{\beta}(\alpha_{ij}n_j-1)$. The integral in Eq.~\eqref{eq:suppl-etahat-rewritten2} can be exactly solved and expressed in terms of \textit{parabolic cylinder functions} $D_p(z)$. In particular, as it is shown in Eq.~9.241 in Ref.~\cite{Gradshteyn7ed}, the integral ($I$) is equivalent to:
\begin{align}
    I=\frac{\Gamma(-p)}{e^{-\frac{z^2}{4}}}D_p(z),
\end{align}
where $\Gamma(-p)$ is the Euler's Gamma function. The first iteration step for the update of the message $\hat\eta_{i\to j}(n_j)$ then corresponds to
\begin{align}\label{eq:suppl-init-cylinderfuncts}
    \hat\eta_{i\to j}(n_j) \propto\beta^{-\frac{\beta\lambda}{2}} \, \Gamma(\beta\lambda)\, e^{\frac{\beta}{4} (\alpha_{ij}n_j-1)^2}D_{-\beta\lambda}\left(\sqrt{\beta}(\alpha_{ij}n_j-1)\right),
\end{align}
which then has to be normalized.

For the following iteration steps, instead, we go back to Eq.~\eqref{eq:suppl-hat-eta2}, which we rewrite as
\begin{align}\label{eq:suppl-hat-eta3}
  \hat\eta_{i \to j}(n_j)\propto\exp\Big\{\frac{\beta}{2}(1-\alpha_{ij} n_j)^2\Big\} \int_{0}^{\infty} dn_i \, n_i^{\beta \lambda - 1} \, \exp \Big\{-\frac{\beta}{2}(n_i-1+\alpha_{ij}n_j)^2 \Big\} \prod_{k \in \partial i^{-} \setminus j} \hat\eta_{k\rightarrow i}(n_i).
\end{align}

In order to analyze the integral part of Eq.~\eqref{eq:suppl-hat-eta3}, which we will denote by $I_{i\rightarrow j}(n_j)$, let us split it at some $\delta>0$:
\begin{align}\label{eq:suppl-I-1}
  I_{i\rightarrow j}(n_j)&= \int_{0}^{\delta} dn_i \, n_i^{\beta \lambda - 1} \, \exp \Big\{-\frac{\beta}{2}(n_i-1+\alpha_{ij}n_j)^2 \Big\} \prod_{k \in \partial i^{-} \setminus j} \hat\eta_{k\rightarrow i}(n_i)+\nonumber\\
  &+ \int_{\delta}^{\infty} dn_i \, n_i^{\beta \lambda - 1} \, \exp \Big\{-\frac{\beta}{2}(n_i-1+\alpha_{ij}n_j)^2 \Big\} \prod_{k \in \partial i^{-} \setminus j} \hat\eta_{k\rightarrow i}(n_i).
\end{align}

For $\delta \ll 1$, the first contribution, i.e. the integral between $0$ and $\delta$, can be approximated as:
\begin{align}\label{eq:suppl-I-delta}
I^\delta_{i\rightarrow j}(n_j)=\exp \Big\{-\frac{\beta}{2}(1-\alpha_{ij}n_j)^2 \Big\}\prod_{k \in \partial i^{-} \setminus j} \hat\eta_{k\rightarrow i}(0) \frac{\delta^{\beta\lambda}}{\beta \lambda},
\end{align}
where we performed the integration of the factor $n_i^{\beta \lambda - 1}$ over the interval $[0,\delta]$, while approximating the remaining part of the integrand by its value at $n_i=0$.

Substituting Eqs.~\eqref{eq:suppl-I-1} and \eqref{eq:suppl-I-delta} into Eq.~\eqref{eq:suppl-hat-eta3}, we obtain the following BP update rule for the cavity messages $\hat\eta_{i\rightarrow j}$:
\begin{align}\label{eq:suppl-BP-etahat}
  \hat\eta_{i\rightarrow j}(n_j)&=\frac{1}{\hat{z}_{i\rightarrow j}}\Big[\frac{\delta^{\beta\lambda}}{\beta \lambda}\prod_{k \in \partial i^{-} \setminus j} \hat\eta_{k\rightarrow i}(0)+\nonumber\\
  &+\int_{\delta}^{\infty} dn_i \, n_i^{\beta \lambda - 1} \, \exp \Big\{-\frac{\beta}{2}(n_i^2-2n_i) \Big\} \exp\Big\{-\beta\alpha_{ij}n_i n_j\Big\} \prod_{k \in \partial i^{-} \setminus j} \hat\eta_{k\rightarrow i}(n_i)\Big].
\end{align}

After having \textit{initialized} the messages following Eq.~\eqref{eq:suppl-init-cylinderfuncts}, the update rule of the continuous BP equations is given by Eq.~\eqref{eq:suppl-BP-etahat}.

This approach is general, but we specifically used it in order to study the transition from a single-equilibrium phase (where BP converges) to a multiple-equilibria one (where BP does not converge) in the presence of homogeneous ($\sigma=0$) and competitive interactions ($\mu>0$). The results are in Subsection IV.C of the main text. In this case, the species abundances will only rarely be larger than the carrying capacities, which in our case are all equal to 1. Some thermal fluctuations could drive species to abundances slightly larger than $1$, but we can safely restrict the integral $\int_\delta^\infty$ in Eq.~\eqref{eq:suppl-BP-etahat} to the interval $[\delta,2]$. In our implementation, we set $\delta= 10^{-4}$ and compute the integral numerically.

Let us specify that in our analysis we used a sequential update, meaning that at each iteration step $k$, each $\hat\eta_{i\to j}^{(k)}(n_j)$ is updated to $\hat\eta_{i\to j}^{(k+1)}(n_j)$ asynchronously. In particular, the order of updates follows a random fixed sequence of directed edges $i \to j$. This sequential approach is essential to avoid convergence inconsistencies, ensuring that BP stops converging only once the multiple-equilibria phase is reached. These issues are also discussed in Appendix C of the main text. As for IBMF and numerical integration, the code is available at Ref.~\cite{GenLotkaVolterra_SparseGraphs}.

%\bibliographystyle{unsrt}
%\bibliography{ref_ecology_2025}
\printbibliography[heading=subbibliography]